\documentclass[11pt]{article}
\pdfoutput=1
\usepackage{jheppub}
\usepackage{graphicx, color, bm, dsfont, enumitem}
\usepackage{tikz,tikz-cd}
\usetikzlibrary{positioning, quotes, arrows}
\usepackage{mathdots}

\graphicspath{{./figures/}}

\newcommand{\be}{\begin{equation}}
\newcommand{\ee}{\end{equation}}
\newcommand{\bp}{\begin{pmatrix}}
\newcommand{\ep}{\end{pmatrix}}
\newcommand{\bsp}{\left(\begin{smallmatrix}}
\newcommand{\esp}{\end{smallmatrix}\right)}

\newcommand{\TNk}{{\CT_{N,\k}}}
\newcommand{\ie}{\emph{i.e.}}

\newcommand{\ol}{\overline}
\newcommand{\wt}{\widetilde}
\newcommand{\mb}{\mathbf}
\newcommand{\cp}{{\mathbb{CP}}}
\newcommand{\ds}{\displaystyle}

\newcommand{\ft}{{\mathfrak{t}}}
\newcommand{\fg}{{\mathfrak{g}}}
\newcommand{\ve}{\varepsilon}
\renewcommand{\k}{\mathbf{k}}

\newcommand{\R}{{\mathbb R}}

\newcommand{\C}{{\mathbb C}}
\newcommand{\Z}{{\mathbb Z}}

\newcommand{\CA}{{\mathcal A}}
\newcommand{\CB}{{\mathcal B}}

\newcommand{\CF}{{\mathcal F}}
\newcommand{\CG}{{\mathcal G}}
\newcommand{\CH}{{\mathcal H}}

\newcommand{\CM}{{\mathcal M}}
\newcommand{\CN}{{\mathcal N}}
\newcommand{\CO}{{\mathcal O}}

\newcommand{\CR}{{\mathcal R}}
\newcommand{\CS}{{\mathcal S}}
\newcommand{\CT}{{\mathcal T}}

\newcommand{\CW}{{\mathcal W}}

\title{Coulomb Branches of Star-Shaped Quivers}
\author[1]{Tudor Dimofte}
\author[2]{Niklas Garner}

\affiliation[1]{Department of Mathematics and Center for Quantum Mathematics and Physics (QMAP), UC Davis, One Shields Ave, Davis, CA 95616, USA}
\affiliation[2]{Department of Physics and QMAP, UC Davis, One Shields Ave, Davis, CA 95616, USA}

\abstract{We study the Coulomb branches of 3d $\CN=4$ ``star-shaped'' quiver gauge theories and their deformation quantizations, by applying algebraic techniques that have been developed in the mathematics and physics literature over the last few years. The algebraic techniques supply an abelianization map, which embeds the Coulomb-branch chiral ring into a vastly simpler abelian algebra $\CA$. Relations among chiral-ring operators, and their deformation quantization, are canonically induced from the embedding into $\CA$. In the case of star-shaped quivers --- whose Coulomb branches are related to Higgs branches of 4d $\CN=2$ theories of Class $\CS$ --- this allows us to systematically verify known relations, to generalize them, and to quantize them. In the quantized setting, we find several new families of relations.}

\begin{document}
\today
\maketitle


\section{Introduction}
\label{intro}

The Coulomb branches of 3d $\CN=4$ gauge theories have long been an object of physical and mathematical interest. Early physical studies \cite{Seiberg-IR, SeibergWitten-3d} led to the discovery of 3d mirror symmetry \cite{IntriligatorSeiberg, dB-MS1, dB-MS2}, and related the Coulomb branch of ADE quiver gauge theories to moduli spaces of monopoles and instantons \cite{ChalmersHanany, HananyWitten}. Unfortunately, non-perturbative corrections make the Coulomb branch difficult to analyze directly in non-abelian gauge theory.
(Calculations of instanton corrections in simple non-abelian theories were carried out in \emph{e.g.}  \cite{DKMTV, FraserTong}, but quickly became impractical.)
This difficulty was recently circumvented in a surprising confluence of physical \cite{CHZ-Hilbert, GMN, BDG2015, VV} and mathematical \cite{Teleman, Nak2016, BFN2016, BFN-quiver, Web2016} work, based on ideas from algebra, representation theory, and topological quantum field theory.

\begin{figure}[htb]
\centering
\includegraphics[width=4.5in]{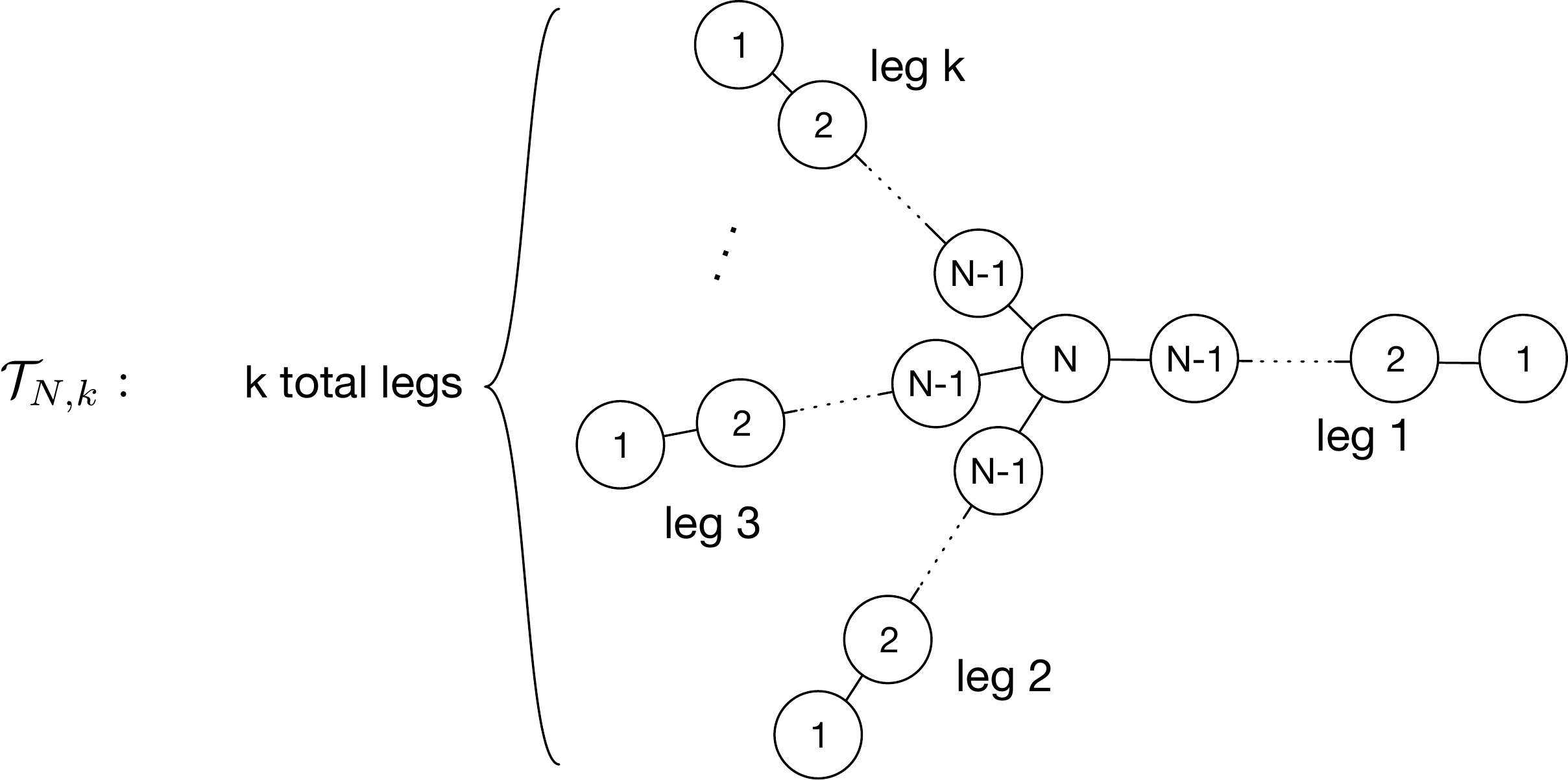}
\caption{The quiver gauge theory $\CT_{N,\k}$, associated with a star quiver with $\k$ legs of length $N$, has gauge group $\big( U(N)\times [U(N-1)\times...\times U(1)]^{\k}\big)/U(1)$, and hypermultiplet matter in a bifundamental representation for each edge in the quiver.}
\label{fig:TNk}
\end{figure}

In this paper, we will apply some of the recent physical and mathematical techniques to study the
Coulomb branch of star-shaped quiver (or simply ``star quiver'') gauge theories $\CT_{N,\k}$, shown in Figure~\ref{fig:TNk}.
This gives a new, concrete perspective on generators and relations for the $\CT_{N,\k}$ Coulomb branch chiral rings, supplementing known physical results and conjectures \cite{Gai09, MT11, BTW10, GMT, Yonekura-twisted, TNRev, HTY, LemosPeelaers, Tachi-review}, as well as the recent geometric analysis in \cite{GK-star, BFN-ring}.
We also explicitly construct natural deformation quantizations of the chiral rings.
\footnote{Another set of examples combining the power of recent Coulomb-branch techniques appeared in \cite{HananyMiketa}, wherein the authors studied balanced quivers of type A and D.}

The 3d theories $\TNk$ first came to prominence due to their relation \cite{BTX10} to 4d $\CN=2$ theories of Class $\CS$ \cite{Gai09, GMN}. 
Let 
 $\CT_N[\Sigma_{0,\k}]$ denote the 4d theory of Class $\CS$ defined by compactifying the 6d $(2,0)$ SCFT of type $A_{N-1}$ on a sphere with $\k$ maximal punctures, and let $\CT_N[\Sigma_{0,\k}\times S^1]$ denote its further compactification to three dimensions, a so-called ``Sicilian'' 3d $\CN=4$ theory.
It was argued by \cite{BTX10} that the star quiver theory $\CT_{N,\k}$ is the 3d mirror of $\CT_N[\Sigma_{0,\k}\times S^1]$ (in the limit of zero $S^1$ radius). This implies several relations among moduli spaces:
\begin{itemize}
\item The Coulomb branch $\CM_C$ of $\CT_{N,\k}$ is isomorphic to the Higgs branch $\CM_H^{\rm 4d}$ of $\CT_N[\Sigma_{0,\k}]$ as a complex symplectic manifold. In particular, the rings of holomorphic functions on the two moduli spaces (which are particular chiral rings of local operators in the supersymmetric QFT's) agree in every complex structure
\be \C[\CM_C] \simeq \C[\CM_H^{\rm 4d}]\,. \label{CH} \ee

\item As hyperk\"ahler manifolds endowed with a Riemannian metric, $\CM_C$ and $\CM_H^{\rm 4d}$ will generally differ. In particular, $\CM_C$ depends on dimensionful parameters --- the gauge couplings of the 3d quiver gauge theory --- while $\CM_H^{\rm 4d}$ does not. However, $\CM_C$ should be isomorphic to $\CM_H^{\rm 4d}$ in the infrared limit where all gauge couplings are sent to infinity.

\item Though not relevant for this paper, one also expects the Higgs branch of $\CT_{N,\k}$ (which is easy to identify from the quiver as the hyperk\"ahler quotient of $\k$ nilpotent cones in $\mathfrak{sl}(N,\C)$ by a diagonal $SU(N)$ isometry) to correspond to a particular decompactification limit of the Coulomb branch of $\CT_N[\Sigma_{0,\k}\times S^1]$ (which is a type-A Hitchin system on the $\k$-punctured sphere \cite{GMN-Hitchin}).

\end{itemize}

The 4d theories $\CT_N[\Sigma_{0,\k}]$ --- and in particular the ``trinion'' theory at $\k=3$, which was simply called $T_N$ in \cite{Gai09} --- are principal building blocks in the gluing construction of Class $\CS$ theories. Their Higgs branches $\CM_H^{\rm 4d}$ were conjecturally used to define a ``2d TQFT valued in holomorphic symplectic varieties'' in \cite{MT11}, now fully constructed by \cite{GK-star} and \cite{BFN-ring}.
However, despite their prominent role, it has been relatively difficult to analyze the Higgs branches of general theories  $\CT_N[\Sigma_{0,\k}]$ directly, because for $N\geq 3$ and $\k\geq 3$ these 4d theories are non-Lagrangian.

Some of what's known about the Higgs branches of $\CT_N[\Sigma_{0,\k}]$ includes their dimension
\be \text{dim}_\C \CM_H^{\rm 4d}=( \k N+2)(N-1)\,, \ee
and the existence of an $SU(N)^\k$ hyperk\"ahler isometry. For low $N$ and $\k$, one has \cite{Gai09, MN96, GNT09}
\be \label{lowNk} \begin{array}{l@{\quad}l} 
N=2,\k=3: & \CM_H^{\rm 4d} \simeq \C^8 \\[.2cm]
N=2,\k=4: & \CM_H^{\rm 4d} \simeq \{\text{minimal nilpotent orbit of $D_4$}\}\,,\\[.2cm] 
N=3,\k=3: & \CM_H^{\rm 4d} \simeq \{\text{minimal nilpotent orbit of $E_6$}\}\,, 
 \end{array} \ee
and for $N=2,3$ and general $\k \geq 3$ the Higgs branches can obtained by a gluing procedure \cite{MT11,GNT09}, as hyperk\"ahler reductions of products of \eqref{lowNk}. More generally, a putative set of generators and (partial) relations for the chiral rings $\C[\CM_H^{\rm 4d}]$ at $\k = 3$ and any $N\geq 3$ were uncovered in a series of papers \cite{BTW10, GMT, Yonekura-twisted, TNRev, HTY, LemosPeelaers}, nicely summarized in the review \cite{Tachi-review}. These putative generators admit a natural generalization to any $\k \geq 3,N \geq 3$.

One of our main motivations was to obtain new information about the structure of the Higgs branches $\CM_H^{\rm 4d}$ 
via a direct analysis of the corresponding Coulomb branches $\CM_C$ of $\CT_{N,\k}$, for general $N$ and $\k$.
The Hilbert series of the chiral ring $\C[\CM_C]\simeq \C[\CM_H^{\rm 4d}]$ was computed in~\cite{CHMZ-Hilbert} with this perspective in mind. However, one can now do much better, producing actual ring elements and relations among them.

To achieve this, we will follow the ``abelianization'' approach of \cite{BDG2015}, which corresponds to fixed-point localization in the equivariant (co)homology of  \cite{Nak2016, BFN2016, VV}.
The basic idea of abelianization is to embed the Coulomb-branch chiral ring $\C[\CM_C]$ of a non-abelian theory into a much larger --- but much simpler --- abelian algebra $\CA$
\be \C[\CM_C] \hookrightarrow \CA\,. \label{embed} \ee
In physical terms, $\CA$ is the local Coulomb-branch chiral ring near a generic point on the Coulomb branch, where the gauge group has been broken to its maximal torus.
The algebra $\CA$ has extremely simple generators and relations. Moreover, it has a simple Poisson structure, a simple deformation quantization, and a simple extension over twister space. Thus,  embedding $\C[\CM_C]\hookrightarrow \CA$ immediately allows one to
\begin{itemize}
\item verify relations among elements of $\C[\CM_C]$ (chiral ring relations)
\item identify the Poisson structure on $\C[\CM_C]$, and its deformation quantization
\item extend the algebra $\C[\CM_C]$ over twistor space,
and thereby access the hyperk\"ahler structure on $\CM_C$\,.
\end{itemize}
In the initial work \cite{BDG2015}, the precise image of the embedding \eqref{embed} was only identified in a handful of examples; however, at least in principle, a complete combinatorial construction of the image has since been described by Webster \cite{Web2016}.

In the case of $\CT_{N,\k}$ theories, we will identify the putative generators of $\C[\CM_C]$ proposed by \cite{BTW10, GMT, TNRev, Yonekura-twisted, HTY} (from a 4d Higgs-branch perspective) as elements of $\CA$. We will show how to explicitly verify and then quantize the conjectured relations among them.

An important insight in the derivation of $\C[\CM_H^{\rm 4d}]$ chiral-ring relations in \cite{TNRev} was that various generators could be ``diagonalized,'' as tensors for the $SU(N)^{\k}$ flavor symmetry. We find that the abelian algebra $\CA$ plays a surprisingly important role in this diagonalization. In particular, the eigenvalues of the generators, which are complicated algebraic functions on the actual moduli space $\CM_C\approx \CM_H^{\rm 4d}$, turn out to be extremely simple monomials in the algebra $\CA$. This allows the entire diagonalization procedure to be deformation-quantized.

From the perspective of 4d Higgs branches, the fact that the chiral ring $\C[\CM_H^{\rm 4d}]$ admits a deformation quantization may not be obvious.
However, this extra structure is completely natural (and physical) in 3d Coulomb branches. Indeed, in the recent mathematical/TQFT constructions of Coulomb branches \cite{BDG2015, VV, Nak2016, BFN2016, BFN-quiver, Web2016}, one typically works with quantized algebras from the very beginning. In physical terms, the Poisson structure in the chiral ring of 3d $\CN = 4$ theories arises from topological descent in the Rozansky-Witten twist \cite{BlauThompson, RozanskyWitten, descent}, and quantization comes from turning on an Omega background \cite{Nek03, Ya14}. (See also \cite{Beem-quant, Pufu-quant}. 
An analogous quantization arising from an Omega background in four dimensions is familiar from \cite{GuW06, DG-refined, NekShat, GMN-framed, DGOT}.)

We note that when $\k=1$ or $\k=2$, the expected relation between the Coulomb branch of $\CT_{N,\k}$ and the Higgs branch of $\CT_N[\Sigma_{0,\k}]$ breaks down. Neither the 3d nor the 4d theories are CFT's in this case. Nevertheless, the Coulomb branch of $\CT_{N,\k}$ is still a well-defined hyperk\"ahler manifold, in fact a smooth manifold. We will see explicitly that the Coulomb-branch chiral rings of $\CT_{N,\k}$ are consistent with
\be \label{k12} \begin{array}{l} \k =1\,:\quad  \CM_C \simeq T^*SL(N,\C)/\!/_\psi\,\mathfrak{N}\,,\\[.1cm]
\k =2\,:\quad  \CM_C \simeq T^* SL(N,\C)\,,\end{array}\ee
where $T^*SL(N,\C)/\!/_\psi\,\mathfrak{N}$ is the Kostant-Whittaker symplectic reduction of the cotangent bundle. 
The spaces in \eqref{k12} agree perfectly with those assigned to 1- and 2-punctured spheres by the Moore-Tachikawa TQFT \cite{MT11}.%
\footnote{Taking some care with scaling limits, the spaces \eqref{k12} can also be related to the Higgs branches of the 6d (2,0) theory compactified on one- or two-punctured spheres, even though they are not Higgs branches of 4d CFT's.}

One of our initial goals was to prove that the finite set of generators proposed by \cite{BTW10, GMT, TNRev, Yonekura-twisted, HTY, LemosPeelaers} really do generate the entire chiral ring $\C[\CM_C]=\C[\CM_H^{\rm 4d}]$. 
Unfortunately, this remains an open question.
It appears that identifying a \emph{finite} set of generators for the Coulomb-branch chiral ring of a nonabelian 3d theory is a rather difficult problem in general. It would be useful to develop methods to address this in the future.

\subsection{Other connections and future directions}

Our work is related to several other ideas that would be interesting to explore. For example:

\begin{enumerate}
\item In upcoming work \cite{GK-star}, Ginzburg and Kazhdan propose a geometric definition for the Higgs-branch chiral ring of 4d $\CT_N[\Sigma_{0,\k}]$ theories.
The proposal was shown in \cite{BFN-ring} to agree with the mathematical structure of the Coulomb branch in $\CT_{N,\k}$ theories. However, the proposed definition is not elementary: it involves the equivariant cohomology of a certain perverse sheaf over the affine Grassmannian for $SL(N,\C)$. It would be interesting to decipher how the relevant cohomology classes match the physically motivated generators and relations of $\C[\CM_C]$ discussed in \cite{BTW10, GMT, TNRev, Yonekura-twisted, HTY} and in this paper.

\item There are many expected relations among geometric structures on the Higgs and Coulomb branches of 3d $\CN=4$ theories --- for example, relations among cohomology rings as in Hikita's conjecture \cite{Hikita, Nak2016, Kam-Yang}, and symplectic duality of module categories associated to the Higgs and Coulomb chiral rings \cite{BLPW, BDGH}. It could be interesting to investigate the structure of these relations for star quivers.

\item The methods in this paper can be extended to 3d quiver gauge theories associated to other punctured spheres in Class $\CS$:  non-maximal punctures in type A, as well as various punctures in type D. 
The 4d $\CN=2$ theories obtained by gluing spheres with more general punctures (and in more general types) participate in an intricate web of dualities, \emph{cf}. \cite{ArgyresSeiberg, Gai09, GNT09, Tachi-D, CD-1, CD-2}; and our methods should allow a comparison of 4d Higgs branches across the dualities.

\item The deformation-quantization $\C_\ve[\CM_C]$ of the Coulomb-branch algebras of star quivers, which as explained above is natural in 3d, should define the basic building blocks for a quantized version of Moore-Tachikawa's ``TQFT valued in holomorphic symplectic varieties'' \cite{MT11}. In particular, one should find a TQFT that assigns a quantum algebra to any punctured 2d surface, with gluing implemented by quantum symplectic reduction.

\end{enumerate}

\subsection{Organization}

Section \ref{sec:rels} is a brief review of known and conjectured relations in $\mathbb{C}[\mathcal{M}_C]$ for the theories $\mathcal{T}_{N,\k}$, as well as a summary of our main results in this paper, including an explicit presentation of a set of ``diagonalized" operators in the abelianized algebra $\CA$ that are expected to produce all relations in $\mathbb{C}[\mathcal{M}_C]$. We present the quantization of these operators and their relations.

Section \ref{sec:review} reviews the general structure of Coulomb branch chiral rings in 3d $\mathcal{N} = 4$ gauge theories, and the algebraic techniques used to analyze them.  (In Appendix \ref{app:BFN} we connect to the mathematical approach of Braverman-Finkelberg-Nakajima.) Of particular interest is the abelianized algebra $\CA$ that contains the chiral ring, as in \eqref{embed}, and its quantization $\CA_\ve$. We also define a subalgebra $\CW_\ve\subset \CA_\ve$ that, due to Webster \cite{Web2016}, helps us characterize the image of the embedding \eqref{embed}.

We then consider some special families of $\TNk$ theories, building our way up to the general case, working almost exclusively with quantum algebras.

Section \ref{sec:2kquivers} analyzes ``small'' star quivers with $N=2$ and arbitrarily many legs. For $\k=3$ legs we observe how the simple Coulomb branch $T^*\C^4$ ( =  the familiar Higgs branch of the $T_2$ trinion theory) is recovered. For all $\k$ we identify the moment maps for the $SU(2)^{\k}$ flavor symmetry and a collection of operators furnishing a $\k$-fold fundamental representation of $SU(2)^{\k}$. The simultaneous diagonalization of the moment maps and the $\k$-fold fundamental, accessible via the the quantum abelianized algebra $\CA_\ve$, results in a dramatic simplification of expressions.

Section \ref{N1quivers} considers the complementary family of linear quivers with $\k=1$ leg but with $N$ arbitrary. A new feature here is the appearance of antisymmetric tensors of the $SU(N)$ flavor symmetry.  We find that the change of basis that diagonalizes the $SU(N)$ moment map vastly simplifies the antisymmetric tensors.

In Section \ref{Nkquivers} we then generalize to arbitrary $\TNk$ star quivers. Many properties of their Coulomb branches may be inferred by combining the results of the previous two sections. In particular, by working with diagonalized operators, relations among moment maps and antisymmetric tensors are easy to determine from the one-legged $\k=1$ analysis.

We conclude with two short Sections \ref{sec:geom12}, \ref{TNrelations} that connect our general results with some important and well-studied examples. Namely, we explain how our characterization of chiral rings for $\k=1$ and $\k=2$ quivers relates to the geometric spaces \eqref{k12} (Kostant-Whittaker reduction and the cotangent bundle of $SL(N,\C)$); and for $\k=3$ we discuss the generalizations we have found of chiral-ring relations in 4d $\CN=2$ $T_N$ trinion theories.

In the appendices, in addition to collecting various computations and interesting examples of quantum chiral-ring relations, we include a summary of the BFN construction $\C_\ve[\CM_C]$ \cite{Nak2016, BFN2016}, from the perspective of physical TQFT.

\section{Summary of results}
\label{sec:rels}

Before delving into the algebraic analysis of 3d $\CN=4$ Coulomb branches, 
we whet the reader's appetite with some results.
We review known and conjectured relations in the chiral rings of $\CT_{N,\k}$ theories for $\k=3$. Then we summarize the general structure found in this paper for arbitrary $\k$, including quantum generalizations of known relations, and a handful of new relations that only appear upon quantization. It is believed that the operators discussed below generate the entire chiral ring, though this has not been proven (and we do not offer any additional proof that this is the case).
It is also still unknown, in general, whether the relations discussed below are complete.

\subsection{$\k=3$, $T_N$ theories}

Much is known about the Coulomb-branch chiral rings of the three-legged $\CT_{N,3}$ quiver gauge theories, due to their relation to the ``trinion'' theories $T_N = \CT_N[\Sigma_{0,3}]$ of Class $\CS$ \cite{BTW10, GMT, TNRev, Yonekura-twisted, HTY, LemosPeelaers, Tachi-review}. 

The quiver gauge theory $\CT_{N,3}$ has an $SU(N)^3$ flavor symmetry acting on the Coulomb branch, which induces a holomorphic $SL(N,\C)^3$ action on the chiral ring $\C[\CM_C]$. As reviewed further below in Section \ref{sec:flavor}, this means that there must exist a triplet of complex moment map operators in the chiral ring,
\be \mu_a \in \mathfrak{sl}(N,\C)^*\,,\qquad a=1,2,3\,. \label{mu-intro} \ee
We denote the components of the moment maps as $(\mu_a)^i{}_j$.
Index considerations suggest that the entire chiral ring is generated by the components of the moment maps as well as a collection of operators
\be Q_{(r)} \in \wedge^r\square\otimes \wedge^r\square\otimes \wedge^r\square\,,\qquad Q^{(r)} \in (\wedge^r\square)^*\otimes (\wedge^r\square)^*\otimes (\wedge^r\square)^* \qquad r=1,2,...,N-1 \label{Q-intro} \ee
in the $r$-th antisymmetric tensor representations of $SL(N,\C)^3$ and their duals. We denote the components of $Q_{(r)}$ and $Q^{(r)}$ as $Q_{(r)}^{IJK}=Q^{[i_1...i_r][j_1...j_r][k_1...k_r]}$ and $Q^{(r)}_{IJK}=Q_{[i_1...i_r][j_1...j_r][k_1...k_r]}$, respectively. (We often drop the `$(r)$' when the choice of representation is unambiguous.) The $Q_{(r)}$ and $Q^{(r)}$ are not independent, obeying
\be \frac{1}{(N-r)!^3}\, \epsilon_{i_1...i_N}\epsilon_{j_1...j_N}\epsilon_{k_1...k_N}\,  Q^{[i_{r+1}...i_N][j_{r+1}...j_N][k_{r+1}...k_N]}
 = Q_{[i_{1}...i_r][j_1...j_r][k_1...k_r]}\,, \ee
or more succinctly $\frac{1}{(r!)^3}\, \epsilon_{II'}\epsilon_{JJ'}\epsilon_{KK'}\, Q_{(r)}^{I'J'K'} =Q^{(N-r)}_{IJK}$. Here $\epsilon$ is the totally antisymmetric tensor of $SL(N,\C)$, normalized so that $\epsilon^{12...N}=\epsilon_{12...N}=1$.

The chiral ring is also graded by charge under a $U(1)$ subgroup of the $SU(2)$ R-symmetry acting on the Coulomb-branch. For a CFT, this R-charge coincides with dimension. The R-charges of the above generators are
\be [\mu_a]=1\,,\qquad  [Q_{(r)}]=[Q^{(r)}] = \tfrac12 r(N-r)\,. \label{Rcharge}     \ee

The most important nontrivial relations among the generators are
\be \text{Tr}[(\mu_1)^n] =  \text{Tr}[(\mu_2)^n] =  \text{Tr}[(\mu_3)^n]\qquad n=2,...,N\,, \label{Tr} \ee
\be \begin{array}{c} (\mu_1)^i{}_{i'}Q^{i'jk}=(\mu_2)^j{}_{j'}Q^{ij'k}=(\mu_3)^k{}_{k'}Q^{ijk'}\,, \\[.1cm]
 (\mu_1)^{i'}{}_{i}Q_{i'jk}=(\mu_2)^{j'}{}_{j}Q_{ij'k}=(\mu_3)^{k'}{}_{k}Q_{ijk'}\,,
\end{array}  \label{muQ}
\ee
and more generally
\be \begin{array}{c} (\mu_1)^{[i_1}{}_{i'}Q^{[i']i_2...i_r]JK}=
(\mu_2)^{[j_1}{}_{j'}Q^{I[j']j_2...j_r]K} =
(\mu_3)^{[k_1}{}_{k'}Q^{IJ[k']k_2...k_r]}\,, \\[.1cm]
(\mu_1)^{i'}{}_{[i_1}Q_{[i']i_2...i_r]JK}=
(\mu_2)^{j'}{}_{[j_1}Q_{I[j']j_2...j_r]K} =
(\mu_3)^{k'}{}_{[k_1}Q_{IJ[k']k_2...k_r]}\,.
\end{array}  \label{muQI}
\ee

The first relation \eqref{Tr} says that all the Casimir operators built from the moment maps are equal. This implies that at generic points on the Coulomb branch, where the moment maps can be diagonalized, the eigenvalues of $\mu_1$, $\mu_2$, and $\mu_3$ will all coincide.
It helps to be somewhat explicit about this: at generic points on the Coulomb branch there should exist three invertible matrices 
 $(S_a)^i{}_j$ such that
\be S_a\mu_a S_a^{-1} = \text{diag}(m_1,...,m_N) \ee
for all $a=1,2,3$, where $(m_1,...,m_N)$ is the common set of eigenvalues, satisfying $\sum_im_i=0$.

The second pair of relations \eqref{muQ} implies that at generic points on the Coulomb branch all the tri-fundamental and tri-antifundamental $Q$'s can be diagonalized, by the same similarity transformation that diagonalizes the moment maps. In other words
\be (S_1)^i{}_{i'}(S_2)^j{}_{j'}(S_3)^k{}_{k'}Q^{i'j'k'} = \begin{cases} q^i & i=j=k \\
0 & \text{otherwise}\,,\end{cases}  \ee
and 
\be Q_{i'j'k'}(S_1^{-1})^{i'}{}_i(S_2^{-1})^{j'}{}_j(S_3^{-1})^{k'}{}_k = \begin{cases} q_i & i=j=k \\
0 & \text{otherwise}\,,\end{cases} \ee
for some ``eigenvalues'' $q^i$ and $q_i$. Due to \eqref{muQI}, the remaining $Q_{(r)}$, $Q^{(r)}$ operators can be simultaneously diagonalized exactly the same way, with eigenvalues that we denote $q_{(r)}^I=q^{[i_1...i_r]}$ and $q^{(r)}_I=q_{[i_1...i_r]}$, respectively. For example,
\be \hspace{-.25in} (S_1)^{[i_1}{}_{[i_1'}(S_1)^{i_2]}{}_{i_2']}(S_2)^{[j_1}{}_{[j_1'}(S_2)^{j_2]}{}_{j_2']}(S_3)^{[k_1}{}_{[k_1'}(S_3)^{k_2]}{}_{k_2']}Q^{[i_1'i_2'][j_1' j_2'][k_1' k_2']} = \begin{cases} q^{[i_1i_2]} & [i_1i_2]=[j_1j_2]=[k_1k_2] \\
0 & \text{otherwise}\,,\end{cases}
\ee
where $[ij]=[kl]$ means these pairs of indices agree modulo the action of the symmetric group.

From the diagonalized perspective, all the information in the chiral ring has been repackaged in the eigenvalues $m_i,q^I,q_I$ and the three similarity transformations $S_a$. Relations in this algebra, upon removing the diagonlization, lead to relations amongst the operators $\mu_a, Q^{(r)}$, and $Q_{(r)}$ and if the algebra of eigenvalues is sufficiently simple then this diagonalization could serve as a convenient avenue for finding chiral-ring relations. This approach is discussed in detail in \cite{HTY,TNRev} and will serve as a motivating principle in much of our analysis for the more complex theories $\TNk$.

The remaining known chiral-ring relations may be found in \cite{TNRev, HTY, LemosPeelaers}. They come in two basic types, contractions that relate $Q^{IJK}Q_{I'J'K}$ to a product of moment maps; and equivalences among products of tri-fundamentals $(Q_{(1)})^r$ and the higher anti-symmetric powers $Q_{(r)}$.
The two simplest relations are
\be Q^{ijk}Q_{i'j'k} = (-1)^N\sum_{\ell=0}^{N-1}c_\ell \sum_{n=0}^{N-\ell-1} (\mu_1^{N-\ell-n-1})^i{}_{i'}(\mu_2^{n})^j{}_{j'}\,, \label{contrac-intro} \ee
where $c_\ell$ are coefficients of the characteristic polynomial $P(x) = \det(x\mb{1}-\mu_a) = \sum_{\ell=0}^N c_\ell x^{N-\ell}$ (due to \eqref{Tr}, these are independent of the choice of $a=1,$ 2, or 3); and\footnote{These expressions agree with (2.8) and (2.9) in \cite{TNRev} upon substituting $Q_{ijk} \mapsto - Q_{ijk}$.} 
\be  \label{Qprod-intro}
\begin{array}{l}
		\frac{1}{(N-1)!}\, Q^{i_1 j_1 k_1} Q^{i_2 j_2 k_2} \dots Q^{i_{N-1} j_{N-1} k_{N-1}} \epsilon_{j_1 j_2 \dots j_{N-1} j} \epsilon_{k_1 k_2 \dots k_{N-1} k}\\ 
		\hspace{1.5in}=  -Q_{i j k} (\mu_1^0)^{(i_1}{}_{i_1'}  (\mu_1)^{i_2}{}_{i_2'}\dots  (\mu_1^{N-2})^{i_{N-1})}{}_{i_{N-1}'} \epsilon^{i_1' i_2' \dots i_{N-1}' i}\,,   \\[.2cm]
		\frac{1}{(N-1)!}\, Q_{i_1 j_1 k_1} Q_{i_2 j_2 k_2} \dots Q_{i_{N-1} j_{N-1} k_{N-1}} \epsilon^{j_1 j_2 \dots j_{N-1} j} \epsilon^{k_1 k_2 \dots k_{N-1} k}\\ 
		\hspace{1.5in} =(-1)^{\frac{(N+1)(N-2)}{2}}  Q^{i j k} (\mu_1^0)^{i_1'}{}_{(i_1}  (\mu_1)^{i_2'}{}_{i_2}\dots (\mu_1^{N-2})^{i_{N-1}'}{}_{i_{N-1})} \epsilon_{i_1' i_2' \dots i_{N-1}' i}\,.
\end{array}
\ee
A more general version of \eqref{contrac-intro} appears in \cite[App. A]{HTY}. A generalization of~\eqref{Qprod-intro} was discovered%
\footnote{This was found by studying 2d chiral algebras embedded in 4d trinion theories \cite{chiral1,chiral2}, which generalize the 4d Higgs-branch chiral ring in a different, extremely interesting way.} %
by \cite{LemosPeelaers} for $N=4$, and extended to all $N$ in \cite{Tachi-review}: $Q^{(i_1[j_1[k_1}Q^{i_2)j_2]k_2]} = -\frac{1}{2} (\mu_1)^{(i_1}{}_{i_1'} Q^{[i_2)i_1'][j_1j_2][k_1k_2]}$ and similarly for the antifundamentals. This can be written more suggestively as
 \be \begin{array}{l} 2!\, Q^{(i_1[j_1[k_1}Q^{i_2)j_2]k_2]} = - \delta^{(i_1}{}_{i_1'} (\mu_1)^{i_2)}{}_{i_2'} Q^{[i_1' i_2'][j_1j_2][k_1k_2]}  \\[.2cm]
2!\, Q_{(i_1[j_1[k_1}Q_{i_2)j_2]k_2]} = Q_{[i_1' i_2'][j_1j_2][k_1k_2]} \delta^{i_1'}{}_{(i_1}(\mu_1)^{i_2'}{}_{i_2)}\,.  \label{LP-intro} \end{array}  \ee

In App. \ref{tensorcomps} we provide a list of miscellaneous relations computed for small $N, \k$ which include variants of \eqref{Qprod-intro} for different rank tensors as well as variants of \eqref{LP-intro} for $N \neq 4$. 
We use these computations to predict relations of the very general form
\be \label{LPgen-intro}\begin{array}{l} r! \,Q^{(i_1[j_1[k_1}Q^{i_2j_2k_2} ... Q^{i_r)j_r]k_r]}= \pm  \delta^{(i_1}{}_{i_1'} (\mu_1)^{i_2}{}_{i_2'} ... (\mu_1^{r-1})^{i_r)}{}_{i_r'} Q^{[i_1' i_2' ... i_r'] [j_1 j_2 ... j_r] [k_1 k_2... k_r]} \\[.2cm]
r!\, Q_{(i_1[j_1[k_1}Q_{i_2 j_2 k_2} ... Q_{i_r)j_3]k_r]} = \pm  Q_{[i_1' i_2' ... i_r'] [j_1 j_2 ... j_r] [k_1 k_2 ... k_r]} \delta^{i_1'}{}_{(i_1}(\mu_1)^{i_2'}{}_{i_2} ... (\mu_1^{r-1})^{i_r'}{}_{i_r)} \end{array}  \ee
for any $N$ and $1\leq r\leq N$.

Near generic points on the Coulomb branch where diagonalization is possible, it has been conjectured \cite{Tachi-review} that all possible relations reduce to
\begin{subequations} \label{qrels}
\be \frac{1}{(N-r)!}\, \epsilon_{i_1...i_N}q^{[i_{r+1}...i_N]} = q_{[i_1...i_r]}\, \sigma \,,  \label{qe} \ee
\be q^{i_1}q^{i_2}\cdots q^{i_r} = q^{[i_1...i_r]} \prod_{1 \leq n < m \leq r}(m_{i_n}-m_{i_m})\,,\ee
 \be q_{i_1}q_{i_2}\cdots q_{i_r} = q_{[i_1...i_r]} \prod_{1 \leq n < m \leq r}(m_{i_n}-m_{i_m})\,,  \ee
\be q^i q_i = \prod_{j\neq i}(m_i-m_j) \qquad \text{(for any fixed $i$)}\,, \ee
\end{subequations}
where $\sigma = \det(S_1)\det(S_2)\det(S_3)$ is the product of determinants of the similarity matrices. Typically it is assumed that $\det(S_a)=1$, though we will find it convenient to keep the determinants generic.

As mentioned in the introduction, the construction of the 3d Coulomb-branch chiral ring $\C[\CM_C]$ will proceed by embedding the ring into an abelianized algebra $\CA$, which is much larger but has canonical generators and relations (see Section \ref{sec:A}). Somewhat miraculously, we will be able to find explicit similarity transformations $S_a$ whose entries belong to $\CA$, and eigenvalues $m_i,q^i,q_i$ that are simple monomials in $\CA$. 

\subsection{Extending to general $\k$, and quantizing}
\label{sec:summary-q}

For $\k=3$ we will verify that there exist operators in the Coulomb-branch chiral ring of $\CT_{N,3}$ with all the expected properties and relations outlined above. More so, we will extend the (putative) generators and relations to general $\k$, and deformation-quantize them.
The key to extending the analysis to general $\k$ rests in implementing a uniform diagonalization procedure.

Recall that the \emph{quantized} Coulomb-branch chiral ring $\C_\ve[\CM_C]$ of $\CT_{N,\k}$ is an associative algebra. 
For general $\k$, 
it has an $SL(N,\C)^\k$ symmetry generated by taking commutators with moment maps $\{\mu_a\}_{a=1}^\k$.%
\footnote{In the quantum setting, saying a symmetry is generated by the moment maps means that the infinitesimal action of a Lie algebra generator $T$ on any operator $\CO \in \C_\ve[\CM_C]$ is given by the commutator $\frac{1}{\ve}[\langle T,\mu_a\rangle,\CO]$. As $\ve\to 0$, this reduces to a Poisson bracket. See Section \ref{sec:review} for more details.} %
In analogy with $\k=3$ above, we also prove that there exist operators $Q_{(r)}^{I_1...I_\k}$ and $Q^{(r)}_{I_1...I_\k}$ in $\C_\ve[\CM_C]$ that transform in the $\k$-fold anti-symmetric tensor representations and their duals,
\be \begin{array}{l}
  Q_{(r)} \in \wedge^r\square_1\otimes \wedge^r\square_2\otimes\cdots\otimes \wedge^r\square_\k\, \\[.2cm]
  Q^{(r)} \in (\wedge^r\square_1)^*\otimes (\wedge^r\square_2)^*\otimes\cdots\otimes (\wedge^r\square_\k)^*\, 
\end{array}\qquad  r=1,2,...,N-1\,. \label{Q-k-intro} \ee
Generalizing the expectations from $\k = 3$, we strongly suspect (but do not prove) that the moment maps and the $Q_{(r)}$, $Q^{(r)}$ are a complete set of generators.

Some of the basic relations among these operators are easy to guess and to verify, even in the quantum setting. 
In particular, if we define $\ve$-shifted moment map operators $\tilde\mu_a = \mu_a-\big(\tfrac{N-1}{2}\ve\big)\mathds{1}$, the relations \eqref{Tr}--\eqref{muQI} become
\be \text{Tr}[(\tilde{\mu}_1)^n] =  \text{Tr}[(\tilde{\mu}_2)^n] = \ldots = \text{Tr}[(\tilde{\mu}_\k)^n]\qquad n=1,...,N\,, \label{quant-Tr-k} \ee \be \begin{array}{c} (\tilde{\mu}_1)^{[i_{1,1}}{}_{i'}Q^{[i']i_{1,2}...i_{1,r}]I_2...I_\k}=\ldots=
	(\tilde{\mu}_\k)^{[i_{\k,1}}{}_{i'}Q^{I_1...I_{\k-1}[i']i_{\k,2}...i_{\k,r}]}\,, \\[.1cm]
	(\tilde{\mu}_1)^{i'}{}_{[i_{1,1}}Q_{[i']i_{1,2}...i_{1,r}]I_2...I_\k}=\ldots=
	(\tilde{\mu}_\k)^{i'}{}_{[i_{\k,1}}Q_{I_1...I_{\k-1}[i']i_{\k,2}...i_{\k,r}]}\,.
\end{array}  \label{quant-muQI-k}
\ee
Thus, in a suitable quantum sense, the ``eigenvalues'' of the $\mu_a$ are independent of the choice of leg, and we may expect to diagonalize the moment maps and all the $Q_{(r)}$, $Q^{(r)}$ operators simultaneously.

We explicitly perform the diagonalization by constructing similarity transformations $S_a$, one for each leg. The entries of the $S_a$ and their inverses $S_a^{-1}$ take values in the \emph{abelianized} quantum algebra $\CA_\ve$, which contains operators that exist at generic points of the Coulomb branch. Applying the similarity transformations in the right order we obtain
\be S_a\tilde \mu_a S_a^{-1} = \text{diag}(m_1,...,m_N) \qquad\text{(independent of $a=1,...,\k$)} \ee
with $\sum_i m_i=0$, as well as 
\be
\begin{array}{l} (S_1)^{i_1}{}_{j_1} (S_2)^{i_2}{}_{j_2} ... (S_{\k})^{i_{\k}}{}_{j_{\k}} Q^{j_1 j_2 ... j_{\k}} = \delta^{i_1}{}_i \delta^{i_2}{}_i ... \delta^{i_{\k}}{}_i q^i\,, \\[.1cm]
Q_{j_1 j_2 ... j_{\k}} (S_1^{-1})^{j_1}{}_{i_1} (S_2^{-1})^{j_2}{}_{i_2} ... (S_{\k}^{-1})^{j_{\k}}{}_{i_{\k}} = \delta^i{}_{i_1} \delta^i{}_{i_2} ... \delta^i{}_{i_{\k}} q_i\,.
\end{array}
\ee
The eigenvalues $m_i$, $q^i$ and $q_i$ are again elements of the abelianized operator algebra $\CA_\ve$. All the higher antisymmetric tensors $Q_{(r)}$ and $Q^{(r)}$ get diagonalized in a similar way, with eigenvalues $q^{[i_1...i_r]},q_{[i_1...i_r]}\in \CA_\ve$.

The eigenvalues $m_i,q^I,q_I$ generate an especially simple quantum algebra. Commutation with the moment-map eigenvalues measures charges of the $q^i$ and $q_i$,
\be [m_i,m_j]=0\,,\qquad [m_i,q^j] = (\delta_i^j-\tfrac1N)\ve q^j\,,\qquad [m_i,q_j] = -(\delta_{ij}-\tfrac1N)\ve q_j\,, \ee
while products of fundamental $q$'s are related to higher tensor powers via
\begin{subequations}\label{qrels-q}
\be q^{i_1} q^{i_2} ... q^{i_r} = (-1)^{r+1} \bigg[\prod \limits_{1 \leq n < m \leq r} \frac{(m_{i_m} - m_{i_n})^{\k-1}}{(m_{i_n} - m_{i_m} - \varepsilon)}\bigg] q^{[i_1 ... i_r]}\,, \label{qprod-q} \ee
\be q_{i_1} q_{i_2} ... q_{i_r} = (-1)^{(N-1)(r+1)} \bigg[ \prod \limits_{1 \leq n < m \leq r} \frac{(m_{i_n} - m_{i_m} + \ve)^{\k-1}}{(m_{i_m} - m_{i_n})}\bigg] q_{[i_1 ... i_r]}\,. \label{qprod-q2}  \ee
In addition, we find
\be \frac{1}{(N-r)!}\, \Bigg\{ \begin{array}{c c} \epsilon_{i_1...i_N} q^{[i_{r+1}...i_N]} & \k \textrm{ odd}\\  \Delta_{i_1...i_N}  q^{[i_{r+1}...i_N]}& \k \textrm{ even} \end{array}\Bigg\} = q_{[i_1...i_r]} \sigma\,, \ee
\be q^i q_i =  (-1)^N \prod_{j\neq i}\frac{(m_i-m_j)^{\k-1}}
  {m_i-m_j-\ve}\,,\qquad q_i q^i = (-1)^N\prod_{j\neq i}\frac{(m_i-m_j+\ve)^{\k-1}}
  {m_i-m_j}
 \qquad \text{(for any fixed $i$)}\,, \ee
 \end{subequations}
where  $\Delta_{i_1 ... i_N} = -|\epsilon_{i_1 i_2... i_N}|$ is a fully symmetric $N$-index tensor, and we have also introduced $\sigma = \det S_1\det S_2... \det S_\k =\Big[\prod_{1\leq i<j\leq N} (m_i-m_j)^{1-\k}(m_j-m_i+\ve)\Big] q^1q^2\cdots q^N$, which in our conventions is a nontrivial operator.%
\footnote{The element $\sigma$ is almost central in the algebra of eigenvalues. It may also be written as $\sigma = w_+ \prod_{1\leq i<j\leq N} (m_i-m_j)^{-\k}$, where $w_+ := \Big[\prod_{1\leq i<j\leq N} (m_i-m_j)(m_j-m_i+\ve)\Big] q^1q^2\cdots q^N$ and  $w_- := (-1)^N\Big[\prod_{1\leq i<j\leq N} (m_i-m_j+\ve)(m_j-m_i)\Big]^{1-\k} q_1q_2\cdots q_N$ are both central, with $w_+w_-=1$.} %

Note that the set of relations \eqref{qrels-q} reduce to the classical expressions in \eqref{qrels} as $\ve \rightarrow 0$, for $\k=3$. Moreover, they are consistent with the R-charge assignments
\be \label{rcharge} \begin{array}{l} [\mu_a]=[m_i] = [\ve]=1\,,\\[.1cm] [Q_{(r)}]=[q_{(r)}]=[Q^{(r)}]=[q^{(r)}] = \tfrac12(\k-2)r(N-r)\,,\\[.1cm]      [\sigma]=0\,.\end{array} \ee
We also emphasize that the $q$'s do not commute with each other for generic $\k$ (though the $q_i$ do commute amongst themselves when $\k = 1$). Instead, their commutators are determined by \eqref{qprod-q} and \eqref{qprod-q2}. For example,
\be [q^{i_1}, q^{i_2}] = -\frac{2 \ve(m_{i_2}-m_{i_1})^{\k-1}}{(m_{i_1}-m_{i_2})^2-\ve^2}q^{[i_1 i_2]}\,.\ee

In principle, relations among the actual chiral-ring operators $\mu_a,Q_{(r)}, Q^{(r)}$ could be obtained by judiciously applying $S_a$ and $S_a^{-1}$ to ``un-diagonalize'' the simple relations \eqref{qrels-q} above. In practice, this is quite difficult to do --- it is known to even be difficult for $\k=3$, in the classical $\ve\to 0$ limit. Nevertheless, we do identify a handful of nontrivial un-diagonalized relations. 

For example, the quantum version of  \eqref{contrac-intro} for $\k=3$ is a straightforward generalization
\begin{equation}
	Q^{i j k} Q_{i' j' k} = (-1)^N \sum\limits_{\ell=0}^{N-1} c_{\ell} \sum \limits_{m = 0}^{N-l-1} (\tilde{\mu}_1^{N-l-m-1})^i_{i'} (\tilde{\mu}_2^{m})^j_{j'}\,.
	\label{quant-contrac-intro}
\end{equation} 
We prove that this quantum relation holds in Section \ref{TNrelations}. Similarly, the quantum version of the $\k=3,N=4$ relation \eqref{LP-intro} is obtained by replacing $\mu_1$ with $\tilde{\mu}_1$: \be \begin{array}{l} 2!\, Q^{(i_1[j_1[k_1}Q^{i_2)j_2]k_2]} = - \delta^{(i_1}{}_{i_1'} (\tilde{\mu}_1)^{i_2)}{}_{i_2'} Q^{[i_1' i_2'][j_1j_2][k_1k_2]}  \\[.2cm]
2! \,Q_{(i_1[j_1[k_1}Q_{i_2)j_2]k_2]} = Q_{[i_1' i_2'][j_1j_2][k_1k_2]} \delta^{i_1'}{}_{(i_1}(\tilde{\mu}_1)^{i_2'}{}_{i_2)} \end{array}.  \ee We find evidence that, for general $N$ and $r$, the quantized version of \eqref{LPgen-intro} is given by \be \label{LPgenquant-intro}\begin{array}{l} r! Q^{(i_1[j_1[k_1}Q^{i_2j_2k_2} ... Q^{i_r)j_r]k_r]}= \pm \delta^{(i_1}{}_{i_1'} (\tilde{\mu}_1)^{i_2}{}_{i_2'} ... (\tilde{\mu}_1^{r-1})^{i_r)}{}_{i_r'} Q^{[i_1' i_2' ... i_r'] [j_1 j_2 ... j_r] [k_1 k_2... k_r]} \\[.2cm]
r! Q_{(i_1[j_1[k_1}Q_{i_2 j_2 k_2} ... Q_{i_r)j_3]k_r]} = \pm Q_{[i_1' i_2' ... i_r'] [j_1 j_2 ... j_r] [k_1 k_2 ... k_r]} \delta^{i_1'}{}_{(i_1}(\tilde{\mu}_1)^{i_2'}{}_{i_2} ... (\tilde{\mu}_1^{r-1})^{i_r'}{}_{i_r)} \end{array}  \ee

Some further relations among the $Q$'s and $\mu$'s for small $N$ and $\k$, generalizing the products \eqref{Qprod-intro} to the quantum setting and to $\k\neq 3$, are summarized in Appendix \ref{tensorcomps}. 
As a simple example, at $N=2$ the relation between first and second-order antisymmetric tensors (which are just products of Levi-Civita tensors when $N=2$) un-diagonalizes to
\begin{equation}
	\label{quant-undiag-N2}
	q^{i_1}q^{i_2} = \frac{(m_{i_2}-m_{i_1})^{\k-1}}{m_{i_1}-m_{i_2}-\ve}\; q^{[i_1 i_2]} \mapsto \begin{cases}
		(\tilde{\mu}Q)^{[i_1} Q^{i_2]} = \frac{1}{2}\epsilon^{i_1 i_2} &  \k = 1\\
		(\tilde{\mu}Q)^{[i_1[j_1} Q^{i_2] j_2]} = \frac{1}{4}\ve \epsilon^{i_1 i_2}\epsilon^{j_1 j_2} & \k = 2\\
		(\tilde{\mu}Q)^{[i_1[j_2 [k_3} Q^{i_2] j_2] k_2]} = \frac{1}{8} \big(4 c_2 + \varepsilon^2 \big)\epsilon^{i_1 i_2}\epsilon^{j_1 j_2}\epsilon^{k_1 k_2} & \k = 3\\
	\end{cases}
\end{equation} where $(\tilde{\mu}Q)^{i_1 i_2 ... i_\k}$ is the contraction of $\tilde{\mu}_a$ with the corresponding index of $Q^{i_1 i_2 ... i_\k}$ (which by  \eqref{quant-muQI-k} is independent of $a$), and the $c_{\ell}$'s are as in \eqref{quant-contrac-intro}. A natural generalization of the conjecture in \cite{Tachi-review} to $\TNk$ is that all relations stem from relations in this algebra of eigenvalues, although we do not have a general proof that this is the case. 

\subsection{New quantum relations}
\label{sec:newq}

Working in the quantized chiral ring also leads to some new identities whose classical limits vanish.
The simplest of these relate commutators of $\k$-fold fundamental tensor operators to higher tensor powers.

In Appendix  \ref{app:index} we prove that when $\k\leq 3$ the commutators of fundamental tensors are anti-symmetric under any exchange of indices%
\footnote{This relation does \emph{not} hold in such a simple form when $\k>3$. See Appendix \ref{app:index} for a counterexample.}
 \begin{equation}
	[Q^{i_1 i_2 ... i_{\k}}, Q^{j_1 j_2 ... j_{\k}}] = - [Q^{j_1 i_2 ... i_{\k}}, Q^{i_1 j_2 ... j_{\k}}] = ... = - [Q^{i_1 i_2 ... j_{\k}}, Q^{j_1 j_2 ... i_{\k}}]\,,
\end{equation}
and similarly for $(Q^{i_1 i_2 ... i_{\k}},Q^{j_1 j_2 ... j_{\k}}) \mapsto (Q_{i_1 i_2 ... i_{\k}},Q_{j_1 j_2 ... j_{\k}})$.
When $\k=3$, this suggests a very simple relation 
\begin{equation} \label{Q-comm-intro}
	[Q^{i_1 j_1 k_1},Q^{i_2 j_2 k_2}] = \varepsilon Q^{[i_1 i_2] [j_1 j_2] [k_1 k_2]}
\end{equation}
between tri-fundamental and tri-antisymmetric-tensor operators. We have verified \eqref{Q-comm-intro} by direct computation for $N\leq 4$. In the $\ve\to0$ limit, \eqref{Q-comm-intro} clearly reduces to a Poisson bracket $\{Q^{i_1 j_1 k_1},Q^{i_2 j_2 k_2}\} = Q^{[i_1 i_2] [j_1 j_2] [k_1 k_2]}$.

For higher-rank tensors, the commutators must be generalized. An obvious choice would be to consider a full 3-fold antisymmetrization of $r$ copies of $Q$ to get rank-$r$ tensor operators. Alternatively, based on the general features described in Appendix \ref{app:cyc},
it is natural to consider  the recursive definition
\begin{equation} \label{Cyc-intro}
	\widetilde{Q}^{[i_1 j_1 ... k_1] [i_2 j_2 ... k_2] [i_3 j_3 ... k_3]} := \textrm{Cyc}^r_1 \circ \textrm{Cyc}^r_2 \circ \textrm{Cyc}^r_3\Big(Q^{i_1 i_2 i_3} \widetilde{Q}^{[j_1 ... k_1] [j_2 ... k_2] [j_3 ... k_3]}\Big)\,,
\end{equation} where \begin{equation}
\textrm{Cyc}^r_a\Big(A^{i_1 i_2 i_3} B^{[j_1 ... k_1] [j_2 ... k_2] [j_3... k_3]}\Big) := \sum \limits_{n = 1}^r(-1)^{n(r-1)} \sigma^n_a\Big(A^{i_1 i_2 i_3} B^{[j_1 ... k_1] [j_2 ... k_2] [j_3... k_3]}\Big)\,,
\end{equation} $\sigma$ is the $r$-cycle (1 2...r) and, $\sigma^n_a$ means apply $\sigma^n$ to the set $\{i_a, j_a, ..., k_a\}$.%
\footnote{This is analogous to writing the symmetric group $S_n$ as a union of cosets of $\mathbb{Z}_n$: $S_r = \bigcup_{i = 1}^{r} \sigma^i S_{r-1}$.
} 
We have checked directly for $N\leq 4$ that both full-anti-symmetrization and \eqref{Cyc-intro} agree with the form of higher-rank tensors $Q^{I_1 I_2 I_3}$ given in the main text, weighted by an appropriate number of $\ve$'s to soak up the R-charge,
\be \qquad \widetilde{Q}^{I_1I_2I_3}\;  \sim\; \ve^\# Q^{I_1I_2I_3} \qquad (\k=3)\,. \ee

\section{Review: the Coulomb branch chiral ring}
\label{sec:review}

In this section, we review the construction of the Coulomb branch chiral ring $\C[\CM_C]$ of a 3d $\CN=4$ gauge theory, following recent advances in the math and physics literature.
In particular, we will incorporate mathematical results of Webster's \cite{Web2016} into the physical understanding of the chiral ring.

We keep much of the discussion general. We assume that the gauge theory is defined by a renormalizable Lagrangian, with compact gauge group $G$ coupled to linear matter (hypermultiplets) in a quaternionic representation~$\CR$. We further assume that $\CR$ is a direct sum of unitary representations
\be \CR = R \oplus R^* \simeq T^*R\,.  \label{RR} \ee
The only additional parameters that the theory may depend on are real gauge couplings, masses, and FI parameters.

\subsection{Generalities}
\label{sec:MC}

Recall that the Coulomb branch of a 3d $\CN=4$ gauge theory is a component of the moduli space of vacua on which all hypermultiplet vevs vanish, and on which vectormultiplet scalars generically acquire diagonal vevs, breaking the gauge symmetry $G$ to its maximal torus~$T$.
The Coulomb branch is a noncompact hyperk\"ahler manifold \cite{Rocek, SeibergWitten-3d}, possibly singular, of dimension
\be \text{dim}_\C \CM_C = 2\,\text{rank}(G)\,.\ee
In a 3d gauge theory, the Coulomb branch has an exact $SU(2)_C$ metric isometry that rotates its $\cp^1$ of complex structures. Thus it essentially looks \emph{the same} in every complex structure.
The $SU(2)_C$ is part of the R-symmetry of the 3d $\CN=4$ theory, and shows up classically as a rotation of the triplet of $\fg$-valued scalar fields in the vectormultiplet.

For example, in a $\CT_{N,\k}$ quiver gauge theory the gauge group is%
\footnote{The overall $U(1)$ quotient is standard in quivers with no ``flavor nodes'' or ``framing''; it makes sense because none of the bifundamental hypermultiplets are charged under the diagonal $U(1)$.}
\be G=\big(U(N)\times [U(N-1)\times \cdots \times U(1)]^{\k}\big)/U(1)\,. \label{GNk} \ee
The dimension of the Coulomb branch is therefore easily computed as
\be \CT_{N,\k}\,:\quad \text{dim}_\C \CM_C =  2N + 2\k\Big(\sum_{i=1}^{N-1} i\Big)  -2  = (\k N+2)(N-1)\,.\ee

In any fixed complex structure, the Coulomb branch is a holomorphic symplectic manifold, \ie\ a K\"ahler manifold, possibly singular, whose smooth part is endowed with a non-degenerate holomorphic two-form $\Omega$.
For every choice of complex structure, there is a chiral ring of half-BPS local operators whose vevs are holomorphic functions on the Coulomb branch.
We simply denote this ring
\be \C[\CM_C]\,, \ee
suppressing the dependence on complex structure. The holomorphic symplectic form $\Omega$ endows the chiral ring with a Poisson bracket, thus turning $\C[\CM_C]$ into a Poisson algebra. Physically, the Poisson bracket of operators may be computed by topological descent~\cite{descent}.

\subsection{Fibration: scalars and monopoles}
\label{sec:fib}

In a fixed complex structure, the Coulomb branch moreover has the structure a complex integrable system.%
\footnote{This integrable system is a degeneration of the Seiberg-Witten integrable system familiar from 4d $\CN=2$ gauge theory \cite{SW1, SW2, Donagi-SW}.} %
Specifically, the Coulomb branch is a singular fibration
\be \label{fib} \begin{array}{ccc}
 T_\C^\vee &\dashrightarrow\!\! & \CM_C \\
  && \;\;\downarrow\pi \\
  && \ft_\C/W\,, \end{array}
\ee
where $\ft_\C$ denotes the complexified Cartan subalgebra of $G$, $W$ the Weyl group, and $T_\C^\vee$ the complexified dual of the maximal torus. 
Roughly speaking, the base $\ft_\C/W \simeq \C^{\text{rank}(G)}$ is parameterized by the `diagonal' expectation value of a complex vectormultiplet scalar
\be  \varphi \in \ft_\C\subset  \fg_\C\,. \ee
The complex scalar $\varphi$ combines two of the three real vectormultiplet scalars, as dictated by the choice of complex structure. Classically, it is forced to take a diagonal vev due to vacuum equations $[\varphi,\varphi^\dagger]=0$.
Global coordinates on the base come from $G$-invariant polynomials (Casimir operators) in~$\varphi$, which are the true gauge-invariant operators in a non-abelian theory.

We call a point $\varphi$ on the base \emph{generic} if 1) it fully breaks gauge symmetry to the torus (making all W-bosons massive) and 2) gives a nonzero effective mass to every hypermultiplet.
Algebraically, these criteria mean that, respectively
\be \begin{array}{l}M_\alpha :=\langle \alpha,\varphi\rangle \neq 0 \\[.1cm] \hspace{.4in} \forall\; \alpha\in\text{roots}(G)\end{array} \qquad\text{and}\qquad \begin{array}{l} M_\lambda:=\langle \lambda,\varphi\rangle \neq 0 \\[.1cm] \hspace{.4in} \forall\; \lambda\in\text{weights}(R) \end{array}\,. \label{generic} \ee
Mathematically, one would say that a generic point of $\mathfrak t_\C/W$ is in the complement of all weight and root hyperplanes.

The fiber of the integrable system \eqref{fib} above any generic point of the base is a complex dual torus $T_\C^\vee\simeq (\C^*)^{\text{rank}(G)}$. It is a holomorphic Lagrangian torus with respect to the holomorphic symplectic structure. The coordinates on the fibers are vevs of chiral monopole operators.
Locally, near a generic point on the base where $G$ is broken to $T$, one may define half-BPS abelian monopole operators as (\emph{cf.} \cite{SeibergWitten-3d, IntriligatorSeiberg, BKW02})
\be v_A \sim e^{\frac{1}{g^2}(A,\sigma+i\gamma)} \label{vA} \ee
where $g$ is the gauge coupling, $A\in \ft$ is a cocharacter (satisfying $e^{2\pi i A}=I$),
$\sigma\in \ft$ is the third real vectormultiplet scalar, $\gamma\in \ft$ are the dual photons (with periodicity $2\pi g^2$), and $(\;,\;)$ is the Cartan-Killing form. The OPE of monopole operators satisfies $v_A v_B\sim v_{A+B}$, for any cocharacters $A$ and $B$, so their vevs are just right to produce global functions on $T_\C^\vee\simeq (\C^*)^{\text{rank}(G)}$.

The way that the $T_\C^\vee$ fibers vary over the base of the Coulomb branch depends qualitatively on locations of the root and weight hyperplanes. Roughly speaking,
\begin{itemize}
\item The fibers blow up (their volume diverges) above root hyperplanes, where W-bosons become massless and gauge symmetry is enhanced.
\item The fibers collapse above weight hyperplanes, where hypermultiplets become massless.
\end{itemize}
The precise hyperk\"ahler metric on the fibration acquires non-perturbative quantum corrections that are extremely difficult to compute directly.

\subsection{TQFT and non-renormalization}

Nevertheless, if one ignores the hyperk\"ahler metric and focuses on $\CM_C$ as a complex symplectic manifold, many computations become tractable. In particular, the computation of the chiral ring $\C[\CM_C]$ and its Poisson structure (as well as its deformation quantization) reduces to a \emph{relatively} simple algebra problem.

There are two ways to think about this simplification. In \cite{BDG2015} it was argued that the chiral ring of a 3d $\CN=4$ gauge theory is independent of the gauge coupling, and thus cannot receive nonperturbative quantum corrections, or perturbative corrections beyond one loop. 

Alternatively, one may recognize that the chiral ring $\C[\CM_C]$ belongs to a topological subsector of the 3d gauge theory. Specifically, the chiral-ring operators are in the cohomology of a topological supercharge $Q$, which was discussed long ago in \cite{BlauThompson}, and may equivalently be characterized as (\emph{cf.} \cite{Nak2016, VV, descent}) \vspace{-.2cm}
\begin{itemize}
\setlength\itemsep{-.1cm}
\item[-] the 3d reduction of the 4d $\CN=2$ Donaldson supercharge

\item[-] one of the scalars under a diagonal subgroup of $SU(2)_{\text{Lorentz}}\times SU(2)_{H}$ (where $SU(2)_{H}$ is the R-symmetry that rotates hypermultiplet scalars)

\item[-] the ``twisted Rozansky-Witten'' supercharge, as it plays the same role on the Coulomb branch that the Rozansky-Witten twist plays for the Higgs branch\,.
\end{itemize}
\vspace{-.2cm}
Then the product of chiral-ring operators is topologically protected, and may be computed using standard TQFT methods. Perhaps surprisingly, the Poisson bracket and deformation quantization (via Omega background) of chiral-ring operators are also topological in nature~\cite{descent}.

The TQFT perspective motivated the initial mathematical work \cite{Nak2016, BFN2016} on Coulomb branches. In Appendix \ref{app:BFN}, we explain how the mathematical characterization of Coulomb-branch operators relates to the physics of 3d $\CN=4$ theories. The TQFT perspective has some important computational consequences, which we draw on in what follows.

\subsection{The abelianized algebra $\CA$}
\label{sec:A}

The TQFT derivation of the ring $\C[\CM_C]$ (in Appendix \ref{app:BFN}) proceeds via reduction to 1d quantum mechanics, where $\C[\CM_C]$ is identified as the equivariant cohomology (or more technically, Borel-Moore homology) of a certain moduli space. Fixed-point localization embeds the chiral ring into a much simpler ``abelianized'' algebra $\CA$,
\be \C[\CM_C] \hookrightarrow \CA\,. \label{embed2} \ee
Physically speaking, one may think of $\CA$ as a local algebra of operators near generic points on the Coulomb branch, where the gauge theory is effectively abelian; this is how the abelian algebra $\CA$ arose in \cite{BDG2015}.%
\footnote{This perspective is directly analogous to abelianization/non-abelianization in 4d $\CN=2$ theories \cite{GMN-Hitchin, GMN-framed}, and to localization computations of algebras of line/loop operators therein, \emph{cf}. \cite{Pestun, GOP, IOT}.} %
Similarly, in an Omega-background both $\C[\CM_C]$ and $\CA$ are deformation-quantized, and one finds an embedding of associative algebras
\be \C_\ve[\CM_C] \hookrightarrow \CA_\ve\,. \label{embede} \ee

All the computations in this paper will take place in $\CA$ or $\CA_\ve$. We review their structure here. Since  $\CA$ can be recovered from $\CA_\ve$ by sending $\ve\to 0$, it would be sufficient to describe $\CA_\ve$. However, some relations are simpler and more intuitive for $\CA$, so we shall start with the commutative case.

The algebra $\CA$ can be defined as the local chiral ring, in the neighborhood of a generic point $\varphi$ on the base of the Coulomb branch, in the sense of \eqref{generic}.
To make this precise, we denote the loci on the base of the Coulomb branch were W-bosons and hypermultiplets become massless as
\be \Delta  = \bigcup_{\text{roots $\alpha$}}\{M_\alpha(\varphi) = 0\} \;\subset \ft_\C\,,\qquad \Delta_R  = \bigcup_{\text{weights $\lambda$ of $R$}}\{M_\lambda(\varphi) = 0\} \;\subset \ft_\C\,. \ee
Then we define
\be \CM_C^{\rm abel} = \pi^{-1}\big((\ft_\C\backslash\Delta\cup \Delta_R)/W\big)\subset \CM_C \ee
as the open subset of the Coulomb branch sitting above the complement of $\Delta$ and $\Delta_R$ in the fibration \eqref{fib}; and define $\wt \CM_C^{\rm abel}$ to be the trivial $W$-cover of $\CM_C^{\rm abel}$ (undoing the quotient by the Weyl group on the base). Then
\be \CA := \C\big[ \wt\CM_C^{\rm abel} \big]\,. \ee
This definition of $\CA$ makes it obvious that there is an embedding \eqref{embed2}, since any global function on $\CM_C$ defines a $W$-invariant local function on $\wt \CM_C^{\rm abel}$.

The algebra $\CA$ has two types of generators:
\begin{enumerate}[leftmargin=.5cm]
\item Rational functions in the components of the abelian complex scalar $\varphi\in \ft_\C$, whose denominators vanish only on $\Delta$ and $\Delta_R$.

In other words, there are polynomials in $\varphi$ and in the inverted generators $(M_\alpha)^{-1}, (M_\lambda)^{-1}$.

\item Abelian monopole operators $v_A$ as in \eqref{vA}, for every cocharacter
\be A \in \text{Hom}(U(1),T) \simeq \Z^{\text{rank}(G)}\,. \label{cochar} \ee
\end{enumerate}
These operators satisfy relations that are essentially the expected product relations $v_Av_B \sim v_{A+B}$ among monopole operators, with one-loop corrections from hypermultiplets and W-bosons.

To write down the relations, we first recall that there is a natural integer-valued product
\be \langle \lambda,A\rangle \in \Z \ee
between weights $\lambda$ and cocharacters.
Then the classical relation $v_Av_{-A}=1$ among abelian monopole operators is corrected by hypermultiplets and W-bosons to
\be v_Av_{-A} = \frac{\ds\prod_{\text{weights $\lambda$ of $R$}}  \big(M_\lambda\big)^{|\langle \lambda,A\rangle|}}
{\ds\prod_{\text{roots $\alpha$ of $G$}}  \big(M_\alpha\big)^{|\langle \alpha,A\rangle|}}\,; \label{vAA}
\ee
and more generally
\be v_A v_B = v_{A+B}\; \frac{\ds\prod_{\substack{\text{weights $\lambda$ of $R$} \\ \text{s.t. $\langle \lambda,A\rangle\langle \lambda,B\rangle<0$}}}  \big(M_\lambda\big)^{\text{min}(|\langle \lambda,A\rangle|,|\langle \lambda,B\rangle|)}}
{\ds\prod_{\substack{\text{roots $\alpha$ of $G$} \\ \text{s.t. $\langle \lambda,A\rangle\langle \lambda,B\rangle<0$}}}  \big(M_\alpha\big)^{\text{min}(|\langle \alpha,A\rangle|,|\langle \alpha,B\rangle|)}}\,. \label{vAB}
\ee
The abelianized algebra $\CA$ simply contains polynomials in $\varphi$, $1/M_\alpha$, and $v_A$, modulo these relations:%
\footnote{Technically, there are also the obvious relations $\langle \alpha,\varphi\rangle \cdot M_\alpha^{-1}=1$, $\langle \lambda,\varphi\rangle \cdot M_\lambda^{-1}=1$  that follow from the definitions of $M_\alpha,M_\lambda$.}
\be \boxed{ \CA = \C\big[\varphi,\{M_\alpha^{-1}\}_{\alpha\in\text{roots}},\{M_\lambda^{-1}\}_{\lambda\in\text{wts(R)}},\{v_A\}_{A\in\text{cochars}}\big]\Big/\raisebox{-.1cm}{(\text{relations \eqref{vAB}})}}\,.\ee

\subsubsection{Quantization}

The quantized algebra $\CA_\ve$ is similar. It is generated by
\begin{enumerate}[leftmargin=.5cm]
\item The components of $\varphi$, and $\ve$.
\item The inverted masses $(M_\alpha+n\ve)^{-1}$ and $(M_\lambda+n\ve)^{-1}$ for all $n\in \Z$.
  
(The shifted quantities  $M_\alpha+n\ve$ may be understood physically as complex masses of all the various \emph{modes} of W-bosons in the presence of an Omega-background, noting that the Omega-background couples to angular momentum. Similarly, $M_\lambda+n\ve$ are masses of the modes of hypermultiplets.)

\item The abelian monopole operators $v_A$.

\end{enumerate}
The parameter $\ve$ is central; and the components of $\varphi$ (and the $(M_{\alpha,\lambda}+n\ve)^{-1}$) all commute with each other.
Otherwise, the generators satisfy two basic sets of relations:

First, note that the components of $\varphi$ can all be picked out by contraction with weights, \emph{e.g.} $\langle \lambda,\varphi\rangle$. All linear functions in $\varphi$ arise this way. The commutator of any such linear function and a monopole operator is
\be 	[\langle \lambda, \varphi \rangle, v_A] = \ve \langle \lambda, A \rangle v_A\,. \label{comm1} \ee
For example, if $G=U(N)$, one would customarily write $\varphi = \text{diag}(\varphi_1,...,\varphi_N)$. Both weights $\lambda=(\lambda_1,...,\lambda_N)$ and cocharacters $A=(A_1,...,A_N)$ are elements of a lattice $\Z^N$. The entries of $\varphi$ are picked out by contractions $\langle (0,...,\overset{i}{1},...,0),\varphi\rangle=\varphi_i$, so \eqref{comm1} says
\be [\varphi_i,v_A] = \ve A_i v_A\,.\ee

\noindent It follows from \eqref{comm1} that the inverted masses also satisfy (\emph{e.g.})  $v_A\, \frac{1}{M_\alpha+n\ve} = \frac{1}{M_\alpha+(n-\langle \alpha,A\rangle)\ve} \,v_A$\,.

Second, the product of two abelianized monopole operators is given by an appropriately ordered and shifted version of \eqref{vAB}\,: 
\be \label{comm3}
	v_{A} v_{B} = \frac{\ds\prod_{\substack{\lambda \in \text{weights}(R)\,\text{s.t.}\\ |\langle \lambda, A\rangle| \leq |\langle \lambda, B \rangle | \\ \langle \lambda, A\rangle \langle \lambda, B \rangle < 0}} 
	[M_{\lambda}+\tfrac\ve2]^{-\langle \lambda, A \rangle}}
	{\ds\prod_{\substack{\alpha \in \text{roots}(G)\,\text{s.t.}\\ |\langle \alpha, A\rangle| \leq |\langle \alpha, B \rangle | \\ \langle \alpha, A\rangle \langle \alpha, B \rangle < 0}} [M_{\alpha}]^{-\langle \alpha, A \rangle}} 
	\,v_{A+B}\,
	\frac{\ds\prod_{\substack{\lambda \in \text{weights}(R)\,\text{s.t.} \\ |\langle \lambda, A\rangle| > |\langle \lambda, B \rangle | \\ \langle \lambda, A\rangle \langle \lambda, B \rangle < 0}}
	 [M_{\lambda}+\tfrac\ve2]^{\langle \lambda, B \rangle}}
	 {\ds\prod_{\substack{\alpha \in \text{roots}(G)\,\text{s.t.} \\ |\langle \alpha, A\rangle| > |\langle \alpha, B \rangle | \\ \langle \alpha, A\rangle \langle \alpha, B \rangle < 0}} [M_{\alpha}]^{\langle \alpha, B \rangle}}\,,
\ee
where
\be
	[a]^b := \begin{cases} 
		\prod\limits_{k=0}^{b-1} (a+k\varepsilon) & b>0\\ 
		\prod\limits_{k=1}^{|b|} (a-k\varepsilon) & b<0\\1 & b = 0\\ 
	\end{cases}
\ee
is a quantum-corrected power. These relations were derived using abelian mirror symmetry in~\cite{BDG2015}, but also follow from an equivariant cohomology (TQFT) computation \cite{Nak2016, VV}.

Altogether, the quantized algebra is
\be \boxed{\CA_\ve = \C\big[\varphi,\{(M_\alpha+n\ve)^{-1},(M_\lambda+n\ve)^{-1}\}_{\alpha\in\text{roots},\lambda\in\text{wts(R)},n\in\Z},\{v_A\}_{A\in\text{cochars}}\big]\big/\raisebox{-.1cm}{\big(\text{rel's \eqref{comm1},\eqref{comm3}}\big)}}\,. \label{Ae} \ee

\subsection{The image of $\C[\CM_C]$ and the algebra $\CW_\ve$}
\label{sec:image}

Once the Coulomb-branch chiral ring $\C[\CM_C]$ (resp $\C_\ve[\CM_C]$) is mapped to the abelianized algebra $\CA$ (resp. $\CA_\ve$), many computations become straightforward. 
In particular, expected relations among elements of $\C_\ve[\CM_C]$ can be checked using the simple relations \eqref{vAB} in $\CA_\ve$. 
Nevertheless, the precise image of $\C_\ve[\CM_C]$ in $\CA_\ve$ can be  tricky to identify. 

A few structural properties of the embedding map were discussed in \cite{BDG2015}. For example:
\begin{itemize}
\item The image of $\C_\ve[\CM_C]$ must sit in the Weyl-invariant subalgebra $\CA_\ve^W\subset \CA_\ve$, since local operators in the full non-abelian gauge theory are gauge invariant.
\item In $\CA_\ve$ one finds arbitrarily large negative powers of the masses $M_{\alpha,\lambda}+n\ve$.
In the case of W-boson masses, this is unavoidable, due to denominators in the products $v_Av_{-A}$.
 
In contrast, the image of $\C_\ve[\CM_C]$ in $\CA_\ve$ cannot contain any of the elements $\frac{1}{M_{\alpha,\lambda}+n\ve}$ themselves, since operators in $\C_\ve[\CM_C]$ must define (as $\ve\to 0$)  global functions on the Coulomb branch that extend smoothly across the discriminant locus.

\end{itemize}

It is also known how a basis for $\C[\CM_C]$ as an \emph{infinite-dimensional vector space} should be indexed \cite{CHZ-Hilbert}.
Physically, one expects that the elements of $\C[\CM_C]$ are monopole operators $V_{A,p(\varphi)}$ labelled by \emph{dominant} cocharacters $A$ (equivalently, by Weyl orbits in the cocharacter lattice) and dressed by polynomials $p(\varphi)$ of $\varphi\in \ft_\C$ that are invariant under the stabilizer $W_A$ of $A$ in the Weyl group. For example, if $A=0$, the ``dressing factors'' are just standard Weyl-invariant polynomials $\C[\varphi]^W$. Formally, we might write
\be \C[\CM_C] \;\overset{\text{as a vector space}}\simeq   \bigoplus_{\text{dominant $A$}} \overbrace{\C[\varphi]^{W_A}}^{\text{dressing factors}}\langle V_A\rangle\,.  \label{vec} \ee

It is unclear whether these structural properties alone can determine how elements of $\C[\CM_C]$ (or $\C_\ve[\CM_C]$) embed in $\CA$ (or $\CA_\ve$). However, much stronger constraints on the embedding come from the mathematical/TQFT perspective.
In fact, the TQFT construction of the chiral ring gives --- in principle --- a complete answer to the embedding problem. Namely, elements of $\C_\ve[\CM_C]$ are identified with equivariant cohomology classes on a certain moduli space; and the embedding $\C_\ve[\CM_C]\hookrightarrow \CA_\ve$ just expresses these classes in terms of equivariant fixed points.

It can still be very difficult to explicitly analyze equivariant cohomology classes in practice. Fortunately, Webster \cite{Web2016} recently outlined a combinatorial calculus that accomplishes this task for Coulomb branches.
We will discuss the physical meaning of Webster's calculus in \cite{lineops}.
 In the current paper, we take a pragmatic approach and use one simple consequence of Webster's combinatorics: the image of $\C_\ve[\CM_C]$ in $\CA_\ve$ must always contain a particular subalgebra $\CW_\ve$ (defined momentarily),
\be  \CW_\ve \subseteq \C_\ve[\CM_C] \subset \CA_\ve\,. \ee
In the case of star quivers $\TNk$, we will identify all expected generators of $\C_\ve[\CM_C]$ as elements of $\CW_\ve$. We in fact suspect that 
\be \label{WCM} \hspace{1in} \CW_\ve \simeq \C_\ve[\CM_C] \qquad \text{(for star quivers)}\,,\ee
though this is not guaranteed.%
\footnote{Unfortunately, some rather complicated combinatorics are required to make up the difference between $\CW_\ve$ and $\C_\ve[\CM_C]$ in general. Nevertheless, there are known examples where $\CW_\ve=\C_\ve[\CM_C]$. In any pure gauge theory, this equality follows from results of \cite{BFM, Ginzburg-Miura}. In linear quiver gauge theories, all the generators and relations of $\C_\ve[\CM_C]$ are known explicitly \cite{BDG2015, BFN-quiver, KWWY}, and equality $\CW_\ve = \C_\ve[\CM_C]$ is also easy to establish. We thus have some hope that equality may hold for star quivers as well.}

The algebra $\CW_\ve$ is defined as follows. One begins with a subalgebra of $\CA_\ve$
generated by polynomials in $\varphi$ and by rescaled monopole operators
\be \label{defu}  u_A \;:= \prod_{\substack{\alpha\in \text{roots}(G) \\ \text{s.t.}\;\langle \alpha,A\rangle < 0}} [M_\alpha]^{-\langle \alpha,A\rangle} \,v_A  \;\; =
  \prod_{\substack{\alpha\in \text{roots}(G) \\ \text{s.t.}\; \langle \alpha,A\rangle < 0}}  v_A\, [M_\alpha]^{\langle \alpha,A\rangle}\,.  \ee
These $u_A$ monopole operators, carrying additional factors associated to the W-boson masses, have the nice property that their products never generate denominators: we simply have
\be \label{commu}
 u_A u_B = \prod_{\substack{\lambda \in \text{weights}(R)\,\text{s.t.}\\ |\langle \lambda, A\rangle| \leq |\langle \lambda, B \rangle | \\ \langle \lambda, A\rangle \langle \lambda, B \rangle < 0}} 
	[M_{\lambda}+\tfrac\ve2]^{-\langle \lambda, A \rangle}
	\;\;\;u_{A+B}\,
	\prod_{\substack{\lambda \in \text{weights}(R)\,\text{s.t.} \\ |\langle \lambda, A\rangle| > |\langle \lambda, B \rangle | \\ \langle \lambda, A\rangle \langle \lambda, B \rangle < 0}}
	 [M_{\lambda}+\tfrac\ve2]^{\langle \lambda, B \rangle}\,,
\ee
with one-loop corrections from the hypermultiplets alone. Otherwise, the usual relations
\be \big[\langle \lambda,\varphi\rangle, u_A\big] = \ve \langle \lambda, A\rangle u_A \label{commu2} \ee
continue to hold for any weight $\lambda$ and cocharacter $A$.

In addition, for each root $\alpha$, let $s_\alpha \in W$ denote the corresponding simple reflection. Recall that the 
Weyl group is generated by the $s_\alpha$'s.
We may adjoin the $s_\alpha$ to the algebra of $\varphi$'s and $u_A$'s, in such a way that the $s_\alpha$'s satisfy the standard Weyl-group relations among themselves, and natural commutation relations
\be s_\alpha u_A = u_{A^\alpha}\, s_\alpha\,,\qquad s_\alpha\, f(\varphi) = f(\varphi^\alpha)\, s_\alpha\,, \label{srel} \ee
where $A^\alpha$ is the reflected cocharacter, and $\varphi^\alpha$ is the reflected element of $\ft_\C$. Finally, for each $\alpha$, introduce the BGG-Demazure operator%
\footnote{The ``BGG'' stands for Bernstein, Gelfand, and Gelfand. The operators $\theta_\alpha$ generate the $G$-equivariant cohomology of the flag variety (known as the nil-Hecke algebra in representation theory), which is a large clue to their physical meaning. Another, related, clue is the appearance of the $\theta_\alpha$ in the work of Gukov and Witten on surface operators in 4d \cite{GuW06}. We will tie these clues together in \cite{lineops}.}
\be \theta_\alpha = \frac{1}{M_\alpha}(s_\alpha-1)\,. \label{Dem} \ee
The algebra $\CW_\ve$ is defined as the Weyl-invariant part of an algebra generated by 1) polynomials in $\varphi$; 2) the $u_A$ monopole operators; and 3) the BGG-Demazure operators:
\be \boxed{ \CW_\ve = \C\big[\varphi,\{u_A\}_{A\in\text{cochars}},\{\theta_\alpha\}_{\alpha\in \text{roots}}\big]^W \;\subset\,\CA_\ve}\,. \label{We} \ee
The relations, which we leave implicit, are of the form \eqref{commu}, \eqref{commu2}, \eqref{srel}. Notice that once Weyl-invariance is imposed, all the $s_\alpha$'s are all projected out, so $\CW_\ve$ does become an actual subalgebra of $\CA_\ve$.

Practically speaking, the role of the Demazure operators $\theta_\alpha$ is to introduce \emph{a few} denominators $\frac{1}{M_\alpha}$, in a controlled way, so that the structural properties of the Coulomb branch discussed above are actually satisfied. In Appendix \ref{app:PSU2} we will work through how \eqref{We} reproduces the chiral ring $\CW_\ve=\C_\ve[\CM_C]$ in the elementary example of pure $G=PSU(2)$ gauge theory.

\subsection{Flavor symmetry and R-symmetry}
\label{sec:flavor}

We finally comment briefly on symmetries of 3d $\CN=4$ theories, in particular those applicable to star quivers.

Flavor symmetries act either on the Higgs branch or on the Coulomb branch, as tri-Hamiltonian isometries. The symmetry group $F$ acting on the Higgs branch is easy to identify in a gauge theory, as the normalizer of $G$ in $USp(R)$
\be F = N_{USp(R)}(G)/G\,, \ee
\emph{i.e.} the group that acts on hypermultiplets independently of $G$. For star quivers, there is no Higgs flavor symmetry at all, $F=1$. In general, complex mass parameters associated to the Higgs flavor symmetry (scalars in the $F$ vector multiplet) can deform the Coulomb-branch chiral ring; but for star quivers such deformations are absent, and Coulomb branches are rigid.

In contrast, star quivers have a rich Coulomb-branch flavor symmetry. In the UV, the Coulomb-branch flavor group $K$ is the Pontryagin dual of $\pi_1(G)$
\begin{align} K &= \text{Hom}(\pi_1(G),U(1)) \simeq U(1)^{\text{rank(Z(G))}}\,, \label{Pont} \end{align}
which is an abelian group with the same rank as the center of $G$.
In the case of star quivers, we easily find
\be K \simeq U(1)^{\k(N-1)}\qquad\text{for $\TNk$}\,. \ee
In the IR the group $K$ may undergo a nonabelian enhancement, controlled by the ``balanced'' nodes in a given quiver \cite{BKW02, GW08a, GW08b}, \emph{i.e.} nodes 
$\tikz[baseline=(char.base)]{
            \node[shape=circle,draw,inner sep=2pt] (char) {$N_c$};}$
that are coupled to exactly $N_f=2N_c$ hypermultiplets.\footnote{It is worth noting that there can be yet further enhancement beyond the naive consideration of balanced nodes. For example, in the theory $\CT_{2,3}$ discussed below there is an obvious $SU(2)^3$ Coulomb-branch flavor symmetry. However, this theory is 3d mirror to a theory of 8 free half-hypermultiplets with Higgs-branch flavor symmetry $USp(4)$, which should be equal to the Coulomb-branch flavor symmetry of $\CT_{2,3}$. Indeed, the Coulomb branch of $\CT_{2,3}$ is $T^*\C^4 \simeq \C^8$ which has a full $USp(4)$ worth of hyperk\"ahler isometries.} For star quivers, the nodes on the legs are always balanced, so there is an IR enhancement
\be K \leadsto SU(N)^{\k}\,. \label{starK} \ee
In addition, in two special cases the central node is balanced as well, leading to
\be \hspace{.5in} \begin{array}{l}
(N,\k) = (2,4) \qquad K = SO(8) \\[.2cm]
(N,\k) = (3,3) \qquad K = E_6 \,.
\end{array}  \label{specialK}\ee

Since the chiral ring $\C[\CM_C]$ is insensitive to RG flow, the fully enhanced IR symmetry group $K$ will act on it. More so, since $\C[\CM_C]$ is a holomorphic object, the complexification $K_\C$ will actually act. This action is generated by complex moment map operators $\mu \in \C[\CM_C]\otimes  \text{Lie}(K)^*$, which are related to the $K$ currents by supersymmetry. In particular, for star quivers, \eqref{starK} implies that one is guaranteed to find $\k$ separate $\mathfrak{sl}(N,\C)^*$-valued moment maps $\mu=(\mu_1,\mu_2,...,\mu_\k)$ in the chiral ring. They generate the action via Poisson brackets.

The $K_\C$ action extends to the quantized chiral ring $\C_\ve[\CM_C]$, where it is generated by taking commutators (rather than Poisson brackets) with moment maps. Explicitly, if $T\in \mathfrak{sl}(N,\C)^\k$ is a generator of the Lie algebra, and we denote by $\mu_T = \langle T,\mu\rangle \in \C_\ve[\CM_C]$ the contraction of $T$ and $\mu$, there must be commutation relations
\be [\mu_T,\mu_{T'}] = \ve\,\mu_{[T,T']}\,, \ee
and the infinitesimal $T$ action on any other operator $\CO$ is
\be T\cdot \CO = \tfrac1\ve [\mu_T,\CO]\,. \ee

In addition to flavor symmetries, 3d $\CN=4$ gauge theories with linear matter also have an $SU(2)_C\times SU(2)_H$ R-symmetry. The two factors act on the Coulomb and Higgs branches, respectively, but in a way that rotates the hyperk\"ahler  $\cp^1$'s of complex structures rather than as tri-holomorphic isometries. The $SU(2)_C$ acting on the Coulomb branch is important to us. Any fixed complex structure on the Coulomb branch is preserved by a $U(1)_R$ subgroup of $SU(2)_C$, which induces into a complexified $\C^*$ action on the chiral ring $\C[\CM_C]$. The $\C^*$ action extends to the quantized $\C_\ve[\CM_C]$, in such a way that the quantization parameter $\ve$ and all moment maps canonically have charge%
\footnote{We are working in conventions where the minimal charge of a $\C^*$ representation is $\frac12$.}
\be [\mu] = [\ve] =1\,.\ee
The R-charges of some other expected operators were summarized in \eqref{rcharge}.

In the abelianized chiral ring $\CA_\ve$, the complex $\varphi$ scalars also necessarily have $[\varphi]=1$. It then follows from monopole products \eqref{comm3} (or in fact the simpler commutative \eqref{vAA}) that
\be [v_A] = \frac12\Big( \sum_{\text{weights $\lambda$ or $R$}} |\langle \lambda,A\rangle| - \sum_{\text{roots $\alpha$ of $G$}} |\langle \alpha,A\rangle|\Big)\,. \label{R-vA}\ee
This is consistent with physical expectations for monopole charges \cite{BKW02, GW08a, GW08b}.

If a 3d $\CN=4$ gauge theory flows to a CFT, the $\C^*$ charges of chiral-ring operators coincide with their conformal dimensions, and must therefore be strictly positive. Star quivers flow to CFT's when $N\geq 2$ and $\k \geq 3$; in this case the positivity of R-charges is manifest in \eqref{rcharge}.

\section{Short quivers ($N=2$)}
\label{sec:2kquivers}

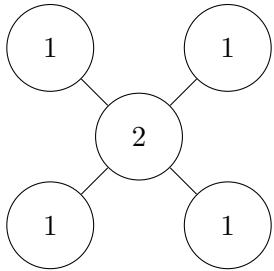
\begin{figure}[htb]
	\centering
		\begin{tikzpicture}[scale=1]
		\begin{scope}[auto, every node/.style={minimum size=3em,inner sep=1, scale=1},node distance=0.5cm]
		\node[draw, circle](v1) at (0,0){2};
		\node[draw, circle, below right=of v1] (v2) {1};
		\node[draw, circle, above right=of v1] (v3) {1};
		\node[draw, circle, below left=of v1] (v4) {1};
		\node[draw, circle, above left=of v1] (v5) {1};
		
		\end{scope}
		\draw (v1)--(v2);
		\draw(v1)--(v3);
		\draw(v1)--(v4);
		\draw(v1)--(v5);
		\end{tikzpicture}
		\caption{The $(N,\k) = (2,4)$ quiver.}
		\label{24quiver}
\end{figure}

We now begin chiral-ring computations in earnest.
Many features of general $\TNk$ star quiver theories already appear in ``short'' quivers that have $N=2$, \emph{i.e.} an arbitrary number $\k$ of legs of length one surrounding a central node. (See  Figure \ref{24quiver} for $\k = 4$.)
These theories are especially computationally friendly, and we work through them in detail as a warm-up for later material.%
\footnote{Coulomb branches of $\CT_{2,\k}$ theories were also recently studies in \cite{HananyZajac}, mainly using Hilbert series. There the authors investigated the effect of gauging discrete global symmetries.}

\subsection{$\k = 3$: $\CM_C\simeq \C^8$}
\label{sec:T23}

We begin with the three-legged quiver $\CT_{2,3}$. In this case the dual 4d theory of Class $\CS$ is the basic $A_1$ trinion theory, \emph{i.e.} a theory of free half-hypermultiplets in the tri-fundamental representation of the $SU(2)^3$ flavor symmetry \cite{Gai09}.\footnote{The theory of eight free (half-)hypermultiplets, the 3d mirror of $\CT_{2,3}$, actually has a larger Higgs flavor symmetry group than this naive $SU(2)^3$. Indeed, the full symmetry group is $USp(4)$, corresponding to the hyperk\"ahler isometries of $T^*C^4 \simeq \C^8$. The 36 generators of (the complexification of) $USp(4)$ fit into a (complex) moment map built out of all the independent bilinears in the coordinates of $T^*\C^4$. This enhancement is not a general feature and only appears because the dual theory in this case is free.} Correspondingly, we expect to find a simple 3d Coulomb branch
\be \CM_C \simeq \C^8\,. \ee
The way this arises from a 3d perspective turns out to be rather nontrivial.

Naively, the gauge group of $\CT_{2,3}$ is $U(2)\times U(1)^3$. The hypermultiplets sit in three fundamental representations of $U(2)$, each charged under a separate $U(1)$.
As discussed in Section \ref{sec:MC}, the a diagonal $U(1)_{\textrm{diag}}$ subgroup of $U(2)\times U(1)^3$ acts trivially on the hypermultiplets, so the true gauge group is actually a quotient
\be G = \big[U(2)\times U(1)^3\big]/U(1)_{\textrm{diag}}\,. \ee

Correspondingly, the cocharacter lattice that will label monopole charges is
\be \text{cochar}(G)= \text{Hom}(U(1), T) = \Z^5/\Z_{\textrm{diag}}\,,\ee
which we may understand as 5-tuples of integers
\be A = (\overbrace{A_1,A_2}^{U(2)};\overset{U(1)_1}{B_1},\overset{U(1)_2}{B_2},\overset{U(1)_3}{B_3}) \;\in\; \text{cochar}(U(2))\times \text{cochar}(U(1)^3) \label{T23-cochar} \ee
modulo the 1-dimensional sublattice generated by $A_{\textrm{diag}}=(1,1;1,1,1)$. In other words, two cocharacters $A,A'\in \Z^5$ are equivalent if they differ by an integer multiple of $A_{\textrm{diag}}$. Dually, the weight lattice of $G$ may be identified with 5-tuples of integers that sum to zero
\be \begin{array}{ll} \text{weights}(G) &= \text{Hom}(T,U(1)) \\[.1cm]
 & = \big\{\lambda = (\lambda_1,\lambda_2;\lambda_1',\lambda_2',\lambda_3')\in \Z^5\text{ s.t. }{\textstyle \lambda_1+\lambda_2+\lambda_1'+\lambda_2'+\lambda_3'=0}\big\}\,.\end{array}
\ee
Note that there is a well-defined product $\langle\;,\;\rangle:\text{weights}(G)\times \text{cochar}(G)\to \Z$. In particular, $\langle \lambda,A_{\textrm{diag}}\rangle=0$ for any weight $\lambda$.
The matter representation may now be written as $\CR=R\oplus R^*$, with weights of $R$ chosen to be
\be \label{T23-wts} \text{weights$(R)$}\;=\;\; \bigg\{ \begin{array}{c@{\quad}c@{\quad}c} (1,0;-1,0,0) & (1,0;0,-1,0) & (1,0;0,0,-1) \\[.1cm]
(0,1;-1,0,0) & (0,1;0,-1,0) & (0,1;0,0,-1) \end{array} \bigg\}\,.
\ee

\subsubsection{The $\CA_\ve$ and $\CW_\ve$ algebras}
\label{sec:T23-AW}

Our first step in constructing the Coulomb branch is to identify the abelianized algebra $\CA_\ve$ from Section \ref{sec:A}, which contains all putative Coulomb-branch operators. We work from the outset with its quantized version. As described in Section \ref{sec:A}, there are three types of generators:
\begin{enumerate}
\item Polynomials in Omega-background parameter $\ve$, the complex scalars $\varphi^a_1$, $a = 1,2,3$ corresponding to the $U(1)$ factors in $G$, and the diagonal components $(\varphi_{21}, \varphi_{22})$ of the complex scalar corresponding to the $U(2)$ factor.

Due to the $U(1)_{\textrm{diag}}$ quotient, we should restrict to polynomials that are invariant under a simultaneous translation of all the $\varphi$'s.
It is natural to think of such polynomials as generated by weights of $G$, \ie\ by the linear functions
\be \langle \lambda,\varphi\rangle = \lambda_1\varphi_{21}+\lambda_2\varphi_{22} + \lambda_1'\varphi^1_1+ \lambda_2'\varphi^2_1+ \lambda_3'\varphi^3_1\, \qquad \lambda\in\text{weights}(G)\,.
\label{phi-const} \ee
The constraint $\lambda_1+\lambda_2+\lambda_1'+\lambda_2'+\lambda_3'=0$ guarantees that $\langle \lambda,\varphi\rangle$ is invariant under translations.

\item The inverted masses $(M_\alpha+n\ve)^{-1}$ for all roots $\alpha$ of $G$ and all $n\in \Z$. Here the only nonzero roots are $\alpha=\pm(1,-1;0,0,0)$, corresponding to the $U(2)$ factor, so we adjoin elements of the form
\be \frac{1}{\varphi_{21}-\varphi_{22}+n\ve}\,. \ee
Similarly, we adjoin inverted hypermultiplet masses $(M_\lambda+n\ve)^{-1}$ for all weights \eqref{T23-wts}.

\item The abelian monopole operators $v_A$ labelled by cocharacters $A\in \text{cochar}(G)$ as above. All monopole operators with diagonal cocharacter $v_{n A_{\rm diag}}=v_{(n,n;n,n,n)}$ are central in the algebra, and we impose the relations
\be v_{(n,n;n,n,n)} = 1\quad \forall\; n\,.\ee
\end{enumerate}

The next intermediary step is to construct the subalgebra $\CW_\ve\subset \CA_\ve$ from Section \ref{sec:image}. It will help us decide which elements of $\CA_\ve$ are actual chiral-ring operators.

To this end, we introduce the rescaled monopole operators $u_A$ as in \eqref{defu}, whose products contain no denominators. For example, we have
\be 
u_{(\pm 1, 0; B_1,B_2,B_3)} = \pm(\varphi_{22}-\varphi_{21})v_{(\pm1,0;B_1,B_2,B_3)} \qquad u_{(0,\pm 1; B_1,B_2,B_3)} = \pm(\varphi_{21}-\varphi_{22})v_{(0,\pm1;B_1,B_2,B_3)} \ee
for any $B_1,B_2,B_3$.
etc. We also introduce the single Weyl reflection $s$ that generates the Weyl group $\Z_2$. It satisfies $s^2=1$ and acts on monopoles by reflecting their cocharacters:
\begin{equation} 
	s v_{(A_1, A_2; B_1, B_2, B_3)} = v_{(A_2, A_1; B_1, B_2, B_3)} s \hspace{0.5 cm} s u_{(A_1, A_2; B_1, B_2, B_3)} = u_{(A_2, A_1; B_1, B_2, B_3)} s.
\end{equation}
Similarly, $s\varphi_1^a=\varphi_1^as$ and $s\varphi_{21}=\varphi_{22}s$.
The corresponding BGG-Demazure operator is $\theta = \frac{1}{\varphi_{21}-\varphi_{22}}(s-1)$\,. Recall that $\CW_\ve$ is the Weyl-symmetric part of $\C[\varphi,u_A,\theta]$.

Some important elements of $\CW_\ve$, which are assured to belong to the the full chiral ring $\C_\ve[\CM_C]$, are%
\footnote{Here we use $[...]_W$ to denote a sum over the Weyl group, proportional to the projection of $[...]$ to Weyl-invariant operators.}
\be  [\pm \theta u_{(\pm 1,0;B_1,B_2,B_3)}]_W = v_{(\pm 1,0;B_1,B_2,B_3)} + v_{(0,\pm 1;B_1,B_2,B_3)}\,. \label{T23-v} \ee
These are the undressed \emph{nonabelian} monopole operators labelled by a fundamental cocharacter on the central $U(2)$ node. The dressed nonabelian monopoles are simply
\be [ u_{(\pm 1,0;B_1,B_2,B_3)}]_W = u_{(\pm 1,0;B_1,B_2,B_3)} + u_{(0,\pm 1;B_1,B_2,B_3)}\,. \label{T23-u} \ee
In addition, $\CW_\ve$ contains monopoles charged only under the legs (which are trivially Weyl-invariant)
\be [ u_{(0,0;B_1,B_2,B_3)}]_W = u_{(0,0;B_1,B_2,B_3)} = v_{(0,0;B_1,B_2,B_3)}\,, \label{T23-0} \ee
and all Weyl-invariant polynomials in the $\varphi$'s. These are \emph{all} the operators we will need to generate $\C_\ve[\CM_C]$!

\subsubsection{Moment maps}

The theory $\CT_{2,3}$ has an $SU(2)^3$ flavor symmetry acting on its Coulomb branch (described in Section \ref{sec:flavor}), and a corresponding $SL(2,\C)^3$ symmetry in the chiral ring. This symmetry should be generated by three $\mathfrak{sl}(2,\C)^*$-valued complex moment maps $\mu_a$, $a=1,2,3$.

Each of these moment maps is associated to a leg of the quiver. Each leg
\be \label{TSU2} \raisebox{-.4cm}{$ \begin{tikzpicture}[scale=1]
	\begin{scope}[auto, every node/.style={inner sep=1}, node distance=1cm]
	\node (v3) {};
	\node[right=of v3] (v4) {};
	\end{scope}
	\begin{scope}[auto, every node/.style={minimum size=3em,inner sep=1},node distance=0.5cm]
	\node[draw, circle, left=of v3] (v2) {$1$};
	\node[draw, rectangle, left=of v2] (v1) {$2$};
	\end{scope}
	\draw(v1)--(v2);
	\end{tikzpicture}  $}\,
\ee
looks like a copy of $T[SU(2)]$ theory, and effectively treats the central node as a flavor symmetry. We may therefore import well known results from the chiral ring of $T[SU(2)]$ (studied \emph{e.g.} in \cite{GW08b, BDG2015}) to identify the moment maps.

The raising and lowering operators in the moment maps turn out to be instances of \eqref{T23-0}
 \begin{equation}
	V^{1\pm}_1 := v_{(0,0;\pm 1, 0,0)}\,, \qquad V^{2\pm}_1 := v_{(0,0;0, \pm 1, 0)} \,,\qquad  V^{3\pm}_1 = v_{(0,0;0,0\pm 1)}\,.
\end{equation}
We may check expected $\mathfrak{sl}(2,\C)$ commutation relations. 
A quick application of  \eqref{comm3} yields
\begin{equation}
	\begin{array}{rl}
		[V_1^{a \pm}, V_1^{b \pm}] & = 0\,,
	\end{array}
\end{equation} as well as \begin{equation}
	\begin{array}{rl}
		[V_1^{a+}, -V_1^{b-}] & =  \delta^{ab}\big[(\varphi_{21} - \varphi^a_1 - \varepsilon/2)(\varphi_{22} - \varphi^a_1 - \varepsilon/2) - (\varphi_{21} - \varphi^a_1 + \varepsilon/2)(\varphi_{22} - \varphi^a_1 + \varepsilon/2)\big]\\[.1cm]
		& = \varepsilon \delta^{ab}(2 \varphi^a_1 - \varphi_{21} - \varphi_{22})\,.\\
	\end{array}
\end{equation}
Similarly,  \eqref{comm1} implies
\begin{equation}
	[(2 \varphi^a_1 - \varphi_{21} - \varphi_{22}), \pm V^{b\pm}_1] = \pm 2\delta^{ab} \varepsilon (\pm V^{a\pm}_1)
\end{equation}
therefore $\{V_1^{a+}, -V_1^{a-}, 2 \varphi^a_1 - \varphi_{21} - \varphi_{22}\}_{a=1}^3$ can be identified as three mutually commuting $\mathfrak{sl}(2,\C)$ triples. These operators fit into moment maps as
\begin{equation}
	\tilde{\mu}_a := \mu_a - \frac{\ve}{2} \mathds{1} = \left( \begin{array}{cc} \varphi^a_1 - \varphi_{21}/2 - \varphi_{22}/2-\varepsilon/2 & V_1^{a+} \\ -V_1^{a-} & \varphi_{21}/2 + \varphi_{22}/2 - \varphi^a_1 -\varepsilon/2\end{array} \right).
	\label{T23-mu}
\end{equation} 
The shift by $\frac{\ve}{2}$ is included for later convenience. It does not affect the action generated by the moment map; in particular, letting $H = \bsp 1&0\\0&-1\esp\in \mathfrak{sl}(2,\C)$ be the Cartan element we find
$\langle H,\mu_a\rangle = \langle H,\tilde\mu_a\rangle = \text{Tr} (H \mu_a^T) = \text{Tr}(H\tilde \mu_a^T)= 2 \varphi^a_1 - \varphi_{21} - \varphi_{22}$\,.

Note that, using the general R-charge formula \eqref{R-vA}, we have
\be [\mu_a] = [V_a^{a\pm}] = [2 \varphi^a_1 - \varphi_{21} - \varphi_{22}] = 1 \ee
as required for moment maps.

\subsubsection{Tri-fundamentals}

Having identified the moment maps, we may organize the chiral ring into $SL(2,\C)^3$ representations. It is easy to check using 
 \eqref{comm3} that the operator
\be Q^{222} := v_{(1,0;0,0,0)} + v_{(0,1;0,0,0)}\,, \ee
which is of type \eqref{T23-v}, is a ``tri''-lowest-weight vector. Namely,
\be [-V_1^{a-},Q^{222}]= 0\quad \forall\, a=1,2,3\,.\ee
By acting with raising operators on $Q^{222}$ we then produce an entire eight-dimensional tri-fundamental representation. For example,
\be \begin{array}{rl} Q^{122} &:=\tfrac1\ve [V_1^{1+},Q^{222}] = v_{(1,0;1,0,0)} + v_{(0,1;1,0,0)}\,, \\[.1cm]
 &\phantom{:=}\tfrac1\ve[V_1^{1+},Q^{122}] = 0\,,\\[.1cm]
Q^{112} &:=\tfrac1\ve [V_1^{2+},Q^{122}] = v_{(1,0;1,1,0)} + v_{(0,1;1,1,0)}\,, \\[.1cm]
Q^{111} &:= \tfrac1\ve [V_1^{3+},Q^{112}] = v_{(1,0;1,1,1)} + v_{(0,1;1,1,1)}\,,
\end{array} \ee
etc. The complete list of operators in this representation is summarized in Table \ref{quantumT2}.

\begin{table}[htb]
	\centerline{
		\begin{tabular}{c | c | c | c}
			Operator & Expression 1 & Expression 2 & Expression 3\\
			\hline
			$Q^{222} = Q_{111}$ & $(\theta u_{(1,0;0,0,0)})_W$ & $v_{(1,0;0,0,0)} + v_{(0,1;0,0,0)}$ & $v_{(-1,0;-1,-1,-1)} + v_{(0,-1;-1,-1,-1)}$\\
			$Q^{122} = -Q_{211}$ & $(\theta u_{(1,0;1,0,0)})_W$ & $v_{(1,0;1,0,0)} + v_{(0,1;1,0,0)}$ & $v_{(-1,0;0,-1,-1)} + v_{(0,-1;0,-1,-1)}$\\
			$Q^{212} = -Q_{121}$ & $(\theta u_{(1,0;0,1,0)})_W$ & $v_{(1,0;0,1,0)} + v_{(0,1;0,1,0)}$ & $v_{(-1,0;-1,0,-1)} + v_{(0,-1;-1,0,-1)}$ \\
			$Q^{221} = -Q_{112}$ & $(\theta u_{(1,0;0,0,1)})_W$ & $v_{(1,0;0,0,1)} + v_{(0,1;0,0,1)}$ & $v_{(-1,0;-1,-1,0)} + v_{(0,-1;-1,-1,0)}$\\
			$Q^{211} = Q_{122}$ & $-(\theta u_{(-1,0;-1,0,0)})_W$ & $v_{(1,0;0,1,1)} + v_{(0,1;0,1,1)}$ & $v_{(-1,0;-1,0,0)} + v_{(0,-1;-1,0,0)}$\\
			$Q^{121} = Q_{212}$ & $-(\theta u_{(-1,0;0,-1,0)})_W$ & $v_{(1,0;1,0,1)} + v_{(0,1;1,0,1)}$ & $v_{(-1,0;0,-1,0)} + v_{(0,-1;0,-1,0)}$\\
			$Q^{112} = Q_{221}$ & $-(\theta u_{(-1,0;0,0,-1)})_W$ & $v_{(1,0;1,1,0)} + v_{(0,1;1,1,0)}$ & $v_{(-1,0;0,0,-1)} + v_{(0,-1;0,0,-1)}$\\
			$Q^{111} = -Q_{222}$ & $-(\theta u_{(-1,0;0,0,0)})_W$& $v_{(1,0;1,1,1)} + v_{(0,1;1,1,1)}$ & $v_{(-1,0;0,0,0)} + v_{(0,-1;0,0,0)}$\\
	\end{tabular}}
	\caption{Expressions for the eight operators furnishing a tri-fundamental representation of the $SL(2,\C)^3$ action on the chiral ring of the $\CT_{2,3}$ star quiver. These eight operators generate the entire chiral ring. The first expression of the operator is in terms of the Weyl symmetrized image of a rescaled monopole operator under the BGG-Demazure operator $\theta$. The second two expressions are related to one another by adding a diagonal cocharacter $A_{\text{diag}}$.}
	\label{quantumT2}
\end{table}

Alternatively, we could have observed that $Q_{222}:=-(v_{(-1,0;0,0,0)} + v_{(0,-1;0,0,0)})$ is a tri-highest-weight vector, which generates an eight-dimensional tri-antifundamental representation. However, it is equivalent to the tri-fundamental above. In particular, since cocharacters \eqref{T23-cochar} that differ by a multiple of $A_{\text{diag}}$ are equivalent, we actually have $Q_{222}=-Q^{111}$, and more generally 
\begin{equation}
\epsilon_{i_1 j_1} \epsilon_{i_2 j_2} \epsilon_{i_3 j_3} Q^{j_1 j_2 j_3} = Q_{i_1 i_2 i_3}\,. \label{T23-Q=Q}
\end{equation} 

We also note that the R-charge formula \eqref{R-vA} quickly implies that
\be [Q^{i_1i_2i_3}] = [Q_{i_1i_2i_3}] = \tfrac12\,.\ee

\subsubsection{Relations}

Using the above expressions for the tri-fundamental operators, it is a straightforward application of \eqref{comm1} and \eqref{comm3} to find additional relations satisfied by the $Q$'s and $\mu$'s.

For example, there are commutation relations \begin{equation}
	\begin{array}{rl}
		[Q^{122},Q^{222}] & = [v_{(1,0;1,0,0)} + v_{(0,1;1,0,0)}, v_{(1,0;0,0,0)} + v_{(0,1;0,0,0)}]\\
		& = \bigg[\frac{\varphi_{21}-\varphi^1_1 + \varepsilon/2}{(\varphi_{21}-\varphi_{22})(\varphi_{22}-\varphi_{21} - \varepsilon)}- \frac{\varphi_{21}-\varphi^1_1 - \varepsilon/2}{(\varphi_{21}-\varphi_{22})(\varphi_{22}-\varphi_{21} + \varepsilon)} + (\varphi_{21} \leftrightarrow \varphi_{22})\bigg]v_{(1,1;1,0,0)}\\
		& = 0\,;\\
	\end{array}
\end{equation} similarly, \begin{equation}
\begin{array}{rl}
[Q^{111},Q^{222}] & = [v_{(-1,0;0,0,0)} + v_{(0,-1;0,0,0)}, v_{(1,0;0,0,0)} + v_{(0,1;0,0,0)}]\\
& = \bigg[\frac{\prod \limits_{a = 1}^3 \varphi_{21}-\varphi^a_1 + \varepsilon/2}{(\varphi_{21}-\varphi_{22})(\varphi_{22}-\varphi_{21} - \varepsilon)}- \frac{\prod \limits_{a = 1}^3 \varphi_{21}-\varphi^a_1 - \varepsilon/2}{(\varphi_{21}-\varphi_{22})(\varphi_{22}-\varphi_{21} + \varepsilon)} + (\varphi_{21} \leftrightarrow \varphi_{22})\bigg]\\
& = - \varepsilon\,;\\
\end{array}
\end{equation} and more generally \begin{equation}
	[Q^{i_1 i_2 i_3}, Q^{j_1 j_2 j_3}] = -\varepsilon \epsilon^{i_1 j_1} \epsilon^{i_2 j_2} \epsilon^{i_3 j_3}\,.
\end{equation}
Sending $\ve\to 0$, these recover the Poisson brackets $\{Q^{i_1 i_2 i_3},Q^{j_1 j_2 j_3}\} = - \epsilon^{i_1 j_1} \epsilon^{i_2 j_2} \epsilon^{i_3 j_3}$ expected from the duality of $\CT_{2,3}$ with free half-hypermultiplets in 4d.

We may also consider contractions between moment maps and $Q$'s.
Schematically writing \begin{equation}
	Q^{i_1 i_2 i_3} = v_{(1,0;B_{i_1 i_2 i_3})} + v_{(0,1;B_{i_1 i_2 i_3})}\,,
\end{equation} we use \eqref{comm3} to find \begin{equation}
	(\tilde{\mu}_1)^{i_1}{}_{i'} Q^{i' i_2 i_3} = (\tilde{\mu}_2)^{i_2}{}_{i'} Q^{i_1 i' i_3} = (\tilde{\mu}_3)^{i_3}{}_{i'} Q^{i_1 i_2 i'} = \frac{\varphi_{21} - \varphi_{22}}{2}\big[v_{(1,0;B_{i_1 i_2 i_3})} - v_{(0,1;B_{i_1 i_2 i_3})}\big]\,. \label{T23-muQ}
\end{equation} More generally, for all $n\geq 0$, we may contract with powers of the moment maps to get \begin{equation}
	(\tilde{\mu}_1^n)^{i_1}{}_{i'} Q^{i' i_2 i_3} = (\tilde{\mu}_2^n)^{i_2}{}_{i'} Q^{i_1 i' i_3} = (\tilde{\mu}_3^n)^{i_3}{}_{i'} Q^{i_1 i_2 i'} = \Big(\frac{\varphi_{21} - \varphi_{22}}{2}\Big)^n \big[v_{(1,0;B_{i_1 i_2 i_3})} + (-1)^n v_{(0,1;B_{i_1 i_2 i_3})}\big].
\end{equation}
Note that the RHS of \eqref{T23-muQ} contains an alternative dressed version of the fundamental nonabelian monopole operators.

Finally, we can recover the moment maps themselves as contractions of $Q$'s,
\be \frac{1}{2}\bigg(Q^{i_1i_2i_3}Q_{i_1i_2'i_3} + \ve \delta^{i_1}{}_{i_1'}\bigg) = (\mu_1)^{i_1}{}_{i_1'}\,, \ee
and similarly for $(\mu_2)^{i_2}{}_{i_2'}$ and $(\mu_3)^{i_3}{}_{i_3'}$.

It is straightforward but tedious to show that the $Q^{i_1i_2i_3}$ operators generate all of $\CW_\ve$. In this case we know from duality with free half-hypermultiplets in 4d that these tri-fundamental operators really generate the entire chiral ring $\C_\ve[\CM_C]$. Since $\CW_\ve$ is necessarily contained in $\C_\ve[\CM_C]$, there is no choice but to have
\be \CW_\ve = \C_\ve[\CM_C]\,, \ee
as desired in \eqref{WCM}.

\subsection{Diagonalization for $\k=3$}

The $\CT_{2,3}$ theory provides a first, nontrivial example of the diagonalization procedure that was previewed in Section \ref{sec:summary-q}.

\subsubsection{Moment maps}

Let's first consider the moment maps associated to each leg of the quiver. In the ``classical'' limit $\ve\to 0$, we expect the Casimir operators built out of any of the moment maps to depend only on the $\varphi_{2i}$ scalars on the central node. Indeed, viewing each leg as a copy of $T[SU(2)]$ theory as in \eqref{TSU2}, we know from \cite{GW08b,BDG2015} that the Casimirs depend only on the mass parameters associated to the $\boxed{2}$ flavor node, and these masses become the $\varphi_{2i}$ scalars upon gauging the $U(2)$ symmetry to tie the quiver together. In particular, in the $\ve\to 0$ limit we expect
\be \text{Tr}(\mu_1^n) = \text{Tr}(\mu_2^n)  = \text{Tr}(\mu_3^n)\qquad \forall\;n\geq 0\,. \label{T23-Cas0} \ee
This relation is also well known from the dual perspective of 4d $\CN=2$ trinion theories. It implies in particular that at generic points on the Coulomb branch all the moment maps should have coincident eigenvalues.

In the current quantum setting, we can similarly compute traces of powers of the quantum moment-map operators \eqref{T23-mu}. Using the $\ve$-shifted moment maps $\tilde\mu_a$, we find
\be \begin{array}{l} \text{Tr}(\tilde\mu_1)=\text{Tr}(\tilde\mu_2)=\text{Tr}(\tilde\mu_3) = -\ve \\[.2cm]
 \text{Tr}(\tilde\mu_1^2)=\text{Tr}(\tilde\mu_2^2)=\text{Tr}(\tilde\mu_3^2) = \tfrac12(\varphi_{21}-\varphi_{22})^2\,.
\end{array} \ee
This similarly suggests that we may ``generically'' be able to diagonalize the quantum moment maps, and find coincident eigenvalues. In the quantum setting, ``generically'' will mean working in the abelianized algebra $\CA_\ve$.

In order to diagonalize $\tilde{\mu}_a$, it is helpful to find its eigenvectors. Using  \eqref{comm3} we find that the vectors%
\footnote{Here and below we use the convention that a rational function $\frac\alpha\beta$ of noncommutative operators is meant to be read with the denominator to the \emph{left} of the numerator, \emph{i.e.} $\frac{1}{\beta} \alpha$.\label{foot:rat}}
\begin{equation} \label{T23-evecs}
	\eta^a_1 = \left( \begin{array}{c}
		 \frac{V_1^{a+}}{\varphi_{21}-\varphi^a_1+\varepsilon/2}\\  1 
	\end{array}\right)\,, \qquad \eta^a_2 = \left( \begin{array}{c}
		 \frac{V_1^{a+}}{\varphi_{22}-\varphi^a_1+\varepsilon/2}\\  1 
	\end{array}\right),
\end{equation} which are related by the $\mathbb{Z}_2$ Weyl symmetry, satisfy (with no summation over $a$)
\begin{equation}
	\tilde{\mu}_a \eta^a_1 = \frac{1}{2}(\varphi_{21} - \varphi_{22}) \eta^a_1\,, \qquad  \tilde{\mu}_a \eta^a_2 = \frac{1}{2}(\varphi_{22} - \varphi_{21}) \eta^a_2\,.
\end{equation}
We thus arrange them into a matrix suggestively denoted $S_a^{-1} = \left(\eta^a_1 \eta^a_2 \right)$. With a little bit of work, it is possible to find a (two-sided) inverse of this matrix: \begin{equation}
	S_a = \left( \begin{array}{c c}
		\frac{-V_1^{a-}}{\varphi_{21} - \varphi_{22}} & \frac{\varphi_{21} - \varphi^a_1 - \varepsilon/2}{\varphi_{21} - \varphi_{22}} \\
		\frac{-V_1^{a-}}{\varphi_{22} - \varphi_{21}} & \frac{\varphi_{22} - \varphi^a_1 - \varepsilon/2}{\varphi_{22} - \varphi_{21}} \\
	\end{array} \right)\,,
\end{equation} which, not too surprisingly, takes the form of two row vectors that are related by action of the Weyl group. With these matrices, it follows that \begin{equation}
	 S_a \tilde{\mu}_a S_a^{-1} = \left( \begin{array}{c c}
		\frac{1}{2}(\varphi_{21} - \varphi_{22}) & 0 \\
		0 & \frac{1}{2}(\varphi_{22} - \varphi_{21})\\
	\end{array} \right) =: \tilde{\mu}_{\textrm{diag}}\,. \label{T23-mudiag}
\end{equation} 

It is worth noting that by choosing an ordering of the eigenvectors $\eta^a_1, \eta^a_2$ the Weyl symmetry is explicitly broken and so any objects obtained by acting with $S_a$ or $S_a^{-1}$ should not be expected to be Weyl invariant. It is also worth noting that the components of $S_a$ and $S_a^{-1}$ commute with $S_b$, $S^{-1}_b$ and $\tilde{\mu}_b$ for $a \neq b$, as well as with $\tilde{\mu}_{\textrm{diag}}$.

An immediate application of this diagonalization is to obtain the polynomial invariants of the moment map $\textrm{Tr}[(\tilde{\mu}_a)^n]$ for all $n \geq 0$. Using the $S$ matrices we now have \begin{equation}
	\textrm{Tr}[(\tilde{\mu}_a)^n] = \textrm{Tr}[(S_a^{-1}\tilde{\mu}_{\textrm{diag}} S_a)^n] =  \textrm{Tr}[S_a^{-1}(\tilde{\mu}_{\textrm{diag}})^n S_a]\,.
\end{equation} Unfortunately, when $\varepsilon \neq 0$, the trace is no longer cyclic. Nonetheless, it is straightforward to compute the diagonal components: \begin{equation} \begin{array}{rl}
	(S_a^{-1}(\tilde{\mu}_{\textrm{diag}})^n S_a)^1{}_1 &= -2^{-n}({\varphi_{21} - \varphi_{22}})^{n-1} \left[(\varphi_{22} - \varphi^a_1 + \varepsilon/2) - (-1)^n (\varphi_{21} - \varphi^a_1 + \varepsilon/2) \right]\,, \\[.2cm]
	(S_a^{-1}(\tilde{\mu}_{\textrm{diag}})^n S_a)^2{}_2 &= 2^{-n}({\varphi_{21} - \varphi_{22}})^{n-1} \left[(\varphi_{21} - \varphi^a_1 - \varepsilon/2) - (-1)^n (\varphi_{22} - \varphi^a_1 - \varepsilon/2) \right]\,, \end{array} \notag
\end{equation} quickly leading to \begin{equation}
	\textrm{Tr}[(\tilde{\mu}_a)^n] = \begin{cases}
		\frac{1}{2^{n-1}}({\varphi_{21} - \varphi_{22}})^n& n \hspace{1mm} \textrm{even}\\
		-\varepsilon\frac{1}{2^{n-1}}({\varphi_{21} - \varphi_{22}})^{n-1}& n \hspace{1mm} \textrm{odd}\,.\\
	\end{cases}
\end{equation}
These explicitly only dependent on $\ve$ and the scalars on the central node for all $n$ and $a$, so \begin{equation}
\textrm{Tr}[(\tilde{\mu}_1)^n] = \textrm{Tr}[(\tilde{\mu}_2)^n] = \textrm{Tr}[(\tilde{\mu}_3)^n]\qquad \forall\; n\geq 0\,. \label{T23-Cas}
\end{equation}

\subsubsection{Tri-fundamentals}
\label{sec:T23-Qdiag}

Having diagonalized the moment maps, we may hope to diagonalize the tri-(anti)fundamentals $Q^{i_1i_2i_3}$ and $Q_{i_1i_2i_3}$ with the same $S_a,S_a^{-1}$ matrices. Indeed, the quantum relation \eqref{T23-muQ}, which generalizes the well-known 4d trinion relations \eqref{muQ}, strongly suggests that this is possible.

Let us therefore define
\begin{equation}
	q^{i_1 i_2 i_3} := (S_1)^{i_1}{}_{j_1} (S_2)^{i_2}{}_{j_2} (S_3)^{i_3}{}_{j_3} Q^{j_1 j_2 j_3}\,, \qquad q_{i_1 i_2 i_3} := Q_{j_1 j_2 j_3} (S_1^{-1})^{j_1}{}_{i_1} (S_2^{-1})^{j_2}{}_{i_2} (S_3^{-1})^{j_3}{}_{i_3}.
\end{equation}
Somewhat amazingly, a straightforward computation reveals
\begin{equation}
	q^{i_1 i_2 i_3} = \delta^{i_1}{}_{i} \delta^{i_2}{}_{i} \delta^{i_3}{}_{i} q^{i}
\end{equation} where \begin{equation}
	q^1 = v_{(1,0;0,0,0)}\,, \qquad q^2 = v_{(0,1;0,0,0)}
\end{equation}
are simple \emph{abelian monopole operators} in $\CA_\ve$, charged only under the central node! This parallels the fact that the moment-map eigenvalues \eqref{T23-mudiag} were scalars on the central node.
Similarly, \begin{equation}
	q_{i_1 i_2 i_3} = \delta^{i}{}_{i_1} \delta^{i}{}_{i_2} \delta^{i}{}_{i_3} q_{i}\,,
\end{equation} where \begin{equation}
	q_1 = -\frac{(\varphi_{21}-\varphi_{22} - \varepsilon)^3}{\prod\limits_{a=1}^3 (\varphi_{21}-\varphi_1^a + \varepsilon/2)}v_{(-1,0;0,0,0)}\,, \quad\;\; q_2 = -\frac{(\varphi_{22}-\varphi_{21} - \varepsilon)^3}{\prod\limits_{a=1}^3 (\varphi_{22}-\varphi_1^a + \varepsilon/2)}v_{(0,-1;0,0,0)}\,. \label{T23-qaf}
\end{equation}

Notice that the magnetic charge of $q^i,q_i$ under the central $U(2)$ node matches the charges of the $Q^{i_1i_2i_3}$, $Q_{i_1i_2i_3}$ operators as representations of the $SL(2,\C)^3$ flavor symmetry. In particular, the $q^i$'s have magnetic charges corresponding to weights of the fundamental representation of $U(2)$, and the $q_i$'s have magnetic charges corresponding to the antifundamental representation.%
\footnote{Being more careful, we should talk about magnetic charge under $U(2)$ as a representation (or weight spaces) of the Langlands dual group. However, $U(2)$ is self-dual, so there is no distinction.}

The lack of perfect symmetry between $q^i$ and $q_i$ (in particular, the more complicated prefactors for $q_i$) is due to our particular choice of similarity matrices $S_a$, $S_a^{-1}$. In our conventions, the determinants
are nontrivial,%
\footnote{The determinant of a matrix with non-commutative entries is not canonically defined. Here we use \emph{row}-determinants, defined for a general $n\times n$ matrix as $\det X = \epsilon_{i_1...i_n}X^{i_1}{}_1... X^{i_n}{}_n$.\label{foot:rowdet}}
 \begin{equation}
	\begin{array}{rl l}
		\det S_1 &:= \epsilon_{ii'}(S_1)^i{}_1(S_1)^{i'}{}_2  &=  \ds \frac{1}{(\varphi_{22} - \varphi_{21})} v_{(0,0;-1,0,0)}\\[.2cm]
		\det S_1^{-1} &:= \epsilon_{ii'}(S_1^{-1})^i{}_1(S_1^{-1})^{i'}{}_2 &=  \ds \frac{(\varphi_{22} - \varphi_{21})}{\prod\limits_{\alpha = 1}^2(\varphi_{2 \alpha}-\varphi^1_{1}+\varepsilon/2)} v_{(0,0;1,0,0)}\\
	\end{array},
\end{equation} and similarly for the other $S_a$ and $S_a^{-1}$. We could have chosen the determinants to be $1$, but only at the expense of introducing roots in the matrix entries. We strongly prefer instead to use matrix entries that are manifestly elements of the algebra $\CA_\ve$.

Despite the determinants being nontrivial, they do satisfy
\begin{equation}
	\det S_a \hspace{1mm} \det S_a^{-1} = \det S_a^{-1} \hspace{1mm} \det S_a = 1\,.
\end{equation} 
Moreover, the products of determinants 
\begin{align} \sigma &= \det S_1 \, \det S_2 \, \det S_3 = \frac{v_{(1,1;0,0,0)}}{(\varphi_{22}-\varphi_{21})^3}\,, \\
 \sigma^{-1} &= \det S_1^{-1} \, \det S_2^{-1} \, \det S_3^{-1} = \frac{(\varphi_{22}-\varphi_{21})^3}{\prod \limits_{i = 1}^2 \prod \limits_{a = 1}^3 (\varphi_{2i} - \varphi^a_1 +\ve/2)} v_{(-1,-1;0,0,0)}\,.
\end{align}
are relatively simple expressions, only involving monopoles charged under the central node.

\subsubsection{Algebra of eigenvalues}

Let us end the discussion of $\CT_{2,3}$ by explicitly describing some relations in the algebra of eigenvalues, giving a simple example of the structure in Section \ref{sec:summary-q}.

We have found moment map and tri-fundamental eigenvalues
\be m_1 = - m_2 = (\varphi_{21}-\varphi_{22})/2\,, \qquad q^1=v_{(1,0;0,0,0)}\,, \qquad  q^2 = v_{(0,1;0,0,0)}\,,\ee
and antifundamentals $q_1\sim v_{(-1,0;0,0,0)}$, $q_2\sim v_{(0,-1;0,0,0)}$ given by \eqref{T23-qaf}. These satisfy
\be [m_i,q^j]=(\tfrac12\delta_i^j-1)\ve q^j\,,\qquad [m_i,q_j]=-(\tfrac12\delta_{ij}-1)\ve q_j\,,\ee
as well as
 \begin{equation}
	\begin{array}{r c l}
		q^1 q_1 = \ds\frac{(m_1 - m_2)^2}{m_1 - m_2 - \ve}\,, & \qquad & q_1 q^1  = \ds\frac{(m_1 - m_2+\ve)^2}{m_1 - m_2}\,,\\[.4cm]
		q^2 q_2 = \ds\frac{(m_2 - m_1)^2}{m_2 - m_1 - \ve}\,, & \qquad & q_2 q^2  = \ds\frac{(m_2 - m_1+\ve)^2}{m_2 - m_1}\,.
	\end{array}
\end{equation}
Fundamentals and antifundamentals are related by $q_1 \sigma  = v_{(0,1;0,0,0)} = q^2$ and $q_2 \sigma  = -v_{(1,0;0,0,0)} = -q^1$, or more simply 
\be	\epsilon_{ij} q^j = q_i \sigma\,. \label{T23-q=q} \ee

The product of determinants $\sigma$ is not quite central, but it is very close. Namely, the operator $w^+=v_{(1,1;0,0,0)}$ \emph{does} commute with all the eigenvalues $m_i,q^i,q_i$\,, and we have $\sigma= (\varphi_{22}-\varphi_{21})^{-3} \,w^+$.

We emphasize that \eqref{T23-q=q} is a diagonalized form of the chiral-ring relation \eqref{T23-Q=Q}. In the diagonalized basis, the product of determinants $\sigma$ is required to pass between fundamental and antifundamental operators.

\subsection{$(U(2) \times U(1)^{\k})/U(1)$ Theory}

The structures we found above at $\k=3$ generalize in a surprisingly straightforward way to general $\k$. At $\k=3$, the algebra $\C_\ve[\CM_C]$ was \emph{freely} generated by the tri-fundamental operators $Q^{i_1i_2i_3}$ (in particular, moment maps could be recovered as products of $Q$'s), in correspondence with the fact that the Coulomb branch itself is a flat space $\CM_C=\C^8$. For $\k<3$ there are additional relations among the $Q$'s; while for $\k>3$ we will need both $Q$'s and moment maps as independent generators. However, the embedding of $\C_\ve[\CM_C]$ into the abelian algebra $\CA_\ve$ and the process of diagonalization look almost identical.%

We note that, in contrast, the Higgs branches of the corresponding 4d Class $\CS$ theories $\CT_2[\Sigma_{0,\k}]$ are \emph{not} usually described in a manner that is uniform for all $\k$. By decomposing the $\k$-punctured sphere into $\k-2$ pairs of pants,
\be \Sigma_{0,\k} \simeq \bigcup_{\alpha=1}^{\k-2} \Sigma_{0,3}^{(\alpha)}\,, \label{pants-k} \ee
one finds that the 4d Higgs branch is a holomorphic symplectic quotient
\be \CM_H^{\rm 4d} \simeq \Big[\prod_{\alpha=1}^{\k-2}\C^8\Big]/\!/SL(2,\C)^{\k-3}\,. \label{symp-k} \ee
However, there are many different ways to choose a decomposition \eqref{pants-k}, and so many ways to express the same quotient \eqref{symp-k}; none is canonical, and none makes the generators and relations of the chiral ring manifest.
(The combinatorics of the spaces \eqref{symp-k} were studied carefully by \cite{HM-trivertices}.)
 The current 3d-Coulomb-branch perspective thus has some marked advantages.

The gauge group of the $\CT_{2,\k}$ theory is a diagonal quotient $(U(2) \times U(1)^{\k})/U(1)_{\rm diag}$. Correspondingly, the cocharacter and weight lattices are
\begin{align} \text{cochar}(G) &= \big\{ A=(A_1,A_2;B_1,...,B_\k)\in \Z^{\k+2}\big\}/(1,1;1,...,1) \simeq \Z^{\k+2}/\Z\,,\\[.1cm]
 \text{weights}(G) &= \big\{\lambda = (\lambda_1,\lambda_2;\lambda_1',...,\lambda_\k')\in \Z^{\k+2}\;\text{s.t.}\;
{\textstyle \lambda_1+\lambda_2+\sum_a\lambda_a'=0}\big\} \simeq \Z^{\k+1}\,. 
\end{align}
As before, we denote by $(\varphi_{21},\varphi_{22})$ the diagonal vectormultiplet scalars on the central $U(2)$ node, and by $\varphi_1^a$ the $U(1)$ scalar on the $a$-th leg. They may be collected into a vector $\varphi = (\varphi_{21},\varphi_{22};\varphi_1^1,...,\varphi_1^\k)$. We split the hypermultiplet representation as $\CR=R\oplus R^*$, where
\be \text{weights$(R)$}\;=\;\; \bigg\{ \begin{array}{c@{\quad}c@{\quad}c} (1,0;-1,0,...,0) & (1,0;0,-1,...,0)\;\;\ldots &  (1,0;0,0,...,-1) \\[.1cm]
(0,1;-1,0,...,0) & (0,1;0,-1,...,0)\;\;\ldots &  (0,1;0,0,...,-1) \end{array} \bigg\}\,.
\ee

The abelian algebra $\CA_\ve$ is then generated by
\be \frac{1}{\varphi_{21}-\varphi_{22}+n\ve}\,, \frac{1}{\varphi_{2\alpha}-\varphi^a_{1}+n\ve}\,,\qquad \varphi\,,\qquad v_A\,, \ee
modulo the usual relations \eqref{comm1}-\eqref{comm3}. The monopoles of diagonal cocharacter $v_{nA_{\rm diag}} = v_{(1,1;1,...,1)}$ are central and are all set equal to $1$.
The subalgebra $\CW_\ve$ that is contained in the chiral ring (and which we conjecture is equal to the chiral ring) is generated just as in Section \ref{sec:T23-AW} by the rescaled monopoles $u_A$, the Weyl reflection $s$ ($s^2=1$) and the BGG-Demazure operator $\theta = \frac{1}{\varphi_{21}-\varphi_{22}}(s-1)$. In particular, we have
\be u_{(\pm1,0;\vec B)} = \pm(\varphi_{22}-\varphi_{21})v_{(\pm 1,0;\vec B)}\,,\qquad u_{(0,\pm1;\vec B)} = \pm(\varphi_{21}-\varphi_{22})v_{(0,\pm 1;\vec B)}\,, \ee
\be u_{(0,0;\vec B)} =  v_{(0,0;\vec B)} \notag \ee
for any $\vec B=(B_1,...,B_\k)$.

\subsubsection{Moment maps and $\k$-fold fundamentals}

There is now an $SL(2,\C)^{\k}$ action on the Coulomb-branch chiral ring, generated by moment maps associated to each ``$T[SU(2)]$'' leg \eqref{TSU2}. Since the moment maps on separate legs are decoupled, they take exactly the same form \eqref{T23-mu} as for $\k=3$ (but now with $a=1,...,\k$). In particular, the $\mathfrak{sl}(2,\C)$ triplets are
\be E^a = V^{a+}_1 := v_{(0,0;0,...,\underset{a}{1},...,0)}\,,\quad F^a = -V^{a-}_1 := -v_{(0,0;0,...,\underset{a}{-1},...,0)}\,,\quad
H^a = 2 \varphi^a_1 - \varphi_{21} - \varphi_{22}\,. \ee
Also in analogy with the $\k=3$ case, we find that the Weyl-symmetrized operators
\be \begin{array}{l} Q^{22...2}:= (\theta u_{(1,0;0,...,0)})_W = v_{(1,0;0,...,0)}+v_{(0,1;0,...,0)}\,,\\[.2cm]
 Q_{22...2} := (-1)^{\k-1}(\theta u_{(-1,0;0,...,0)})_W = (-1)^{\k}[v_{(-1,0;0,...,0)}+v_{(0,-1;0,...,0)}]
\end{array}\ee
are lowest-weight and highest-weight, respectively, for every $\mathfrak{sl}(2,\C)$ triple. They generate a $\k$-fold fundamental representation $Q^{i_1,...,i_\k}$ and a $\k$-fold antifundamental $Q_{i_1,...,i_\k}$, with indices $i_1,...,i_\k$ taking the values $1$ and $2$. The operators in these representations are given by straightforward generalizations of the expressions in Table \ref{quantumT2}. Explicitly,
\be Q^{i_1i_2...i_\k} = (\theta u_{(1,0;2-i_1,2-i_2,...,2-i_\k)}) = v_{(1,0;2-i_1,2-i_2,...,2-i_\k)}+v_{(0,1;2-i_1,2-i_2,...,2-i_\k)}\,.\ee
and similarly for $Q_{i_1i_2...i_\k}$. The fundamental and antifundamental operators are manifestly related by
\begin{equation}
	\epsilon_{i_1 j_1} ...  \epsilon_{i_k j_\k}Q^{j_1 ...  j_{\k}} = Q_{i_1 ...  i_{\k}}\,. \label{T2k-Q=Q}
\end{equation} 

\subsubsection{Diagonalization}

Since the moment maps for general $\k$ look identical to those at $\k=3$, we easily generalize \eqref{T23-Cas} to 
\begin{equation}
	\textrm{Tr}[(\tilde{\mu}_1)^n] = ... = \textrm{Tr}[(\tilde{\mu}_{\k})^n] = \begin{cases}
		\frac{1}{2^{n-1}}({\varphi_{21} - \varphi_{22}})^n & n \textrm{ even}\\
		-\frac{\ve}{2^{n-1}}({\varphi_{21} - \varphi_{22}})^{n-1} & n \textrm{ odd}\,.\\
	\end{cases}
\end{equation}
More explicitly, for each leg $a=1,...,\k$ we may introduce pairs of matrices
\be S_a^{-1} = \left( \begin{array}{c c}
		 \frac{V_1^{a+}}{\varphi_{21}-\varphi^a_1+\varepsilon/2} &  \frac{V_1^{a+}}{\varphi_{22}-\varphi^a_1+\varepsilon/2}\\ 
		 1 &  1 
	\end{array}\right)\,,\qquad 
 S_a = \left( \begin{array}{c c}
	\frac{-V_1^{a-}}{\varphi_{21} - \varphi_{22}} & \frac{\varphi_{21} - \varphi^a_1 - \varepsilon/2}{\varphi_{21} - \varphi_{22}} \\
	\frac{-V_1^{a-}}{\varphi_{22} - \varphi_{21}} & \frac{\varphi_{22} - \varphi^a_1 - \varepsilon/2}{\varphi_{22} - \varphi_{21}} \\
	\end{array} \right)\,,
\ee
satisfying $S_aS_a^{-1} = S_a^{-1}S_a=\mathds{1}$ and the components of $S_a^{\pm}$ commute with the components of $S_b^{\pm}$, $\tilde{\mu}_b$ and $\tilde\mu_{\rm diag}$ for $a\neq b$. These matrices diagonalize the moment maps to
\be  \tilde\mu_{\rm diag} = S_a \tilde{\mu}_a S_a^{-1} = \left( \begin{array}{c c}
		\frac{1}{2}(\varphi_{21} - \varphi_{22}) & 0 \\
		0 & \frac{1}{2}(\varphi_{22} - \varphi_{21})\\
	\end{array} \right) \quad \forall \;a=1,...,\k\,.
\ee

Now consider applying the operator $S_1 \otimes ... \otimes S_{\k}$ to $Q^{i_1 ... i_{\k}}$ (resp. $S_1^{-1} \otimes ... \otimes S_{\k}^{-1}$ to $Q_{i_1 ... i_{\k}}$).
In complete analogy with the $\k = 3$ result, we find
\be \begin{array}{cc} (S_1)^{i_1}{}_{j_1} ... (S_{\k})^{i_{\k}}{}_{j_{\k}} Q^{j_1 ... j_{\k}} &= \delta^{i_1}{}_{i} ... \delta^{i_k}{}_{i} q^{i}\,,\\[.2cm]
 Q_{j_1 ... j_{\k}} (S_1^{-1})^{j_1}{}_{i_1} ... (S_{\k}^{-1})^{j_{\k}}{}_{i_{\k}} &= \delta^{i'}{}_{i_1} ... \delta^{i'}{}_{i_{\k}} q_{i'}\,, \end{array}
\ee
with eigenvalues
\begin{equation}
q^1 = v_{(1,0;\vec{0})}\,, \qquad q^2 = v_{(0,1;\vec{0})}\,,
\end{equation}
\begin{equation}
	q_1 = -\frac{(\varphi_{21} - \varphi_{22} +\varepsilon)^{\k}}{\prod \limits_{a=1}^{\k}(\varphi_{21} - \varphi^a_1 + \varepsilon/2)} v_{(-1,0;\vec{0})}\,, \qquad q_2 = -\frac{(\varphi_{22} - \varphi_{21} +\varepsilon)^{\k}}{\prod \limits_{a=1}^{\k}(\varphi_{22} - \varphi^a_1 + \varepsilon/2)} v_{(0,-1;\vec{0})}\,.
\end{equation} 
Again, rather beautifully, the eigenvalues involve abelian monopole operators charged only under the central node.

Let us analyze the subalgebra of $\CA_\ve$ that is generated by the eigenvalues,  as we did for $\CT_{2,3}$ above. 
With moment-map eigenvalues $m_1 = - m_2 = (\varphi_{21}-\varphi_{22})/2$, we find that
\begin{equation}
	\begin{array}{r c l}
		q^1 q_1 = \frac{(m_1 - m_2)^{\k-1}}{m_1 - m_2 - \ve} & \qquad & q_1 q^1  = \frac{(m_1 - m_2+\ve)^{\k-1}}{m_1 - m_2}\\
		q^2 q_2 = \frac{(m_2 - m_1)^{\k-1}}{m_2 - m_1 - \ve} & \qquad & q_2 q^2  = \frac{(m_2 - m_1+\ve)^{\k-1}}{m_2 - m_1}\\
	\end{array}
\end{equation}
and similarly
\begin{equation} \label{T2k-q=q}
	q_1 \sigma  = (-1)^{\k-1} v_{(0,1;\vec{0})} = (-1)^{\k-1}q^2  \qquad q_2 \sigma  = -v_{(1,0;\vec{0})} = -q^1,
\end{equation}
where
\be \sigma = (\det S_1)(\det S_2)...(\det S_\k) = \frac{v_{(1,1;\vec{0})}}{(\varphi_{22}-\varphi_{21})^{\k}}\,, \ee
\be \sigma^{-1} = \frac{(\varphi_{22}-\varphi_{21})^{\k}}{\prod \limits_{i = 1}^2 \prod \limits_{a = 1}^{\k} (\varphi_{2i} - \varphi^a_1 +\ve/2)} v_{(-1,-1;\vec{0})}\,. \notag \ee
For odd $\k$, we may write \eqref{T2k-q=q} as
\begin{subequations}  \label{T2k-q=q-oddeven}
 \begin{equation} 
	\epsilon_{ij} q^j = q_i \sigma \hspace{0.5cm} \textrm {or}  \hspace{0.5cm}  q_i \epsilon^{ij}  = q^j \sigma^{-1}\,;
\end{equation} but for even $\k$ we require a slight modification \begin{equation}
	\Delta_{ij} q^j = q_i \sigma \hspace{0.5cm} \textrm {or}  \hspace{0.5cm}  q_i \Delta^{ij}  = q^j \sigma^{-1}
\end{equation} \end{subequations} where \begin{equation}
	\Delta_{ij} = \Delta^{ij} := \left(\begin{array}{cc} 0 & -1\\ -1 & 0\\ \end{array}\right) = -|\epsilon_{ij}|.
\end{equation}

As before, the product of determinants is very close to central. We may write $\sigma = (\varphi_{22}-\varphi_{21})^\k w^+$, where $w^+ = v_{(1,1;\vec 0)}$ is actually central in the algebra of eigenvalues.

\subsubsection{Un-diagonalization}

The chiral ring relation $\epsilon_{i_1 j_1} ...  \epsilon_{i_k j_\k}Q^{j_1 ...  j_{\k}} = Q_{i_1 ...  i_{\k}}$ and the diagonalized relations \eqref{T2k-q=q-oddeven} are intertwined by acting with the similarity transformations $S_1 \otimes ... \otimes S_{\k}$. This relationship is neatly summarized in the commutative diagram of Figure \ref{2kcommdiag}.

\begin{figure}[htb]
	\begin{center}
		\begin{tikzpicture}
		\tikzset{vertex/.style = {minimum size=1.5em}}
		\tikzset{edge/.style = {->,> = latex}}
		\node[vertex] (1) at  (0,0) {$Q^{i_1 ... i_{\k}}$};
		\node[vertex] (2) at  (5,0) {$Q_{j_1 ... j_{\k}}$};
		\node[vertex] (3) at  (0,-4) {$q^{i'}$};
		\node[vertex] (4) at  (5,-4) {$q_{j'}$};
		
		\draw[edge] (1.350) -- (2.190) node[midway, yshift=0.75cm] {$\bigotimes \limits_{a = 1}^{\k}\epsilon^{j_a i_a}$};
		\draw[edge] (2.170) -- (1.10) node[midway, yshift=-0.75cm] {$\bigotimes \limits_{a = 1}^{\k} \epsilon_{j_a i_a}$};
		
		\draw[edge] (1.260) -- (3.100) node[midway, left] {$\overset{\bigotimes \limits_{a = 1}^{\k} (S_a)^{i'}{}_{i_a}}{\longrightarrow}$};
		\draw[edge] (3.80) -- (1.280) node[midway, right] {$\overset{\bigotimes \limits_{a = 1}^{\k} (S_a^{-1})^{i_a}{}_{i'}}{\longrightarrow}$};
		
		\draw[edge] (3.350) -- (4.190) node[midway, yshift=0.75cm] {$\overset{\sigma \epsilon^{j' i'}}{\longleftarrow}$ or $\overset{\sigma \Delta^{j' i'}}{\longleftarrow}$};
		\draw[edge] (4.170) -- (3.10) node[midway, yshift=-0.75cm] {$\overset{\sigma^{-1} \epsilon_{j' i'}}{\longleftarrow}$ or $\overset{\sigma^{-1} \Delta^{j' i'}}{\longleftarrow}$};
		
		\draw[edge] (2.260) -- (4.100) node[midway, right] {$\overset{\bigotimes \limits_{a = 1}^{\k} (S_a)^{j'}{}_{j_a}}{\longleftarrow}$};
		\draw[edge] (4.80) -- (2.280) node[midway, left] {$\overset{\bigotimes \limits_{a = 1}^{\k} (S_a^{-1})^{j_a}{}_{j'}}{\longleftarrow}$};
		\end{tikzpicture}
		\caption{Commutative diagram relating the $\k$-fold fundamental and antifundmental operators in the standard basis and the ``basis of eigenvectors'' of the moment map. The arrow under $\bigotimes S_a$, $\bigotimes S^{-1}_a$ and the determinants indicate which side to act on. (The distinction is only relevant when $\varepsilon$ is nonzero.) For the bottom arrows, $\epsilon$ is used for odd $\k$ and $\Delta$ is used for even $\k$.}
		\label{2kcommdiag}
	\end{center}
\end{figure}

We may also consider applying $\tilde{\mu}_{\textrm{diag}}$ to $\delta^{i_1}{}_{i} ... \delta^{i_{\k}}{}_{i} q^{i}$ and then un-diagonalizing. By choosing to contact $S^{-1}_1$ with $\tilde{\mu}_{\textrm{diag}}$, we find \begin{equation}
	(S_1^{-1})^{i_1}{}_{j} ... (S_{\k}^{-1})^{i_{\k}}{}_{j_{\k}} (\mu_{\textrm{diag}})^{j}{}_{j_1} \delta^{j_1}{}_{i} ... \delta^{j_{\k}}{}_{i} q^{i} = (\tilde{\mu}_1)^{i_1}{}_{j_1} Q^{j_1 ... i_{\k}}\,.
\end{equation}
However, since contracting $S^{-1}_1$ with $\tilde{\mu}_{\textrm{diag}}$ was an arbitrary choice (\emph{i.e.} we could have contracted any of the $S^{-1}_a$), it follows that
\begin{subequations} \label{T2k-muQ}
 \begin{equation}
	(\tilde{\mu}_1)^{i_1}{}_{j_1} Q^{j_1 ... i_k} = ... = (\tilde{\mu}_k)^{i_k}{}_{j_k} Q^{i_1 ... j_k}\,.
\end{equation} A similar argument can also be used in applying $\tilde{\mu}_a$ to $Q_{i_1... i_{\k}}$ to obtain \begin{equation}
	Q_{j_1 ... i_{\k}}(\tilde{\mu}_1)^{j_1}{}_{i_1} = ... = Q_{i_1 ... j_{\k}} (\tilde{\mu}_{\k})^{j_{\k}}{}_{i_{\k}}\,.
\end{equation} \end{subequations}

When analyzing 4d Higgs branches of trinion theories, equations \eqref{T2k-muQ} are usually a starting point, used to deduce the fact that the $Q$ tensors can be diagonalized. Here, in the 3d Coulomb-branch analysis, it is actually easier to describe the explicit diagonalization first, and then deduce \eqref{T2k-muQ}.

\section{One-legged quivers ($\k=1$)} \label{N1quivers}

With the $\CT_{2,\k}$ quivers well understood, we now consider the other extreme: the $\CT_{N,1}$ star quivers that have just a single leg, of arbitrary length:
\be \raisebox{-.4cm}{$ \begin{tikzpicture}[scale=1]
	\begin{scope}[auto, every node/.style={inner sep=1}, node distance=1cm]
	\node (v3) {};
	\node[right=of v3] (v4) {};
	\end{scope}
	\begin{scope}[auto, every node/.style={minimum size=3em,inner sep=1},node distance=0.5cm]
	\node[draw, circle, left=of v3] (v2) {$N-1$};
	\node[draw,circle, left=of v2] (v1) {$N$};
	\node[draw, circle, right=of v4] (v5) {2};
	\node[draw, circle, right=of v5] (v6) {1};
	\end{scope}
	\draw(v1)--(v2);
	\draw(v2)--(v3);
	\draw[loosely dotted, line width = 1pt] (v3)--(v4);
	\draw(v4)--(v5);
	\draw(v5)--(v6);
	\end{tikzpicture}  $} \label{eq:linquiver}
\ee
This is a linear quiver, with gauge group
\be  G = (U(N) \times U(N-1)\times ... \times U(1))/U(1)_{\rm diag} \ee
and flavor symmetry $SU(N)$. The Coulomb branch of such quivers was studied in \cite{HananyWitten}, where it was related to a moduli space of $PSU(N)$ monopoles. The quantized Coulomb-branch chiral ring of linear quivers was analyzed in detail in \cite{BDG2015}. Our main (novel) goal here is to describe the diagonalization of the moment map associated to the flavor symmetry of this quiver, 
we will then upgrade our analysis to general $\CT_{N,\k}$ theories in Section \ref{Nkquivers}.

We note that the $\CT_{N,1}$ star quiver is a small modification of the $T[SU(N)]$ quiver \cite{GW08b}:
\be \raisebox{-.4cm}{$ \begin{tikzpicture}[scale=1]
	\begin{scope}[auto, every node/.style={inner sep=1}, node distance=1cm]
	\node (v3) {};
	\node[right=of v3] (v4) {};
	\end{scope}
	\begin{scope}[auto, every node/.style={minimum size=3em,inner sep=1},node distance=0.5cm]
	\node[draw, circle, left=of v3] (v2) {$N-1$};
	\node[draw, rectangle, left=of v2] (v1) {$N$};
	\node[draw, circle, right=of v4] (v5) {2};
	\node[draw, circle, right=of v5] (v6) {1};
	\end{scope}
	\draw(v1)--(v2);
	\draw(v2)--(v3);
	\draw[loosely dotted, line width = 1pt] (v3)--(v4);
	\draw(v4)--(v5);
	\draw(v5)--(v6);
	\end{tikzpicture}  $}\,. \label{TSUN}
\ee
Namely, $\CT_{N,1}$ is obtained from $T[SU(N)]$ by gauging the the terminal $\boxed{$N$}$ node. The Coulomb branch of $T[SU(N)]$ is the nilpotent cone in $\mathfrak{sl}(N,\C)$, and is fully parameterized by vevs of the moment-map operators. The Coulomb branch of $\CT_{N,1}$ is just a little bit bigger, and the corresponding chiral ring contains operators $Q^i, Q_i$ in fundamental/antifundamental representations of the $SU(N)$ flavor symmetry in addition to the moment map itself.
This conforms with the general expectations for $\CT_{N,\k}$ theories. 
Indeed, we will also identify operators $Q^{[i_1...i_r]},Q_{[i_1,...,i_r]}$ in arbitrary antisymmetric tensor powers of the fundamental/antifundamental; though when $\k=1$ these can all be generated from the basic $Q^i,Q_i$.

An alternative mathematical description of the Coulomb branch of the $\CT_{N,1}$ theory is as a Kostant-Whittaker symplectic reduction
\be \CM_C \simeq T^*SL(N,\C)/\!/_\psi\, \mathfrak{N} \;\simeq\; \C^{N-1}\times SL(N,\C)\,. \label{Kostant} \ee
In Section \ref{sec:KW}, we will use our understanding of the algebra $\C_\ve[\CM_C]$ to explain how \eqref{Kostant} comes about.

\subsection{Conventions}

Abstractly, the cocharacter lattice of $G$ is $\mathbb{Z}^{\frac{1}{2}N(N+1)}/\mathbb{Z}_{\textrm{diag}}$. We will write cocharacters as
\be A = (\vec{a}_N;\vec a_{N-1},\ldots,\vec{a}_1)\,,\qquad \vec{a}_n \in \mathbb{Z}^n\,,\ee
where each $\vec a_n = (a_{n1},...,a_{nn})$ corresponds to the $U(n)$ cocharacter on the $n$-th node; and we identify any two cocharacters whose difference is a multiple of $A_{\rm diag} = (\vec 1;\vec 1,..., 1)$.

It is also useful to introduce basis elements for the cocharacter lattice (prior to the quotient by $A_{\rm diag}$). For $1\leq n\leq N$ and $1\leq \alpha\leq n$, let
\be e_{n\alpha} = (\vec 0;\ldots;\overbrace{(0,...,\underset{\alpha}{1},...,0)}^n;\ldots;0) \label{TN1-basis} \ee
denote the cocharacter with $a_{n\alpha}=1$ and all other entries set to zero.

The weight lattice is $\Z^{\frac12N(N+1)-1}$, containing elements
\be \lambda = (\vec\lambda_N;\vec\lambda_{N-1},\ldots,\vec\lambda_1)\,,\qquad \vec\lambda_n\in \Z^n\,, \ee
constrained so that the total sum of entries in the $\lambda$ vector is zero. We can use differences of the $e_{n\alpha}$'s from \eqref{TN1-basis}  (now re-interpreted as weights) as a basis for the weight lattice. 
The hypermultiplets sit in bifundamental representations associated to the edges of the $\CT_{N,1}$ quiver. We choose to split the  representation as $\CR=R\oplus R^*$, such that the weights of $R$ are
\be \label{TN1-wts} \text{weights}(R) =  \big\{ e_{n\alpha}-e_{n-1\beta}\big\}\,,\ee
with $ 2\leq n\leq N$, $1\leq \alpha \leq n$, and $1\leq \beta\leq n-1$.

Generalizing the notation of Section \ref{sec:T23}, we also introduce complex scalars $\vec\varphi_n=(\varphi_{n1},...,\varphi_{nn})$ for each $U(n)$ node. They may be assembled in a vector
\be  \varphi = (\vec \varphi_N;\vec\varphi_{N-1},...,\vec\varphi_{1})\, \in \text{cochar}(G)\otimes \C\,. \ee
The contractions of $\varphi$ with weights (or roots) are then given by
\be M_\lambda =  \langle \lambda,\varphi\rangle = \vec\lambda_N\cdot \vec\varphi_N+\vec\lambda_{N-1}\cdot\vec\varphi_{N-1}+\ldots+ \lambda_1\varphi_1 \ee
and are naturally invariant under simultaneous shifts of each component of $\varphi$.

The abelian algebra $\CA_\ve$ is now determined systematically as described in Section \ref{sec:A}. It is generated by the components of $\varphi$ and by abelian monopoles operators $v_A$, as well as the inverted roots
\be \frac{1}{\varphi_{n\alpha}-\varphi_{n\beta} + p\,\ve}\qquad n=1,...,N\,,\;\;  1\leq \alpha<\beta\leq n\,,\;\; p\in \Z\,, \ee
and inverted weights $(M_\lambda+p\,\ve)^{-1}$ with $\lambda$ an element of \eqref{TN1-wts}, \emph{i.e.} 
\be \frac{1}{\varphi_{n\alpha}-\varphi_{n-1\beta}+p\,\ve} \qquad n=2,...,N\,,\;\;  {\alpha=1,...,n \atop \beta=1,...,n-1}\,,\;\; p\in \Z\,. \ee

The subalgebra $\CW_\ve \subset \C_\ve[\CM_C]$, which is actually equal to the chiral ring in this case, %
is built from $\varphi$'s and the rescaled monopoles $u_A$, as well as BGG-Demazure operators corresponding to simple reflections in the Weyl group. For $\CT_{N,1}$ quivers the Weyl group is a product 
\begin{equation}
W = \prod\limits_{n = 2}^{N}S_n
\end{equation}
and the simple reflections will be labelled  $s_{n i}$, with $n=2,...,N$ and $i = 1, ..., n-1$, with corresponding BGG-Demazure operators $\theta_{n i}$.

\subsection{Moment map and diagonalization}
\label{sec:TN1-mu}

The Coulomb-branch flavor symmetry of $\CT_{N,1}$ is $SU(N)$. Correspondingly, the chiral ring must contain a complex moment map $\mu\in \mathfrak{sl}(N,\C)^*$ generating a complexified $SL(N,\C)$ action. Borrowing results of \cite{BDG2015} for the $T[SU(N)]$ quiver \eqref{TSUN}, we find that the Chevalley-Serre generators of $\mathfrak{sl}(N,\C)$ are
\begin{equation}
E_n = V^{+}_n \,,\qquad  F_n = -V^{-}_n \,,\qquad H_n = 2 \Phi_n - \Phi_{n-1} - \Phi_{n+1} \qquad (1\leq n\leq N-1)
\end{equation}
for $n=1,...,N-1$, where
\be V_n^\pm  := \sum_{\alpha=1}^n v_{\pm e_{n\alpha}} = v_{(\vec 0;\ldots;(10...0);\ldots;0)} + \text{Weyl images} \ee
are nonabelian monopoles operators with fundamental (or antifundamental) magnetic charge on the $n$-th node of the quiver, and
\be \Phi_n = \sum_{\alpha=1}^n \varphi_{n\alpha} \ee
are Weyl-invariant sums of the scalars on the $n$-th node. It is clear that $\Phi_n$ belong to $\CW_\ve$; and we can see that $V_n^\pm$ also belong to $\CW_\ve$ by writing them as
\begin{equation}  \label{VnW}
V^{\pm}_n \propto \big[ \theta_{n n-1} ... \theta_{n 2} \theta_{n 1} u_{\pm e_{n1}}\big]_W
\end{equation}
where $[...]_W$ as usual denotes symmetrization over the Weyl group.

The Chevalley-Serre generators can be used to construct the complex moment map operator, which altogether takes the form
\begin{equation}
\label{momentmaps}
\hspace{-.5in}  \begin{array}{l}
\tilde{\mu} := \mu - \left(\frac{N-1}{2}\varepsilon\right) \mathds{1} =  \\[.2cm] \left(\begin{array}{c c c c c c}

\Phi_1 - \frac{1}{N}\Phi_N-\frac{N-1}{2}\varepsilon & V^{+}_1 & V^{+}_{[2:1]} & \ldots & V^{+}_{[N-1:1]}\\
-V^{-}_1 & \Phi_2-\Phi_1- \frac{1}{N}\Phi_N-\frac{N-1}{2}\varepsilon & V^{+}_2 & \ldots & V^{+}_{[N-1:2]}\\ 

V^{-}_{[2:1]} & -V^{-}_2 & \Phi_3-\Phi_2- \frac{1}{N}\Phi_N-\frac{N-1}{2}\varepsilon & \ldots & V^{+}_{[N-1:3]}\\ 
\vdots & \vdots & \vdots & \ddots & \vdots\\ 

(-1)^{N-1}V^{-}_{[N-1:1]} & (-1)^{N-2}V^{-}_{[N-1:2]} & (-1)^{N-3}V^{-}_{[N-1:3]} & \ldots & \frac{N-1}{N}\Phi_N - \Phi_{N-1} -\frac{N-1}{2}\varepsilon
\end{array}\right)
\end{array}.
\end{equation}
We have included the constant shift by $\left(\frac{N-1}{2}\varepsilon\right) \mathds{1}$ for convenience. The general form of the raising and lowering operators is given by
\begin{align} V^{\pm}_{[n:n']} & = \sum\limits_{\alpha = 1}^{n} \sum\limits_{\beta = 1}^{n-1}\dots \sum\limits_{\zeta = 1}^{n'} v_{\pm(e_{n\alpha}+e_{n-1\beta}+...+e_{n'\zeta})} \\
	&= v_{(\vec0;...;\underset{n}{(\pm10...0)};(\pm10...0);...;\underset{n'}{(\pm10...0)};\vec 0;...;0)} + \text{Weyl images}
\end{align}

It turns out that, in the abelianized algebra $\CA_\ve$, the moment map can be explicitly diagonalized. Recall that working in $\CA_\ve$ is the quantum equivalent of working at ``generic'' points in the Coulomb branch. With a bit of work, we find right eigenvectors
\begin{equation}
\eta_i := \left( \begin{array}{c}
\ds \sum\limits_{\alpha = 1}^{N-1} \sum\limits_{\beta = 1}^{N-2}\dots \sum\limits_{\zeta = 1}^{2} \frac{
	v_{e_{N-1\alpha}+e_{N-2\beta}+...+e_{2\zeta}+e_{11}}
}{\varphi_{N i}-\varphi_{(N-1) \alpha} + \varepsilon/2}\\
\ds \sum\limits_{\alpha = 1}^{N-1} \sum\limits_{\beta = 1}^{N-2}\dots \sum\limits_{\zeta = 1}^{2} \frac{
	v_{e_{N-1\alpha}+e_{N-2\beta}+...+e_{2\zeta}}
}{\varphi_{N i}-\varphi_{(N-1) \alpha} + \varepsilon/2} \\ \vdots\\
\ds  \sum\limits_{\alpha = 1}^{N-1} \sum\limits_{\beta = 1}^{N-2} \frac{
	v_{e_{N-1\alpha}+e_{N-2\beta}}
}{\varphi_{N i}-\varphi_{(N-1) \alpha} + \varepsilon/2}\\ 
\ds \sum\limits_{\alpha = 1}^{N-1} \frac{
	v_{e_{N-1\alpha}}
}{\varphi_{N i}-\varphi_{(N-1) \alpha} + \varepsilon/2}\\ 1\\ 
\end{array}\right)\,,
\quad
\tilde \mu\, \eta_i = \big(\varphi_{Ni}-\tfrac1N \Phi_N\big)\eta_i\,
\end{equation}
for $i=1,...,n$, which generalize the $N=2$ formulas \eqref{T23-evecs}. Note that the eigenvectors and their eigenvalues are permuted by the $S_N$ Weyl symmetry associated to the terminal $\tikz[baseline=(char.base)]{\node[shape=circle,draw,inner sep=2pt] (char) {$N$};}$ node of the quiver. 
We assemble these eigenvectors into a similarity transformation
\be S^{-1} = \Big(\, \eta_1\;\eta_2\;\cdots\; \eta_N\,\Big)\,.  \label{TN1-Si} \ee
Its two-sided inverse is given by
\be S = \bp \chi_1 \\\chi_2 \\ \vdots \\ \chi_N \ep\,, \label{TN1-S} \ee
with rows
\begin{equation}
\chi_i^{\textrm{T}} = \left( \begin{array}{c}
\ds (-1)^{N+1}\sum\limits_{\alpha = 1}^{N-1} \sum\limits_{\beta = 1}^{N-2}\dots \sum\limits_{\zeta = 1}^{2} \frac{\left[\prod\limits_{\alpha' \neq \alpha} (\varphi_{N i}-\varphi_{(N-1) \alpha'}-\varepsilon/2)\right]
	v_{-(e_{N-1\alpha}+e_{N-2\beta}+...+e_{2\zeta}+e_{11})}
}{\prod\limits_{\alpha'' \neq i}(\varphi_{N i} - \varphi_{N \alpha''})}\\
\vdots\\
\ds -\sum\limits_{\alpha = 1}^{N-1} \frac{\left[\prod\limits_{\alpha' \neq \alpha} (\varphi_{N i}-\varphi_{(N-1) \alpha'}-\varepsilon/2)\right]
	v_{-e_{N-1\alpha}}
}{\prod\limits_{\alpha'' \neq i}(\varphi_{N i} - \varphi_{N \alpha''})}\\
\ds \frac{\prod\limits_{\alpha = 1}^{N-1}(\varphi_{N i}-\varphi_{(N-1) \alpha}-\varepsilon/2)}{\prod\limits_{\alpha'' \neq i}(\varphi_{N i} - \varphi_{N \alpha''})}
\end{array} \right)\,.
\end{equation}
The diagonal form of the moment map thus becomes
\begin{equation} \tilde \mu_{\rm diag}=
S \tilde{\mu} S^{-1} = \left( \begin{array}{c c c c}
\varphi_{N 1} - \Phi_N/N & 0 & \dots & 0\\
0 & \varphi_{N 2} - \Phi_N/N & \dots & 0\\
\vdots & \vdots & \ddots & \vdots\\
0 & 0 & \dots & \varphi_{N N} - \Phi_N/N\\
\end{array}\right)\,.
\end{equation}

It is worth noting that the determinants of the similarity transformations are given by
\be \label{TN1-detS}
\det S =  \frac{v_{(\vec{0};-\vec 1,\dots,-1)}}{\prod\limits_{1 \leq i < j \leq N}(\varphi_{N j} - \varphi_{N i})}\,,\qquad 
\det S^{-1} =  \frac{\left[\prod\limits_{1 \leq i < j \leq N}(\varphi_{N j} - \varphi_{N i})\right] v_{(\vec{0};\vec{1},\dots,1)}}{\prod\limits_{i = 1}^N\prod\limits_{\alpha = 1}^{N-1}(\varphi_{N i}-\varphi_{(N-1) \alpha}+\varepsilon/2)}\,,
\ee
which generalize the $N=2$ expressions. Just as in Section \ref{sec:T23-Qdiag}, $\det S$ and $\det S^{-1}$ are defined as \emph{row}-determinants, \emph{e.g.} $\det S := \epsilon_{i_1...i_N} S^{i_1}{}_1... S^{i_N}{}_N$. The determinants satisfy
\begin{equation}
\det S \hspace{1mm} \det S^{-1} = \det S^{-1} \hspace{1mm} \det S = 1\,.
\end{equation} 
There seems to be no way to restore symmetry between $\det S$ and $\det S^{-1}$ (or make them unimodular) while keeping $S$ and $S^{-1}$ valued in $\CA_\ve$ --- restoring symmetry would introduce roots.

\subsection{Tensor operators}
\label{sec:TN1-tensor}

As mentioned at the beginning of this section, the chiral ring of $\CT_{1,N}$ contains additional operators in arbitrary antisymmetric tensor powers of the fundamental and antifundamental representations of $SL(N,\C)$. We would like to identify them as elements of   the abelianized algebra $\CA_\ve$ (more precisely, as element of $\CW_\ve$), and then investigate how the similarity matrices act on them.

In building the moment map, we used monopole operators charged under all nodes of the quiver except the first $U(N)$ node. So, let us now consider
\be Q^N :=  V_N^+ = \sum_{\alpha=1}^N v_{e_{N\alpha}} = v_{((10...0);\vec 0;...;0)} + \text{Weyl images}\,,\ee
with fundamental magnetic charge on the $U(N)$ node.
Comparing with the $N=2$ analysis from Section \ref{sec:2kquivers}, it is natural to guess that the operator $Q^N$ generates a fundamental representation.%
\footnote{Without knowledge of the $N=2$ results, another easy way to obtain the fundamental and antifundamental operators is by decomposing the adjoint representation of $SU(N+1)$ into $SU(N)$ irreducibles. By inspection of  \eqref{momentmaps}, it follows that the fundamental (resp. antifundamental) representation is generated by the operator $Q^{N} = V^+_N$ (resp. $Q_N = -V^{-}_N$).} %
We prove in Appendix \ref{app:reps} that it is indeed a lowest-weight vector, in the sense that
\be [F_n,V^+_N]=0\quad \forall\; n \ee
The remaining operators in the fundamental representation may be expressed as Weyl-symmetric sums
\begin{align} Q^i  = V^+_{[N:i]} &=  \sum_{\alpha=1}^N\sum_{\beta=1}^{N-1}\cdots \sum_{\zeta=1}^{i} v_{e_{N\alpha}+e_{N-1\beta}+...+e_{i\zeta}} \\
	&=v_{(\underset{N}{(10...0)};\underset{N-1}{(10...0)},...,\underset{i}{(10...0)},\underset{i-1}{\vec 0},...,0)}  +(\text{Weyl images})\,. \notag
\end{align}
In other words, the $Q^i$ are nonabelian monopole operators with fundamental magnetic charge on the $U(N)$ node through the $U(i)$ node.

The antifundamental representation is similar. It is generated by the highest-weight vector $Q_N=-V^-_N$, and more generally contains Weyl-symmetric sums
\begin{align} Q_i  &= (-1)^{N-i+1}\sum_{\alpha=1}^N\sum_{\beta=1}^{N-1}\cdots \sum_{\zeta=1}^{i} v_{-(e_{N\alpha}+e_{N-1\beta}+...+e_{i\zeta})} \label{TN1-af} \\
	&= (-1)^{N-i+1}v_{(\underset{N}{(-10...0)};\underset{N-1}{(-10...0)},...,\underset{i}{(-10...0)},\underset{i-1}{\vec 0},...,0)}  +(\text{Weyl images})\,. \notag
\end{align}

In order to construct higher antisymmetric tensor representations $\wedge^r\square$, we consider monopole operators whose magnetic charges on the various nodes are those of antisymmetric tensors. For example, a collection of operators $Q^{[i_1i_2]}$ ($1\leq i_1 < i_2 \leq N$) furnishing the representation $\wedge^2\square$ may be constructed as Weyl-symmetric sums
\begin{align} Q^{[i_1i_2]} &= - \sum_{1\leq \alpha 1 <\alpha_2\leq N} \cdots \sum_{1\leq \gamma_1< \gamma_2 \leq i_2}
	\sum_{1\leq \delta \leq i_2-1} \cdots \sum_{1\leq \zeta\leq i_1} v_{e_{N\alpha_1}+e_{N\alpha_2}  + ... + e_{i_2\gamma_1}+e_{i_2\gamma_2} + e_{i_2-1\delta} + ...+e_{i_1\zeta}} \\
	&= 
	-v_{(\underset{N}{(110...0)};\underset{N-1}{(110...0)},...,\underset{i_2}{(110...0)},\underset{i_2-1}{(10...0)},...,\underset{i_1}{(10...0)},\vec 0,...,0)}+ (\text{Weyl images})\,.
\end{align}
The operators $Q^{[i_1i_2i_3]}$ in the $\wedge^3\square$ representation are sums over the Weyl images of $v_{A_{[i_1i_2i_3]}}$, where
\be A_{[i_1i_2i_3]} = (\underset{N}{(1110...0)};\underset{N-1}{(1110...0)},...,\underset{i_3}{(1110...0)},(110...0),...,\underset{i_2}{(110...0)},(10...0),...,\underset{i_1}{(10...0)},\vec0,...,0)\,, \ee
and more generally the operators $Q^{[i_1...i_r]}$ in the $\wedge^r\square$ representation may be constructed as sums over the Weyl images of $(-1)^{r-1} v_{A_{[i_1...i_r]}}$ (when $1\leq i_1 < ... < i_r \leq N$) with
\begin{equation} \label{A-Qr}
A_{[i_1...i_r]} =  \sum\limits_{n = i_1}^{N} \vec{e}_{n 1} +\sum\limits_{n = i_2}^{N} \vec{e}_{n 2} + \ldots + \sum\limits_{n = i_r}^{N} \vec{e}_{n r}\,.
\end{equation}
The process must stop at $r=N$, where the cocharacter
\be A_{[12...N]} = (\vec 1;\vec1,...,1) \ee
is saturated with $1$'s. In fact, $A_{[12...N]}= A_{\rm diag}$ is equivalent to the trivial cocharacter, so $Q^{[1...N]}= (-1)^{N-1} v_{\rm A_{\rm diag}}= (-1)^{N-1} $.

We prove in Appendix \ref{app:reps} that the operators $Q^{[N,N-1,...,N-r]}$ are indeed lowest-weight vectors for the $SL(N,\C)$ action generated by the moment map. From there it is straightforward to check (by taking commutators with the moment map) that the remaining operators in each $\wedge^r\square$ representation (and its dual) are of the form given here.

Similarly, the dual representations $\wedge^r\ol\square$ contain operators $Q_{[i_1...i_r]}$ that are Weyl sums of the negative cocharacters $-A_{[i_1...i_r]}$ from \eqref{A-Qr}. A beautiful relation between the $\wedge^r\square$ operators and the $\wedge^{N-r}\ol \square$ operators comes from using the fact that two cocharacters that differ by a multiple of the diagonal $A_{\rm diag}=(\vec 1;\vec 1,...,1)$ are equivalent. By subtracting $A_{\rm diag}$ from \eqref{A-Qr}, we obtain a cocharacter $A_{[i_1...i_r]}-A_{\rm diag}$ that is in the same Weyl orbit as $-A_{[\hat i_1,...,\hat i_{N-r}]}$, where $\{i_1,...,i_r\}$ and $\{\hat i_1,...,\hat i_{N-r}\}$ are complementary ordered subsets of $\{1,...,N\}$.
For example, taking $N=5$ and $[i_1i_2]=[24]$ we find
\be \begin{array}{rl} A_{[24]} &= ((11000);(1100),(100),(10),(0)) \\[.1cm]
	A_{[24]}-A_{\rm diag} &= -((00111);(0011),(011),(01),(1)) \\[.1cm] & \overset{\text{Weyl}}\sim -((11100);(1100),(110),(10),(1)) = -A_{[135]}\end{array}\,.\ee
With the overall signs chosen above, this translates to a general operator relation
\begin{equation} \label{TN1-Q=Q}
\frac{1}{r!}\epsilon_{I I'} Q^{I'} = Q_{I}\,,
\end{equation}
where $I, I'$ are multi-indices of size $N-r$ and $r$, respectively.

We note that all the $Q^{[i_1...i_r]}$ and $Q_{[i_1...i_r]}$ operators belong to the subalgebra $\CW_\ve\subset \CA_\ve$. 
This can be shown directly by generalizing \eqref{VnW} to express $Q^{[i_1...i_r]}$ as a Weyl-average of the rescaled monopole operator $u_{A_{[i_1...i_r]}}$, hit with an appropriate number of BGG-Demazure operators. For $n > 1 $, set \be
\Theta^{(k)}_{n} = (\theta_{n k}...\theta_{n 2} \theta_{n 1})... (\theta_{n (n-2)}... \theta_{n (n-k)} \theta_{n (n-k-1)})(\theta_{n (n-1)}...\theta_{n (n-k+1)} \theta_{n (n-k)})\,
\ee and $n = 1$ \be
\Theta^{(k)}_{1} = 1\,;
\ee  then for $1 \leq i_1 < ... < i_r \leq N$
\be \label{rDem}
[ v_{A_{[i_1...i_r]}}]_W \;\propto\;  \big[\Theta^{(r)}_{i_r} ... \Theta^{(2)}_{i_2} \Theta^{(1)}_{i_1} u_{A_{[i_1...i_r]}}\big]_W\,,
\ee from which it follows that $Q^I$ and $Q_I$ belong to $\CW_\ve$.
Alternatively, this follows from the facts that 1) $Q^N=V_N^+$ and $Q_N=-V_N^-$ are elements of $\CW_\ve$ due to \eqref{VnW}; 2) the remaining $Q^i,Q_i$ in the (anti)fundamental representations are elements of $\CW_\ve$ because they are obtained from $Q^N$, $Q_N$ by taking commutators with moment maps; and 3) the higher tensors $Q^{[i_1...i_r]}, Q_{[i_1...i_r]}$ are in $\CW_\ve$ because they can be written as products of the $Q^i,Q_i$ --- see \eqref{fund2tensor} below.

\subsection{New basis for the tensors}

For $\k=1$, the $Q$ operators are already ``diagonalized.'' Nevertheless, it is natural to ask how they are represented in the basis of eigenvectors of the moment map. The results will help us later in Section \ref{Nkquivers} to diagonalize the $Q$'s for general $\TNk$ quivers.

Let us define
\be q^j = S^j{}_i Q^i\,,\qquad q_j = Q_i (S^{-1})^i{}_j\,, \ee
and more generally 
\be \label{TN1qQ1}  q^J = S^{[j_1}{}_{[i_1} \dots S^{j_r]}{}_{i_r]}Q^I\,,\qquad q_J = Q_I (S^{-1})^{[i_1}{}_{[j_1} \dots (S^{-1})^{i_r]}{}_{j_r]}\,,\ee
where $I=[i_1...i_r]$, $J=[j_1...j_r]$ are antisymmetric multi-indices. By direct evaluation of the RHS, we find that%
\footnote{We have checked \eqref{TN1qQ1} and instances of \eqref{TN1qQ} explicitly for $N\geq 5$, though do not yet have a general proof for all $N$.}
\be q^i =   v_{e_{Ni}}\,,\qquad
q_i =  -\frac{\prod\limits_{j \neq i}(\varphi_{N i}-\varphi_{N j}+ \varepsilon)}{\prod\limits_{\alpha = 1}^{N-1}(\varphi_{N i} - \varphi_{(N-1) \alpha} + \varepsilon/2)} v_{-e_{Ni}}\,.
\ee
In this basis, the magnetic charge has been fully shifted to the $U(N)$ node! The (anti)fundamental $q^i$ ($q_i$) are magnetically charged in a (anti)fundamental representation of $U(N)$.

Similarly, we find for higher tensors that
\begin{subequations} \label{TN1qQ}
	\begin{equation}
	q^I = (-1)^{r+1} \frac{v_{e_{Ni_1}+...+e_{Ni_r}}}{\prod \limits_{\substack{i_n,i_m \in I \\ n<m}}(\varphi_{N i_m} - \varphi_{N i_n})}\,,
	\end{equation}
	and, with the same general form but a more complicated prefactor,
	\begin{equation} 
	q_I = (-1)^{N(r-1)+1} \prod \limits_{i_n \in I} \left(\frac{\bigg[\prod \limits_{\substack{i_m \in I \\ n<m}}(\varphi_{N i_m} - \varphi_{N i_n})\bigg] \bigg[ \prod \limits_{i_p \in I_c}(\varphi_{N i_n} - \varphi_{N i_p} + \varepsilon)\bigg]}{\prod \limits_{\alpha = 1}^{N-1}(\varphi_{N i_n} - \varphi_{(N-1) \alpha} + \varepsilon/2)}\right) v_{-(e_{Ni_1}+...+e_{Ni_r})}\,.
	\end{equation} 
\end{subequations}
Once again, all magnetic charge has shifted to the $U(N)$ node, and has been abelianized. The base-changed $\wedge^r\square$ operators now have magnetic charges in the $\wedge^r\square$ representations of the central $U(N)$.%
\footnote{We note again that when referring to magnetic charge (\emph{i.e.} a cocharacter) as a representation, we mean a representation of the Langlands-dual group. Here $U(N)$ is its own Langlands-dual, so there is no confusion.}

Just as in the $N=2$ theories of Section \ref{sec:2kquivers},, the asymmetry in prefactors of \eqref{TN1qQ} is a result of asymmetry in our choice of similarity transformations.
In the new basis, the relation between fundamental and antifundamental powers is expressed as
\begin{equation}
q^{I} =  \frac{1}{(N-r)!} q_{I'} \epsilon^{I'I}  \det S\,,\qquad	\frac{1}{r!} \epsilon_{I' I} q^{I} \det S^{-1} = q_{I}\,,
\end{equation}
where $I$ is a multi-index of length $r$ and $I'$ is of length $N-r$. Thus, the ``undiagonalized'' relation \eqref{TN1-Q=Q} is corrected by determinants.

In the new basis it is fairly easy to relate the fundamental $q^i,q_i$ to higher tensor powers. Letting $I = \{i_1, ..., i_r\}$, we find by direct computation that  \begin{equation}
q^{i_1} ... q^{i_r} = \frac{v_{(\vec{e}_{I},\vec{0}, \dots, 0)}}{\prod \limits_{\substack{i_n,i_m \in I \\ n<m}}(\varphi_{N i_m} - \varphi_{N i_n})(\varphi_{N i_n} - \varphi_{N i_m} - \ve)}
\end{equation} and \begin{equation}
q_{i_1} ... q_{i_r} = (-1)^{r} \prod \limits_{i \in I} \left(\frac{\prod \limits_{j \in I_c}(\varphi_{N i} - \varphi_{N j} + \varepsilon)}{\prod \limits_{\alpha = 1}^{N-1} (\varphi_{N i} - \varphi_{(N-1) \alpha} + \varepsilon/2)}\right) v_{(-\vec{e}_{I},\vec{0}, \dots, 0)}\,,
\end{equation} where $I_c = \{1, ..., N\} \backslash I$ is the complementary subset, and $\vec{e}_{I} \in \mathbb{Z}^N$ is a vector with $1$ entries for positions in $I$ and zeroes elsewhere.%
\footnote{It is interesting to note that the $q_i$'s commute amongst themselves, this feature does not persist for $\k > 1$.} %
Combining these formulas with \eqref{TN1qQ}, we arrive at
\begin{subequations} \label{TN1-qprods}
	\begin{equation}
	q^{i_1} q^{i_2} ... q^{i_r} = (-1)^{r+1} \bigg[\prod \limits_{\substack{i_n, i_m \in I \\ n < m}} \frac{1}{\varphi_{N i_n} - \varphi_{N i_m} - \varepsilon}\bigg] q^{[i_1 ... i_r]}
	\label{qprod-q-N1}
	\end{equation} and \begin{equation}
	q_{i_1} q_{i_2} ... q_{i_r} = (-1)^{(N-1)(r+1)} \bigg[ \prod \limits_{\substack{i_n, i_m \in I \\ n < m}} \frac{1}{\varphi_{N i_m} - \varphi_{N i_n}}\bigg] q_{[i_1 ... i_r]}\,.
	\end{equation}
\end{subequations}
These expressions will be used momentarily to determine how to express $Q^I$ and $Q_I$ in terms of products of $Q^i$ and $Q_i$.

\subsection{Relations}
\label{sec:TN1-rels}

As mentioned at the beginning of this section, we expect the chiral ring $\C_\ve[\CM_C]$ of the one-legged quiver $\CT_{N,1}$ to be generated by the moment map and the fundamental $Q^i$'s alone. In principle, the relations between fundamental $Q^i$'s and higher tensors can be obtained by carefully applying quantum similarity transformations to the simple relations \eqref{TN1-qprods} above.

As a preliminary step, it is helpful to move the denominators on the right side of (say) \eqref{qprod-q-N1} to the left side. By multiplying both sides on the left and commuting the factors $\varphi_{N j} - \varphi_{N i}$ through some of the $q^i$'s, we can bring the relation to the form 
\begin{equation} \label{qmuq}
q^{i_1} \Big[\prod \limits_{i_n > i_1}(\varphi_{N i_1} - \varphi_{N i_n})\Big] q^{i_2} \Big[\prod \limits_{i_n > i_2}(\varphi_{N i_2} - \varphi_{N i_n})\Big] ... q^{i_{r-1}} (\varphi_{N i_{r-1}} - \varphi_{N i_r}) q^{i_r} = (-1)^{r+1} q^I.
\end{equation}
This is \emph{suggestive} of an un-diagonalized relation
\begin{equation}
\label{fund2tensor}
r! (\tilde{\mu}^{r-1}Q)^{[i_1} \dots (\tilde{\mu}Q)^{i_{r-1}} Q^{i_r]} =(-1)^{r+1} Q^I\,,
\end{equation}
where $(\tilde{\mu}^nQ)^i := (\tilde{\mu}^n)^i{}_j Q^j$. At least classically, \eqref{fund2tensor} diagonalized precisely to \eqref{qmuq}. By explicit computation for $N \leq 4$ and $r \leq N$, we find that \eqref{fund2tensor} holds even at the quantum level and we conjecture that this relation holds for all $N$ and $r$. (The explicit computations appear in Appendix \ref{tensorcomps}.)

By working out the explicit commutation relations between the components of $S, S^{-1}, q^i, q_i$, one could imagine deriving \eqref{fund2tensor} from \eqref{qmuq} in full generality. We do not do this here. The analogous formula for antifundamentals is given by
\begin{equation} \label{anti2tensor}
r! Q_{[i_1} (Q \tilde{\mu})_{i_2} \dots (Q\tilde{\mu}^{r-1})_{i_r]} =(-1)^{(N-1)(r+1)}Q_I\,.
\end{equation} where $(Q\tilde{\mu}^n)_i := Q_j (\tilde{\mu}^n)^j{}_i $.

We also expect additional relations obtained from \eqref{fund2tensor} and \eqref{anti2tensor} via (signed) permutations of the $(\tilde{\mu}^n Q)^i$ and $(Q\tilde{\mu}^n)_i$ factors.%
\footnote{To better understand these formulae, consider the contraction of $\tilde{\mu}$ with $Q^i$. Taking the form of $\tilde{\mu}$ given above and diligently using  \eqref{comm3}, one finds that this contraction is a telescoping sum, and \begin{equation}
	(\tilde{\mu}^n Q)^j = \sum\limits_{\alpha=1}^{N!/(j-1)!} \left(\varphi_{N \alpha_{N,j}} - \frac{1}{N} \Phi_N\right)^n v_{A_{\alpha,j}}.
	\end{equation} A nearly identical formula appears for the antifundamental operator $(\tilde{\mu}Q)_{j} := \tilde{\mu}^{j'}{}_j Q_{j'}$; one simply tacks on an appropriate factor of $\left(\varphi_{N \alpha_{N,j}} - \frac{1}{N}\Phi_N - (N-1)\varepsilon\right)$ to each abelianized monopole found in $Q_j$. The main feature of note is that the coefficients in each of these expressions only depend on the scalars on the central node. When moving onto multi-legged quivers this will greatly simplify computations. In particular, the operation of contracting a moment map (with its respective tensor index) will have the same result regardless of which moment map is used.}

\section{The general case}
\label{Nkquivers}

The structure we found for ``short'' $\k$-legged quivers in Section \ref{sec:2kquivers} and ``long'' single-legged quivers in Section \ref{N1quivers} generalizes in a straightforward way to general $\TNk$ star quivers. Indeed, analyzing general $\TNk$ quivers is largely a matter of bookkeeping.

We shall describe the general results here, starting with the identification of moment maps and fundamental/antifundamental tensors (and higher tensors) as elements of the abelianized algebra $\CW_\ve\subset \CA_\ve$; then diagonalizing the tensors; and finally using diagonalized relations to derive (or motivate) a general collection of relations in $\C_\ve[\CM_C]$.

\subsection{Conventions}

The gauge group of $\TNk$ is
\be G = \Big[ U(N) \times \prod_{a=1}^\k U(N-1)_a\times U(N-2)_a \times\cdots\times U(1)_a \Big]/U(1)_{\rm diag}\,. \ee
Correspondingly, the cocharacter lattice is $(\Z^{N+\frac{\k}{2}N(N-1)})/\Z_{\rm diag}$. When needed, we write explicit cocharacters as vectors
\be A = (\vec A_N\,;\, \vec A_{N-1}^1,...,\vec A_{1}^1\,;\, \vec A_{N-1}^2,...,\vec A_{1}^2\,;\, \ldots\,;\, \vec A_{N-1}^\k,...,\vec A_{1}^\k ) \in \text{cochar}(G)\,,\ee
subject to an identification $A\sim A'$ if $A-A'$ is a multiple of $A_{\rm diag} = (\vec 1;....;\vec 1)$ (\emph{i.e.} the vector with every entry equal to $1$). Here $\vec A_N = (A_{N1},...,A_{NN})$ is the cocharacter associated to the central $U(N)$ node of the quiver, while $\vec A_n^a = (A^a_{n1},...,A^a_{nn})$ is the cocharacter associated to the $U(n)$ node on the $a$-th leg, with $1\leq n\leq N-1$ and $1\leq a\leq \k$.
In a similar way, we denote the diagonal gauge scalars as
\be \varphi = (\vec \varphi_N\,;\, \vec \varphi_{N-1}^1,...,\vec \varphi_{1}^1\,;\, \vec \varphi_{N-1}^2,...,\vec \varphi_{1}^2\,;\, \ldots\,;\, \vec \varphi_{N-1}^\k,...,\vec \varphi_{1}^\k ) \in \text{cochar}(G)\otimes \C \ee
where $\vec\varphi_N = (\varphi_{N\alpha})_{\alpha=1}^N$ and $\vec\varphi_{n}^a = (\varphi_{n\alpha}^a)_{\alpha=1}^n$ for $1\leq n\leq N-1$\,.

As before, it is helpful to introduce basis vectors for the unquotiented cocharacter lattice. Let $e_{N\alpha}$ be the cocharacter with $A_{N\alpha}=1$ and all remaining entries ($A_{N\beta}$ and all $\vec A_n^a$) set to zero. Similarly, let $e_{n\alpha}^a$ be the cocharacter with $A_{n\alpha}^a=1$ and all remaining entries set to zero.

The weight lattice of $G$ is  $\Z^{N+\frac{\k}{2}N(N-1)-1}$, consisting of vectors
\be \lambda = (\vec \lambda_N\,;\, \vec \lambda_{N-1}^1,...,\vec \lambda_{1}^1\,;\, \vec \lambda_{N-1}^2,...,\vec \lambda_{1}^2\,;\, \ldots\,;\, \vec \lambda_{N-1}^\k,...,\vec \lambda_{1}^\k ) \in \text{weights}(G)\,,\ee
whose entries sum to zero. We can interpret differences of the $e_{N\alpha}$, $e^a_{n\alpha}$ as elements of the weight lattice. The weights of the hypermultiplet representation $\CR=R\oplus R^*$ are taken to be
\be \text{weights}(R) = \big\{ e_{N\alpha}-e_{N-1\beta}^a\big\}_{\begin{smallmatrix} 1\leq a\leq \k \\1\leq \alpha\leq N\\ 1\leq \beta\leq N-1\end{smallmatrix}} \cup\; 
\big\{ e_{n\alpha}^a-e_{n-1\beta}^a\big\} _{\begin{smallmatrix} 1\leq a\leq \k \\ 2\leq n\leq N-1\\1\leq \alpha\leq n\\ 1\leq \beta\leq n-1\end{smallmatrix}}\,.
\ee

The abelianized algebra $\CA_\ve$ is generated in the usual way by the entries of $\varphi$, monopole operators $v_A$, inverted roots 
\be \frac{1}{\varphi_{N\alpha}-\varphi_{N\beta}+p\,\ve}\,,\qquad \frac{1}{\varphi_{n\alpha}^a-\varphi_{n\beta}^a+p\,\ve}\qquad (\alpha\neq \beta)\,, \ee
and inverted weights
\be \frac{1}{\varphi_{N\alpha}-\varphi_{N-1\beta}^a+p\,\ve}\,,\qquad \frac{1}{\varphi_{n\alpha}^a-\varphi_{n-1\beta}^a+p\,\ve}\,. \ee
The subalgebra $\CW_\ve$, in which we will find all the components of our moment maps and $\k$-fold tensors, is generated by $\varphi$, rescaled monopoles $u_A$, simple Weyl reflections $s_{N\alpha}, s_{n\alpha}^a$ (on the central $U(N)$ node and the $U(n)_a$ nodes on the $a$-th leg, respectively), and corresponding BGG-Demazure operators $\theta_{N\alpha},\theta_{n\alpha}^a$.

\subsection{Moment maps and tensor operators}
\label{sec:Nk-ops}

The chiral ring now has a $\prod_{a=1}^\k SL(N,\C)_a$ action, generated by moment maps $\mu_a$, $a=1,...,\k$. These moment maps are associated to each leg of the star quiver, and are decoupled from each other. Indeed, each $\mu_a$ may be identified as a copy of \eqref{momentmaps} from Section \ref{sec:TN1-mu}. Explicitly, for each $n=1,...,N-1$ and $a=1,...,\k$ we may define nonabelian monopole operators
\be V_n^{a\pm} := \sum_{\alpha=1}^n v_{\pm e_{n\alpha}^a}
\ee
with (anti)fundamental magnetic charge for $U(n)_a$. Since $V_n^{a\pm}\propto \big[\theta^a_{n n-1} ... \theta^a_{n 2} \theta^a_{n 1} u_{\pm e^a_{n1}}\big]_W$, it is an element of $\CW_\ve$. Similarly, we introduce Weyl-symmetric scalars
\be \Phi_N := \sum_{\alpha=1}^N \varphi_{N\alpha}\,,\qquad \Phi_n^a = \sum_{\alpha=1}^n \varphi_{n\alpha}^a\ee
on each node. Then the Chevalley-Serre generators of $SL(N,\C)_a$ are
\be E_n^a = V_n^{a+}\,,\qquad F_n^a = -V_n^{a-}\,,\qquad H_n^a = 2\Phi_n^a-\Phi_{n-1}^a-\Phi_{n+1}^a \qquad (n=1,...,N-1)\,, \ee
with the convention that $\Phi_N^a\equiv \Phi_N$ and $\Phi_0=0$. They fit into the $\ve$-shifted moment maps $\tilde\mu_a$ just as in \eqref{momentmaps}. Since each $\tilde\mu_a$ only contains monopoles (and scalars) charged under nodes on the $a$-th leg, the components of the moment maps commute with each other, $[(\tilde \mu_a)^{i_a}{}_{j_a},(\tilde \mu_b)^{i_b}{}_{j_b}]=0$ ($a\neq b$).

Generalizing Sections \ref{sec:2kquivers} and \ref{N1quivers}, the operators in $\k$-fold fundamental and antifundamental representations now arise from adding magnetic charge on the central node. Let
\be V_N^\pm := \sum_{\alpha=1}^N v_{\pm e_{N\alpha}}\quad = \big[\theta_{NN-1} ... \theta_{N 1} u_{\pm e_{n1}}\big]_W \in \CW_\ve \ee
be the nonabelian monopoles of (anti)fundamental magnetic charge on the central node.
Then we show in Appendix \ref{app:reps} that
\be Q^{NN...N} = V_N^+\,,\qquad Q_{NN...N} = (-1)^\k V_N^- \ee
are \emph{simultaneous} lowest-weight and highest-weight vectors (respectively) for every copy of $SL(N,\C)_a$. By taking repeated commutators with the moment-map components $E_n^a$ (resp. $F_n^a$), they generate $k$-fold fundamental and (resp. antifundamental) representations of $SL(N,\C)^\k$. We denote the remaining operators in these representations as
\be Q^{i_1...i_\k}\,,\qquad Q_{i_1... i_\k}\,,\qquad 1\leq i_a\leq N\,. \ee
Following the pattern of Section \ref{sec:TN1-tensor}, we find that $Q^{i_1...i_\k}$ is a nonabelian monopole operator with fundamental magnetic charge for the central $U(N)$ node as well as the $U(n-1)_a,...,U(i_a)_a$ nodes of each $a$-th leg. The operators $Q_{i_1... i_\k}$ are similar, involving antifundamental magnetic charges instead.

To find the higher $\k$-fold antisymmetric tensor representations, we repeat the same process, starting with nonabelian monopole operators that have the corresponding $\k$-fold antisymmetric magnetic charge on the central node. Namely, for any $1\leq r\leq N$, we may consider the the Weyl-averaged operator
\begin{align} &Q^{[N-r,N-r+1,...,N][N-r,N-r+1,...,N] ... [N-r,N-r+1,...,N]} \\
	&= (-1)^{\k(r-1)}[v_{A_{[N-r,N-r+1,...,N][N-r,N-r+1,...,N] ... [N-r,N-r+1,...,N]}}]_W\,. \notag
\end{align} where 
\be
A_{[N-r,N-r+1,...,N][N-r,N-r+1,...,N] ... [N-r,N-r+1,...,N]} = \sum \limits_{\alpha = 1}^{r} e_{N \alpha} + \sum \limits_{a = 1}^{\k} \sum \limits_{\alpha = 1}^r \sum \limits_{n = N-r+\alpha-1}^{N-1} e^a_{n \alpha}
\ee
This is again a simultaneous lowest-weight vector for every $SL(N,\C)_a$ action, and generates an entire $\k$-fold $r$-th antisymmetric power representation $\wedge^r\square_1\otimes \cdots \otimes \wedge^r\square_k$. The remaining operators in the representation may be denoted
\be Q^{I_1...I_\k}\,, \ee
with antisymmetric multi-indices $I_a = [i^a_1,...,i^a_r]$. They have various combinations of magnetic charge in the $\leq r$-th antisymmetric representations, on various legs nodes, generalizing \eqref{A-Qr}. Explicitly, if $1 \leq i^a_1 < ... < i^a_r \leq N$ and $I_a = [i^a_1 ... i^a_r]$ then \be Q^{I_1 ... I_{\k}} = (-1)^{\k(r-1)}[v_{A_{I_1 ... I_{\k}}}]_W\,. \notag
\ee where %
\be
A_{I_1 ... I_{\k}} = \sum \limits_{\alpha = 1}^{r} e_{N \alpha} + \sum \limits_{a = 1}^{\k} \sum \limits_{\alpha = 1}^r \sum \limits_{n = i^a_{\alpha}}^{N-1} e^a_{n \alpha}
\ee

Similarly, the dual tensors powers of the antifundamental representation $\wedge^r\ol \square_1\otimes \cdots \otimes \wedge^r\ol \square_k$ are defined to satisfy
\be \label{TNk-Q=Q}
\Big( \bigotimes_{a=1}^\k \frac{1}{r!} \epsilon_{J_aI_a} \Big) Q^{I_1I_2...I_\k} = Q_{J_1J_2...J_\k}\,,\quad
Q^{I_1I_2...I_\k} =   Q_{J_1J_2...J_\k}\Big( \bigotimes_{a=1}^\k \frac{1}{(N-r)!} \epsilon^{J_aI_a}\Big)\,.
\ee which is a natural genealization of \eqref{TN1-Q=Q}. By a generalization of \eqref{rDem}, it follows that $Q^{I_1...I_\k}$ and $Q_{I_1...I_\k}$ are all elements of $\CW_\ve$.

We will \emph{assume} that the chiral ring $\C_\ve[\CM_C]$ is entirely generated by the components of the moment maps $\tilde \mu_a$, and the operators in the various tensor representations $Q^{I_1...I_\k}$ and $Q_{I_1...I_\k}$. Since these operators are all in $\CW_\ve$ they must (by the discussion in Section \ref{sec:image}) belong to $\C_\ve[\CM_C]$. We do not have a general proof that they generate.

Let us also take a moment to summarize R-charges of various operators. Using the general formula \eqref{R-vA}, we find that
\be \label{TNk-R} \begin{array}{l}
	[\mu_a] = 1 \\[.2cm]
	[Q^{i_1...i_\k}]=[Q_{i_1...i_\k}] = \frac12(\k-2)(N-1)  \\[.2cm]
	[Q^{I_1...I_\k}]=[Q_{I_1...I_\k}] = \frac12(\k-2)r(N-r) \qquad |I_i|=r\,,
\end{array}
\ee
where in the final row we have the $\wedge^r\square_1\otimes \cdots \otimes \wedge^r\square_k$ operators and their duals.

\subsection{Diagonalization and diagonalized relations}

Since the moment maps $\tilde \mu_a$ on different legs decouple from each other, the quantum similarity transformations $S_a,S_a^{-1}$ that diagonalize $\tilde \mu_a$ are simply obtained by specializing the formulas \eqref{TN1-Si}, \eqref{TN1-S} to the $a$-th leg. (In other words, starting from \eqref{TN1-Si}, \eqref{TN1-S} we simply replace $\varphi_{n\alpha}\to \varphi_{n\alpha}^a$ and $e_{n\alpha}\to e_{n\alpha}^a$ to get $S_a$ and $S_a^{-1}$.) Then for every $a$,
\be \tilde{\mu}_{\textrm{diag}} = S_a \tilde{\mu}_a S^{-1}_a =  \left( \begin{array}{c c c c}
	\varphi_{N 1} - \Phi_N/N & 0 & \dots & 0\\
	0 & \varphi_{N 2} - \Phi_N/N & \dots & 0\\
	\vdots & \vdots & \ddots & \vdots\\
	0 & 0 & \dots & \varphi_{N N} - \Phi_N/N\\
\end{array}\right)\,. \ee
Note that the eigenvalues $m_i := \varphi_{Ni}-\tfrac 1N \Phi_N$ are independent of the choice of leg, as they only contain scalars on the central node.

Just as in the $\CT_{2,\k}$ theories of Section \ref{sec:2kquivers}, the matrix elements of the $S_a$ and $S^{-1}_a$ commute with the components of $\tilde{\mu}_b$, $S_b$ and $S^{-1}_b$ for $b \neq a$ as well as with the components of $\tilde{\mu}_{\textrm{diag}}$.

We can now use the similarity transformations to diagonalize the $\k$-fold tensor operators.
By generalizing explicit computations for low values of $\k$ and $N$, we find that 
\begin{equation}
\begin{array}{rcl}
(S_1)^{i_1}{}_{j_1} (S_2)^{i_2}{}_{j_2} ... (S_{\k})^{i_{\k}}{}_{j_{\k}} Q^{j_1 j_2 ... j_{\k}} &=& \delta^{i_1}{}_i \delta^{i_2}{}_i ... \delta^{i_{\k}}{}_i q^i\,,\\[.2cm]
Q_{j_1 j_2 ... j_{\k}} (S_1^{-1})^{j_1}{}_{i_1} (S_2^{-1})^{j_2}{}_{i_2} ... (S_{\k}^{-1})^{j_{\k}}{}_{i_{\k}} &=& \delta^i{}_{i_1} \delta^i{}_{i_2} ... \delta^i{}_{i_{\k}} q_i\,,
\end{array}
\end{equation}
with
\begin{equation}
\label{Qdiag}
q^i =   v_{e_{Ni}}\,,\qquad 
q_i =  -\frac{\prod\limits_{j \neq i}(\varphi_{N i}-\varphi_{N j}+ \varepsilon)^{\k}}{\prod\limits_{a =1}^{\k}\prod\limits_{\alpha = 1}^{N-1}(\varphi_{N i} - \varphi^a_{(N-1) \alpha} + \varepsilon/2)} v_{-e_{Ni}}\,.
\end{equation}
This is a rather amazing simplification. One could have guessed that diagonalization should be possible from the chiral-ring relations \eqref{muQrel} and  \eqref{Qmurel} below. A nice surprise, however, is that the eigenvalues $q^i,q_i$ are extremely simple elements of the abelianized algebra $\CA_\ve$, containing abelian monopoles charged \emph{only} under the central node.

Similarly, we can use $S_a$ and $S_a^{-1}$ to diagonalize the operators in the $\wedge^r\square_1\otimes\cdots \otimes \wedge^r\square_\k$ and  $\wedge^r\ol\square_1\otimes\cdots \otimes \wedge^r\ol\square_\k$. Letting $I=[i_1...i_r]$ be an antisymmetric multi-index and $I_c=\{1....N\}\backslash I$ its complement (as a set), we find eigenvalues
\begin{equation}
q^I = \frac{(-1)^{r+1} }{\prod \limits_{\substack{i_n, i_m \in I \\ n<m}}(\varphi_{N i_m} - \varphi_{N i_n})^{\k}} v_{e_{Ni_1}+...+e_{Ni_r}}\,,
\end{equation}  \begin{equation}
q_I = (-1)^{N(r-1)+1} \prod \limits_{i_n \in I}\left(\frac{\bigg[\prod \limits_{\substack{i_m \in I \\ n < m}}(\varphi_{N i_m} - \varphi_{N i_n})\bigg]^{\k} \bigg[\prod \limits_{i_p \in I_c}(\varphi_{N i_n} - \varphi_{N i_p} + \varepsilon)\bigg]^{\k}}{\prod \limits_{a = 1}^{\k} \prod \limits_{\alpha = 1}^{N-1}(\varphi_{N i_n} - \varphi^a_{(N-1) \alpha} + \varepsilon/2)}\right) v_{-e_{Ni_1}-...-e_{Ni_r}}.
\end{equation}
Again, these are proportional to abelian monopole operators on the central node alone, with magnetic charges corresponding to a weight space in a \emph{single} copy of the $\wedge^r\square$ or $\wedge^r\ol \square$ representation.

The eigenvalues $m_i, q^I,q_I$ satisfy the relations that were anticipated back in Section \ref{sec:summary-q}. It follows easily from the part of $\CA_\ve$ associated to the central node that
\be [m_i,m_j]=0\,,\qquad [m_i,q^j] = (\delta_i^j-\tfrac1N)\ve q^j\,,\qquad [m_i,q_j] = -(\delta_{ij}-\tfrac1N)\ve q_j\,. \ee
Moreover, introducing the product of row-determinants
\be \sigma = \det S_1 \det S_2\cdots \det S_\k =  \frac{v_{(\vec{1};\vec 0;\dots;\vec 0)}}{\prod\limits_{1 \leq i < j \leq N}(\varphi_{N j} - \varphi_{N i})^\k} =  \frac{v_{(\vec{1};\vec 0;\dots;\vec 0)}}{\prod\limits_{1 \leq i < j \leq N}(m_j-m_i)^\k} \,, \ee
(noting that $\sigma$ is not quite central among the eigenvalues, though $w^+:= v_{(\vec{1};\vec 0;\dots;\vec 0)}$ is), 
we obtain the simple relations 
\be \frac{1}{(N-r)!}\, \Bigg\{ \begin{array}{c c} \epsilon_{i_1...i_N} q^{[i_{r+1}...i_N]} & \k \textrm{ odd}\\  \Delta_{i_1...i_N}  q^{[i_{r+1}...i_N]}& \k \textrm{ even} \end{array}\Bigg\} = q_{[i_1...i_r]} \sigma\,, \ee
with $\Delta_{i_1...i_N}=-|\epsilon_{i_1...i_N}|$. These are the diagonalized versions of \eqref{TNk-Q=Q}. The passage between fundamental and antifundamental representations, before and after diagonalization, is summarized in the commutative diagram of Figure \ref{Nkcommdiag}.

\begin{figure}[htb]
	\begin{center}
		\begin{tikzpicture}
		\tikzset{vertex/.style = {minimum size=1.5em}}
		\tikzset{edge/.style = {->,> = latex}}
		\node[vertex] (1) at  (0,0) {$Q^{I_1 I_2 ... I_{\k}}$};
		\node[vertex] (2) at  (9,0) {$Q_{J_1 J_2 ... J_{\k}}$};
		\node[vertex] (3) at  (0,-5) {$q^{I'}$};
		\node[vertex] (4) at  (9,-5) {$q_{J'}$};
		
		\draw[edge] (1.350) -- (2.190) node[midway, yshift=0.75cm] {$\bigotimes \limits_{a = 1}^{\k} \frac{1}{(N-r)!}\epsilon^{J_a I_a}$};
		\draw[edge] (2.170) -- (1.10) node[midway, yshift=-0.75cm] {$\bigotimes \limits_{a = 1}^{\k} \frac{1}{r!}\epsilon_{J_a I_a}$};
		
		\draw[edge] (1.260) -- (3.100) node[midway, left] {$\overset{\bigotimes \limits_{a = 1}^{\k} \bigwedge^r (S_a)^{I'}{}_{I_a}}{\longrightarrow}$};
		\draw[edge] (3.80) -- (1.280) node[midway, right] {$\overset{\bigotimes \limits_{a = 1}^{\k} \bigwedge^r (S_a^{-1})^{I_a}{}_{I'}}{\longrightarrow}$};
		
		\draw[edge] (3.350) -- (4.190) node[midway, yshift=0.75cm] {$\overset{\frac{1}{(N-r)!}\left[\prod \limits_{a = 1}^{\k} (\det S_a)\right] \epsilon^{J' I'}}{\longleftarrow}$ or $\overset{\frac{1}{(N-r)!}\left[\prod \limits_{a = 1}^{\k} (\det S_a)\right] \Delta^{J' I'}}{\longleftarrow}$};
		\draw[edge] (4.170) -- (3.10) node[midway, yshift=-0.75cm] {$\overset{\frac{1}{r!} \left[\prod \limits_{a = 1}^{\k} (\det S_a^{-1})\right] \epsilon_{J' I'}}{\longleftarrow}$ or $\overset{\frac{1}{r!} \left[\prod \limits_{a = 1}^{\k} (\det S_a^{-1})\right] \Delta_{J' I'}}{\longleftarrow}$};
		
		\draw[edge] (2.260) -- (4.100) node[midway, right] {$\overset{\bigotimes \limits_{a = 1}^{\k} \bigwedge^{N-r}(S_a)^{J'}{}_{J_a}}{\longleftarrow}$};
		\draw[edge] (4.80) -- (2.280) node[midway, left] {$\overset{\bigotimes \limits_{a = 1}^{\k} \bigwedge^{N-r}(S_a^{-1})^{J_a}{}_{J'}}{\longleftarrow}$};
		\end{tikzpicture}
		\caption{Commutative diagram relating the $\k$-fold rank $r$ antisymmetric tensor operators and rank $N-r$ antisymmetric tensor operators in the standard basis and the ``basis of eigenvectors'' of the moment map. $I_a, I'$ should be understood as multi-indices of size $r$ and $J_a, J'$ as multi-indices of size $N-r$. The arrow under $\bigotimes \bigwedge^n S_a$, $\bigotimes \bigwedge^n S^{-1}_a$ and the determinants indicate which side to act on.
			For the bottom arrows, $\epsilon$ is used for odd $\k$ and $\Delta$ is used for even $\k$.}
		\label{Nkcommdiag}
	\end{center}
\end{figure}

Straightforward algebra in $\CA_\ve$ also gives us
\be \label{TNk-qqm}
q^i q_i =  (-1)^N \prod_{j\neq i}\frac{(m_i-m_j)^{\k-1}}
{m_i-m_j-\ve}\,,\qquad q_i q^i = (-1)^N\prod_{j\neq i}\frac{(m_i-m_j+\ve)^{\k-1}}
{m_i-m_j}
\qquad \text{(for any fixed $i$)}\,, \ee
and relates products of (anti)fundamental $q^i,q_i$ to higher tensor powers as
\be \label{TNk-qq}
\begin{array}{l} \ds q^{i_1} q^{i_2} ... q^{i_r} = (-1)^{r+1} \bigg[\prod \limits_{1 \leq n < m \leq r} \frac{(m_{i_m} - m_{i_n})^{\k-1}}{(m_{i_n} - m_{i_m} - \varepsilon)}\bigg] q^{[i_1 ... i_r]}\,, \\[.5cm]  
	\ds q_{i_1} q_{i_2} ... q_{i_r} = (-1)^{(N-1)(r+1)} \bigg[ \prod \limits_{1 \leq n < m \leq r} \frac{(m_{i_n} - m_{i_m} + \ve)^{\k-1}}{(m_{i_m} - m_{i_n})}\bigg] q_{[i_1 ... i_r]}\,.   \end{array}
\ee

We also point out that the R-charges of the eigenvalues $m_i$, $q^I,q_I$ are the same as the R-charges of the original operators $\mu,Q^{I_1...I_\k},Q_{I_1...I_\k}$. This follows from the general computation of R-charges \eqref{R-vA} in $\CA_\ve$, but is not entirely obvious from the formulas above. In particular,
\be [m_i]=1\,,\qquad [q^I]=[q_I] = \tfrac12(\k-2)r(N-r)\qquad (\text{if}\;|I|=r)\,.\ee

\subsection{Un-diagonalized relations}

Many of the relations in the quantum chiral ring $\C_\ve[\CM_C]$ of $\CT_{N,\k}$ follow immediately from the diagonalization above, and from discussions in the previous sections.

First, since each of the moment maps diagonalize to $\tilde{\mu}_{\textrm{diag}}$, it follows that they should satisfy \begin{equation}
\label{traces}
\textrm{Tr} [(\tilde{\mu}_1)^n] = \textrm{Tr} [(\tilde{\mu}_2)^n]  = ... = \textrm{Tr} [(\tilde{\mu}_{\k})^n]
\end{equation} for all $n\geq 0$.%
\footnote{This property holds trivially when $\varepsilon = 0$, but it holds in general despite the fact that the cyclic property of the trace isn't guaranteed for $\varepsilon \neq 0$. In a similar fashion to the $N = 2$ analysis of Section \ref{sec:2kquivers}, these traces must be Weyl-symmetric polynomials in the scalars on the central $U(N)$ node and in $\varepsilon$. Since all the legs are attached to the same central node, these traces must match.} %
Another set of relations that follows immediately from the earlier comments, in particular the derivation of \eqref{T2k-muQ}, is\begin{equation}
\label{muQrel}
\begin{array}{r c l}
(\tilde{\mu}_1{}^n)^{i_1}{}_{i'} Q^{i' i_2 ... i_{\k}} & = ... = & (\tilde{\mu}_{\k}{}^n)^{i_{\k}}{}_{i'} Q^{i_1 i_2 ... i'}\\
(\tilde{\mu}_1{}^n)^{i'}{}_{i_1} Q_{i' i_2 ... i_{\k}} & = ... = & (\tilde{\mu}_{\k}{}^n)^{j_{\k}}{}_{i'} Q_{i_1 i_2 ... i'}\,,
\end{array}
\end{equation} as well as \begin{equation}
\label{Qmurel}
\begin{array}{r c l}
Q^{i' i_2 ... i_{\k}}	(\tilde{\mu}_1{}^n)^{i_1}{}_{i'} & = ... = & Q^{i_1 i_2 ... i'} (\tilde{\mu}_{\k}{}^n)^{i_{\k}}{}_{i'}\\
Q_{i' i_2 ... i_{\k}} (\tilde{\mu}_1{}^n)^{i'}{}_{i_1} & = ... = & Q_{i_1 i_2 ... i'} (\tilde{\mu}_{\k}{}^n)^{i'}{}_{i_{\k}}\,.
\end{array}
\end{equation}
These relations naturally extend to any rank and one finds that
\begin{equation}
\label{muQQrel}
\begin{array}{r c l}
(\tilde{\mu}_1{}^n)^{[i_{1,1}}{}_{i'} Q^{[i'] i_{1,2} ... i_{1,r}] I_2 ... I_{\k}} &  = ... = & (\tilde{\mu}_{\k}{}^n)^{[i_{\k,1}}{}_{i'} Q^{I_1 I_2 ... I_{\k-1} [i'] ... i_{\k,r}]}\\
(\tilde{\mu}_1{}^n)^{i'}{}_{[i_{1,1}} Q_{[i'] i_{1,2} ... i_{1,r}] I_2 ... I_{\k}} &  = ... = & (\tilde{\mu}_{\k}{}^n)^{i'}{}_{[i_{\k,1}} Q_{I_1 I_2 ... I_{\k-1} [i'] ... i_{\k,r}]}\\
\end{array}
\end{equation} and \begin{equation}
\label{QQmurel}
\begin{array}{r c c l}
Q^{[[i' i_{1,2} ... i_{1,r}] I_2 ... I_{\k}} (\tilde{\mu}_1{}^n)^{i_{1,1}]}{}_{i'} &  = ... = & Q^{I_1 I_2 ... I_{\k-1} [[i' ... i_{\k,r}]} (\tilde{\mu}_{\k}{}^n)^{i_{\k,1}]}{}_{i'}\\
Q_{[[i' i_{1,2} ... i_{1,r}] I_2 ... I_{\k}} (\tilde{\mu}_1{}^n)^{i'}{}_{i_{1,1}]} &  = ... = & Q_{I_1 I_2 ... I_{\k-1} [[i' ... i_{\k,r}]}  (\tilde{\mu}_{\k}{}^n)^{i'}{}_{i_{\k,1]}}\\
\end{array}
\end{equation} where $I_a = \{i_{a,1}, ..., i_{a,r}\}$.

Finally, in analogy with \eqref{fund2tensor} at $\k=1$, we expect to be able to relate products of fundamental $Q$'s to higher tensor powers. Heuristically, this should come from un-diagonalizing \eqref{TNk-qq}. We can write (say) the first equation in \eqref{TNk-qq} more suggestively as 
\begin{equation}
\begin{array}{c} 
q^{i_1} \left(\prod \limits_{i_n > i_1}(\varphi_{N i_1} - \varphi_{N i_n})\right) q^{i_2} \left(\prod \limits_{i_n > i_2}(\varphi_{N i_2} - \varphi_{N i_n})\right) ... q^{i_{r-1}} (\varphi_{N i_{r-1}} - \varphi_{N i_r}) q^{i_r}\\  =  \Bigg[\prod \limits_{\substack{i_n, i_m \in I \\ n<m}}(\varphi_{N i_m} - \varphi_{N i_n})^{\k-1}\Bigg] q^I\,.
\end{array}
\end{equation} Just as in the $\k = 1$ case, the LHS classically arises (up to a numerical factor) from diagonalizing%
\footnote{Note that due to \eqref{muQrel}, \eqref{Qmurel}, there is an unambiguous meaning to the operators $(\tilde{\mu}^nQ)^{i_1 i_2 ... i_{\k}}$, $(Q\tilde{\mu}^n)^{i_1 i_2 ... i_{\k}}$, $(\tilde{\mu}^nQ)_{i_1 i_2 ... i_{\k}}$ and $(Q\tilde{\mu}^n)_{i_1 i_2 ... i_{\k}}$ for all $n$.}
\be
(\mu^{r-1}Q)^{[i_1 [i_2  \dots [i_{\k}} (\mu^{r-2}Q)^{j_1 j_2  \dots j_{\k}} \dots  Q^{k_1] k_2] \dots k_{\k}]}
\ee 
while the RHS classically arises (up to a numerical factor) from diagonalizing 
\be
(\mu_1^0)^{[i_1}{}_{[i_1'} (\mu_1^{\k-1})^{j_1}{}_{j_1'} ... (\mu_1^{(r-1)(\k-1)})^{k_1]}{}_{k_1']}  Q^{[i_1' j_1' ... k_1'] I_2 ... I_{\k}}.
\ee
together these imply that, at least classically, there should be a relation of the form \begin{subequations} \label{TNk-Tcl}
	\be
	(\mu^{r-1}Q)^{[i_1 [i_2  \dots [i_{\k}} (\mu^{r-2}Q)^{j_1 j_2  \dots j_{\k}} \dots  Q^{k_1] k_2] \dots k_{\k}]} \propto (\mu_1^0)^{[i_1}{}_{i_1'} (\mu_1^{\k-1})^{j_1}{}_{j_1'} ... (\mu_1^{(r-1)(\k-1)})^{k_1]}{}_{k_1'}  Q^{[i_1' j_1' ... k_1'] I_2 ... I_{\k}}
	\ee and similarly for the antifundamentals \be
	Q_{i_1] i_2] \dots i_{\k}]} \dots (Q\mu^{r-2})_{[j_1 [j_2  \dots [j_{\k}} (Q\mu^{r-1})_{k_1 k_2  \dots k_{\k}} \propto Q_{[i_1' j_1' ... k_1'] I_2 ... I_{\k}} (\mu_1^0)^{[i_1'}{}_{[i_1} (\mu_1^{\k-1})^{j_1'}{}_{j_1} ... (\mu_1^{(r-1)(\k-1)})^{k_1']}{}_{k_1]}
	\ee
\end{subequations} where the constants of proportionality are purely numerical.

The quantum corrections to these relations are highly nontrivial. In the quantum case, we expect a general form
\begin{subequations} \label{TNk-T}
	\begin{equation}
	(\tilde{\mu}^{r-1}Q)^{[i_1 [i_2  \dots [i_{\k}} (\tilde{\mu}^{r-2}Q)^{j_1 j_2  \dots j_{\k}} \dots  Q^{k_1] k_2] \dots k_{\k}]} = (T_1)^{[i_1 \dots k_1]}{}_{[i_1' \dots k_1']}Q^{[i_1' ... k_1'] \dots [i_{\k} \dots k_{\k}]}
	\end{equation} and \begin{equation}
	Q_{[i_1 [i_2 \dots [i_{\k}} \dots (Q\tilde{\mu}^{r-1})_{j_1 j_2  \dots j_{\k}} (Q\tilde{\mu}^{r-1})_{k_1] k_2]  \dots k_{\k}]} = Q_{[i_1' ... k_1'] \dots [i_{\k} \dots k_{\k}]}(\wt{T}_1)^{[i_1' \dots k_1']}{}_{[i_1 \dots k_1]}
	\end{equation} 
\end{subequations}
for some tensors $T_1$,$\wt{T}_1$ constructed out of $\tilde{\mu}_1$, with $R$-charge $\frac{1}{2}(\k -1)r(r-1)$.
The tensors $T_1, \wt{T}_1$ can be quite complicated and we have not yet found a general formula for them. Appendix~\ref{tensorcomps} contains examples of relations \eqref{TNk-T} for several small values of $N$, $\k$ and $r$.  See also \eqref{quant-undiag-N2}.

We do expect several additional relations among the moment maps and tensor operators. They are increasingly difficult to guess; though once a putative relation is written down, it is straightforward to verify using the abelianized algebra $\CA_\ve$. (In principle, \emph{all} relations in $\C_\ve[\CM_C]$ are induced from the embedding of generators into $\CA_\ve$.) One family of additional relations should correspond to un-diagonalizing \eqref{TNk-qqm}. They should relate partial contractions of the multi-index operators $Q^{I_1 ... I_{\k}} Q_{I_1' ... I_{\k}'}$ to moment maps. Specific examples of such relations at $\k=2$ and $\k=3$ are discussed in Sections \ref{sec:geom12} and \ref{TNrelations}. 
For $\k=3$ we will also propose a special set of relations between products of fundamentals and higher antisymmetric tensor operators, anticipated in  \eqref{LPgenquant-intro}.

\section{Geometry of one- and two-legged quivers}
\label{sec:geom12}

As mentioned in the introduction, the Coulomb branches of one- and two-legged star quivers coincide with well known geometric spaces. Namely, the Coulomb branch of $\CT_{N,2}$ is the cotangent bundle of the complex group $SL(N,\C)$,
\be \CT_{N,2}\,:\quad \CM_C \simeq T^*SL(N,\C)\,, \label{k2T*} \ee
while the Coulomb branch of $\CT_{N,1}$ is the so-called Kostant-Whittaker reduction of $T^*SL(N,\C)$,
\be \CT_{N,1}\,:\quad \CM_C \simeq  T^*SL(N,\C)/\!/_\psi \,\mathfrak{N}\,. \label{k1T*N}\ee
These are precisely the spaces assigned to one- and two-punctured spheres by the Moore-Tachikawa TQFT \cite{MT11}. (Taking some care with scaling limits, these spaces can also be related to Higgs branches of the 6d (2,0) theory compactified on one- or two-punctured spheres.)

In this section, we use the generators and relations of chiral rings from Sections \ref{N1quivers}--\ref{Nkquivers} to explain how \eqref{k2T*} and \eqref{k1T*N} come about.

\subsection{Kostant-Whittaker reduction}
\label{sec:KW}

To discuss the geometry of one-legged quivers, we begin by recalling what the Kostant-Whittaker reduction $T^*SL(N,\C)/\!/_\psi \,\mathfrak{N}$ means. In physics, this space has played a major in the analysis of ``punctures'' in 4d $\CN=2$ theories of class S, and in the Nahm pole boundary condition of 4d $\CN=4$ theories.

First, $T^*SL(N,\C)$ is the cotangent bundle of the complex group $SL(N,\C)$. It is canonically a complex symplectic space.
 The cotangent bundle is trivial, so we have $T^*SL(N,\C)\simeq \mathfrak{sl}(N,\C)^*\times SL(N,\C)$. However, it is naturally trivial in not one but \emph{two} different ways, which correspond to identifying the fiber at the identity $1\in SL(N,\C)$ with either left-invariant or right-invariant one-forms on the group. A convenient way to parameterize points of $T^*SL(N,\C)$ is in terms of triples
 \be \label{triples} T^*SL(N,\C) = \bigg\{(\mu_L,g,\mu_R)\;\text{s.t.}\, \begin{array}{c} \mu_{L,R}\in \mathfrak{sl}(N,\C)^*,\, g\in SL(N,\C)\,,\\ \mu_Lg= g\mu_R \end{array}\bigg\}\,.
 \ee
Here $\mu_L$ and $\mu_R$ are the complex moment maps for the action of left and right multiplication of $SL(N,\C)$ on itself, extended to $T^*SL(N,\C)$ as complex Hamiltonian actions. Either moment map can be used to parameterize the fiber at $1$; $\mu_L,\mu_R$ correspond precisely to the left-invariant and right-invariant trivializations of the cotangent bundle.

Under the action of left and right multiplication, the triplet $(\mu_L,g,\mu_R)$ transforms as
\be \begin{array}{ccc}
 (\mu_L,g,\mu_R) & \overset{g_L}\longrightarrow & (g_L\mu_L g_L^{-1},\;g_L g,\;\mu_R)\,, \\
 (\mu_L,g,\mu_R) & \overset{g_R}\longrightarrow & (\mu_L,\; g g_R,\; g_R^{-1}\mu_R g_R)\,.
\end{array}\ee

The group $\mathfrak N\subset SL(N,\C)$ is a subgroup of unipotent matrices, \emph{i.e.} the exponential of a nilpotent subalgebra of $\mathfrak{sl}(N,\C)$. We take it to contain upper-triangular matrices of the form
\be n = \bp 1 & n_{12} & n_{13} & \cdots & n_{1N} \\
0 & 1& n_{23} & \cdots & n_{2N} \\
\vdots & & \ddots & & \vdots \\
0 & 0 & 0 & \cdots & 1 \ep \in \mathfrak{N}\,.
\ee
We assume that $\mathfrak N$ acts on $SL(N,\C)$ on the \emph{right}. Then the moment map for the induced Hamiltonian $\mathfrak N$ action on $T^*SL(N,\C)$ may be identified as the strictly-lower-triangular part of $\mu_R\in \mathfrak{sl}(N,\C)^*$.

The holomorphic symplectic quotient $T^*SL(N,\C)/\!/_\psi \,\mathfrak{N}$ is now constructed in two steps. First, one fixes the complex moment map for the $\mathfrak N$ action to a \emph{generic} character $\psi$. Explicitly, this means fixing the form of $\mu_R$ to be
\be \label{mupsi}
 \mu_R =  \bp * & * & * & \cdots & * \\
\psi_1 & *& * & \cdots &* \\
0 & \psi_2 & * & \cdots & * \\
\vdots & & \ddots & & \vdots \\
0 & 0 & \cdots & \psi_{N-1} & * \ep\,, 
\ee
with fixed nonzero complex numbers $\psi_1,...,\psi_{N-1}$ below the diagonal. Note that the form \eqref{mupsi} is invariant under the right $\mathfrak N$ action, which takes $\mu_R \to n^{-1}\mu_R n$.
Second, one quotients by the $\mathfrak N$ action, obtaining the space
\be \CM_\psi := \Big(T^*SL(N,\C)\big|_{\mu_R=\eqref{mupsi}}\Big)/\mathfrak N = T^*SL(N,\C)/\!/_\psi \,\mathfrak{N}\,. \label{KW-red}\ee
With the restriction \eqref{mupsi}, the $\mathfrak N$ action is free, so $\CM_\psi$ is smooth.

Note that
\be \text{dim}_\C(\CM_\psi) = \text{dim}_\C(T^*SL(N,\C)) - 2\,\text{dim}_\C(\mathfrak N) = (N+2)(N-1)\,.\ee
Moreover, $\CM_\psi$ is complex symplectic, with holomorphic symplectic structure induced from $T^*SL(N,\C)$. And there is still a Hamiltonian $SL(N,\C)$ action on $\CM_\psi$, induced from the left action on $T^*SL(N,\C)$; its moment map is $\mu_L$. 
The complex symplectic geometry of $\CM_\psi$ is independent of the precise value of the $\psi_i$'s as long as they are nonzero, so it is convenient to take $\psi=(\psi_1,...,\psi_{N-1})=(1,...,1)$. We want to check that $\CM_{\psi=1}$ is equivalent to the Coulomb branch of $\CT_{N,1}$.

There are several ways to describe the ring of functions on $\CM_\psi$. Since $\C[\CM_\psi]$ the ring of functions on the quotient \eqref{KW-red}, it is equivalent to $\mathfrak N$-invariant functions on the restriction
\be \C[\CM_\psi] = \C\big[T^*SL(N,\C)|_{\mu_R=\eqref{mupsi}}\big]^{\mathfrak N}\,. \ee
The $\mathfrak N$-invariants are generated by
\begin{itemize}
\item the entries of the moment map $\mu_L$, and
\item the entries in the first column $g_{(1)}$ of the $SL(N,\C)$ matrix $g$ \\(since $g\to gn$ preserves $g_{(1)}$).
\end{itemize}
Now compare this to the quantized chiral ring $\C_\ve[\CM_C]$ of the $\CT_{N,1}$ quiver from Section \ref{N1quivers}. The chiral ring is generated by a moment map $\tilde \mu=\mu - \frac{N-1}{2}\ve\mathds{1}$ for the $SL(N,\C)$ flavor symmetry and a set of operators $\{Q^i\}_{i=1}^N$ in the fundamental representation. (All antifundamental $Q_i$ and higher tensor powers $Q^I,Q_I$ can be constructed from $\mu$ and $Q^i$ by using \eqref{TN1-Q=Q} and \eqref{fund2tensor}.) We identify
\be \mu_L=\mu \,,\qquad g_{(1)} =  Q^\bullet = \bp Q^1\\ \vdots \\ Q^N \ep \qquad \text{(as $\ve\to 0$)}\,. \ee

Dimension-counting shows that there should be a relation among the entries of $\mu$ and $g_{(1)}$. We find it as follows. Note that the first column of every matrix $((\mu_L)^kg)_{(1)}=(g(\mu_R)^k)_{(1)}$ is $\mathfrak N$-invariant. Moreover, due to the form \eqref{mupsi} of $\mu_R$, the first column of $g(\mu_R)^k$ is a linear combination of the $k$-th column of $g$ and some of the previous columns 
\be \label{g-columns} (g(\mu_R)^k)_{(1)} = (\psi_1\psi_2...\psi_k) g_{(k+1)} + \#\,g_{(k)} + \#\,g_{(k-1)}+ ...\qquad (0\leq k \leq N-1)\,.\ee
This implies that if we construct an $N\times N$ matrix $X$ with columns
\be X = \Big( \,g_{(1)} \;\; (g\mu_R)_{(1)}\;\; ... \;\; (g(\mu_R)^{N-1})_{(1)}\,\Big) = \Big( \,g_{(1)} \;\; \mu_L g_{(1)}\;\; ... \;\; (\mu_L)^{N-1}g_{(1)}\,\Big)  \label{Xg} \ee
we will get
\be \det X = (\psi_1^{N-1} \psi_2^{N-2}...\psi_{N-1}) \det g = \psi_1^{N-1} \psi_2^{N-2}...\psi_{N-1}\,.\ee
This is the relation. Specializing to $\psi= 1$, we simply have
\be \det X = \det\Big( \,g_{(1)} \;\; \mu_L g_{(1)}\;\; ... \;\; (\mu_L)^{N-1}g_{(1)}\,\Big)    = 1\,.\ee
Rather beautifully, upon identifying $\mu_L=\mu$ and $g_{(1)}=Q^\bullet$, this coincides with the $\ve\to 0$ limit of the quantum relation \eqref{fund2tensor} we found in Section \ref{sec:TN1-rels}. Namely, \eqref{fund2tensor} with $r=N$ implies
\be \det \Big( \,(\tilde\mu^{N-1}Q)^\bullet\;\;...\;\; (\tilde\mu Q)^\bullet \;\; Q^\bullet \,\Big) = 1\,.\ee

An alternative way to describe the ring of functions on $\CM_\psi$ involves taking a slice through the orbits of the (free) $\mathcal N$ action on $T^*SL(N,\C)|_{\mu_R=\eqref{mupsi}}$, \emph{i.e.} gauge fixing. This description often appears in the literature, particularly in \cite{MT11}. We can completely gauge-fix the $\mathfrak N$ action by forcing the upper-diagonal part of $\mu_R$ in \eqref{mupsi} to take the form
\be \label{muK}
 \mu_R =  \bp 0 & 0 & 0 & \cdots & 0 & x_1 \\
\psi_1 & 0& 0 & \cdots & 0 & x_2 \\
0 & \psi_2 & 0 & \cdots & 0 & x_3 \\
\vdots & & &\;\;\; \ddots &  & \vdots \\
0 & 0 & 0 & \cdots & 0& x_{N-1} \\
0 & 0 & 0 & \cdots & \psi_{N-1} & 0 \ep\,, 
\ee
where most entries above the diagonal, except the first $N-1$ entries of the last column, are set to zero. This is the so-called Kostant slice (or maximal Slodowy slice) in $\mathfrak{sl}(N,\C)^*$. 
The entries $x_1,...,x_{N-1}$ are independent functions on $\CM_\psi$. Using this parameterization of $\mu_R$ and moreover specializing $\psi=1$ we find that the matrix $X$ from \eqref{Xg} simply becomes
\be X=g\,.\ee
Therefore,
\be \CM_{\psi=1} \simeq \{(g,x_1,...,x_{N-1})\} = SL(N,\C)\times \C^{N-1}\,.\ee
Unfortunately, this description makes the holomorphic symplectic structure of $\CM_\psi$ rather obscure.

Comparing to the Coulomb branch of $\CT_{N,1}$ from Section \ref{N1quivers}, some quick computations indicate that the chiral-ring operators given there are \emph{already} adapted to a gauge-fixed version of $\CM_\psi$. In particular, defining the matrix
\be X = \Big(\, Q^\bullet \;\; (\tilde \mu Q)^\bullet \;\; ... \;\; (\tilde \mu^{N-1} Q)^\bullet \,\Big) \ee
(which satisfies $\det X=1$, and which we expect to equal $g$ after gauge-fixing), we find that in the $\ve \to 0$ limit
\be X^{-1}\tilde \mu X  = \bp 0 & 1 & 0 & \cdots & 0 \\
-c_2 & 0& 1 & \cdots & 0 \\
c_3 & 0 & 0 & \cdots & 0 \\
\vdots & & \ddots &  & \vdots \\
(-1)^{N-2} c_{N-1} & 0 & 0 & \cdots & 1 \\
(-1)^{N-1} c_{N}  & 0 & 0 & \cdots & 0 \ep\,, \label{KW-phys}  \ee
where $c_i$ are the Casimir operators of $\tilde \mu$, \emph{cf.} Appendix \ref{sec:Capelli}. The expression \eqref{KW-phys} is not  the Kostant slice \eqref{muK} on the nose, but is related to it by a simple transformation, namely transposing and conjugating with the longest element of the Weyl group. The same transformation appears in the analysis of two-legged quivers below, \emph{cf.} \eqref{muRK}, and accounts for a difference in conventions in defining the right $SL(N,\C)$ action on $T^*SL(N,\C)$.

The computation of $X^{-1}\tilde \mu X$ can be carried out even at $\ve\neq 0$, but the result does not follow an easily recognizable pattern. For small values of $N$ we find:
\be N=2:\quad  X^{-1}\tilde \mu X  = \bp \ve & 1\\
-c_2 - \frac{1}{4}\ve^2& 0 \ep\,, \qquad N=3:\quad X^{-1}\tilde \mu X  = \bp 2 \ve & 1 & 0 \\
-c_2 -\frac{1}{3}\ve^2& 0& 1 & \\
c_3 + \frac{2}{3} \ve c_2 -\frac{10}{27}\ve^3 & 0 & 0 &  \ep\,, \notag  \ee 
\be N=4:\quad  X^{-1}\tilde \mu X  = \bp 3 \ve & 1 & 0 &  0 \\
-c_2 +\frac{29}{8}\ve^2 & 0& 1 &  0 \\
c_3+\frac{5}{2}\ve c_2 - \frac{237}{16}\ve^3 & 0 & 0 &  1 \\
- c_{4} -\frac{3}{4}\ve c_3 - \frac{21}{16}\ve^2 c_2 + \frac{2079}{256}\ve^4  & 0 & 0  & 0 \ep\,. \ee

\subsection{The cotangent bundle}

For two-legged $(\k=2)$ quivers, we expect the Coulomb branch to be even simpler: classically, $\CM_C\simeq T^*SL(N,\C)$. 
We gave a description of this cotangent bundle in \eqref{triples}, in terms of an $SL(N,\C)$ element $g$ and moment maps $\mu_L,\mu_R$ for the left and right actions of $SL(N,\C)$ on itself, extended to $T^*SL(N,\C)$ as Hamiltonian actions. Here we'll explain how holomorphic functions on $T^*SL(N,\C)$ match the structure of the Coulomb-branch chiral ring.

In the quantum chiral ring $\C_\ve[\CM_C]$ of a two-legged quiver, we already have two $SL(N,\C)$ moment maps $\tilde \mu_1,\tilde \mu_2$. It is natural to identify these with $\mu_L,\mu_R$ --- modulo a slight reparameterization that we'll explain below. In addition, the chiral ring contains bi-fundamental tensors $Q^{ij}$ and bi-antifundamental tensors $Q_{ij}$ for left and right $SL(N,\C)$ actions. We expect to identify one or the other of these with the group element $g$. 

The tensor operators in $k=2$ theories turn out to satisfy some special relations that are the key to making this identification. First we note that the general R-charge formula \eqref{TNk-R} implies that the bi-(anti)fundamentals and all other higher antisymmetric tensors have R-charge exactly zero
\be [Q^{ij}]=[Q_{ij}] = [Q^{IJ}]=[Q_{IJ}] = 0\,. \label{k2zero}\ee
This is consistent with the natural $\C^*$ scaling action on $T^*SL(N,\C)$, which acts on $g$ with weight zero and the cotangent fibers $\mu_L$ (or $\mu_R$) with weight 1, also scaling the holomorphic symplectic form with weight $1$.

The R-charges \eqref{k2zero} allow for simple contraction relations on a single index
\be Q^{ij}Q_{i'j} = (-1)^N \delta^i{}_{i'}\,,\qquad Q^{ij}Q_{ij'} = (-1)^N \delta^j{}_{j'} \label{k2contr} \ee
that un-diagonalize \eqref{TNk-qqm} with no need for a moment map on the RHS. In addition, all the higher tensor operators can be built in a simple way from the bi-(anti)fundamentals, again without involving moment maps. Namely, if we let $I=[i_1...i_r]$ and $J=[j_1...j_r]$ denote fully antisymmetric multi-indices, then
\be \label{k2tens} Q^{[i_1[j_1}Q^{i_2j_2}\cdots Q^{i_r]j_r]} = (-1)^{\frac{1}{2} r(r-1)}\frac{1}{r!} Q^{IJ} \,,\qquad Q_{[i_1[j_1}Q_{i_2j_2}\cdots Q_{i_r]j_r]} = (-1)^{\frac{1}{2} r(r-1)}\frac{1}{r!} Q^{IJ}\,,\ee
where the LHS has a full antisymmetrization on each set of indices. We have checked these relations explicitly in the quantum algebra for $N\leq 4$ (see Appendix \ref{app:oneleg}), and infer the general form shown here.

Thanks to \eqref{k2tens} and the universal relations \eqref{TNk-Q=Q}, we find that all antifundamentals $Q_{ij}$ and all higher tensors $Q^{IJ}$, $Q_{IJ}$ can be written in terms of the basic bifundamental operators $Q^{ij}$. Under the general assumption (made throughout this paper) that the chiral ring $\C_\ve[\CM_C]$ is generated by moment maps $\tilde \mu_a$ and the entire collection of tensor operators, this observation implies that for $\k=2$ quivers the generators $\tilde\mu_1,\tilde\mu_2$ and $Q^{ij}$ are actually sufficient.

Now, if we set $r=N$ in \eqref{k2tens}, we get the specialization
\be Q^{[i_1[j_1}Q^{i_2j_2}\cdots Q^{i_N]j_N]} = (-1)^{\frac{1}{2}N(N-1)}\frac{1}{N!} \epsilon^{i_1...i_N}\epsilon^{j_1...j_N}\,. \label{k2e} \ee
In the classical $\ve\to 0$ limit, this ensures that $\det Q^{ij}=(-1)^{\frac{1}{2} N(N-1)}$. 
At $\ve\neq 0$ we must be more careful about defining the determinant. It is convenient to introduce anti-diagonal matrices $K_{ij}=\delta_{i,N+1-j} = K^{ij}$, satisfying $K_{ij}K^{jk}=\delta_i{}^k$,
\be \label{defK} K = \bp 0&\cdots &0&1 \\ 0&\cdots & 1 & 0 \\ \vdots &\reflectbox{$\ddots$} & & \vdots \\ 1 &\cdots & 0 & 0 \ep\,.\ee
(This is the longest element of the $SL(N,\C)$ Weyl group.) Then, defining
\be g^i{}_j := Q^{ij'}K_{j'j}\,, \label{gQK} \ee
we find that $g$ naturally has a unit row-determinant, even at $\ve\neq 0$
\be \det g = \epsilon_{i_1i_2...i_N} g^{i_1}{}_1 g^{i_2}{}_2 \cdots g^{i_N}{}_N = 1\,. \ee
We identify this $g$ with (a quantization of) the $SL(N,\C)$ matrix in the expected Coulomb branch $T^*SL(N,\C)$. We also note that, due to \eqref{k2zero}, $g$ has a two-sided inverse given by
\be (g^{-1})^i{}_j = (-1)^N K^{ii'}Q_{ji'}\,.\ee

It remains to carefully identify the left and right moment maps $\mu_L,\mu_R$ for $T^*SL(N,\C)$ (in the presentation give by \eqref{triples}) in terms of $\tilde \mu_1$ and $\tilde \mu_2$. We propose to set, both for $\ve=0$ and $\ve\neq 0$,
\be \mu_L = \tilde \mu_1\,,\qquad \mu_R = K\tilde \mu_2^T K\,.  \label{muRK} \ee
In other words, $(\mu_R)^i{}_j = \tilde \mu_2^{N+1-j}{}_{N+1-i}$.
The universal relations  \eqref{muQrel}  involving moment-map contractions with $Q$'s then imply
\be (\tilde\mu_1)^i{}_{i'} Q^{i'j} = (\tilde\mu_2)^j{}_{j'} Q^{ij'}\,, \ee
which conveniently translates (using \eqref{gQK}, \eqref{muRK}) to
\be \mu_L g = (\mu_R^T g^T)^T\,, \ee
and in the $\ve\to 0$ limit reproduces the relation $\mu_L g = g\mu_R$ in \eqref{triples}.
For $\ve\neq 0$, we have explicitly recovered a quantization of $T^*SL(N,\C)$.

We remark that the matrix $K$ here, together with transposition, plays the role of intertwining the $SL(N,\C)^\k$ action on $\C_\ve[\CM_C]$ that is most natural for general $\TNk$ quivers (namely, with every $SL(N,\C)$ acting as a left multiplication) with the $SL(N,\C)_L\times SL(N,\C)_R$ action by left and \emph{right} multiplication that is more natural when $\k=2$.

\section{Relations for trinion theories}
\label{TNrelations}

In Section \ref{Nkquivers}, we identified moment maps and antisymmetric tensor operators as elements of the abelianized algebra $\CA_\ve$, for general $\TNk$ star quivers. We then derived and/or verified several universal families of relations among the moment maps and tensor operators, which hold for all $N$ and $\k$. In this section, we specialize to $\k=3$, \emph{i.e.} the star quivers related to trinion theories $T_N$ of Class $\CS$. We quickly compare the universal relations to know $T_N$ relations in the literature, and then discuss how to quantize and generalize several additional relations from the literature that are special for $\k=3$.

We mainly follow \cite{TNRev} and \cite{LemosPeelaers} as references for known $T_N$ relations, at $\ve= 0$.

\subsection{Universal relations}

Relations \eqref{traces}--\eqref{QQmurel} in Section \ref{Nkquivers} give us
\begin{equation}
	\label{TNtraces}
	\textrm{Tr} [(\tilde \mu_1)^n] = \textrm{Tr} [(\tilde \mu_2)^n] = \textrm{Tr} [(\tilde \mu_3)^n]
\end{equation}
as well as
\begin{equation}
	(\tilde\mu_1)^{i}{}_{i'}Q^{i' j k} = (\tilde\mu_2)^{j}{}_{j'}Q^{i j' k} = (\tilde\mu_3)^{k}{}_{k'}Q^{i j k'}\,,\qquad
	Q^{i' j k}(\tilde\mu_1)^{i}{}_{i'} = Q^{i j' k}(\tilde\mu_2)^{j}{}_{j'} = (\tilde\mu_3)^{k}{}_{k'}Q^{i j k'}(\tilde\mu_3)^{k}{}_{k'}\,,	
\notag \end{equation}
\begin{equation}
	\label{mQanti}
	(\tilde \mu_1)^{i'}{}_{i}Q_{i' j k} = (\tilde \mu_2)^{j'}{}_{j}Q_{i j' k} = (\tilde \mu_3)^{k'}{}_{k}Q_{i j k'}\,,\qquad
	Q_{i' j k}(\tilde \mu_1)^{i'}{}_{i} = Q_{i j' k}(\tilde \mu_2)^{j'}{}_{j} = (\tilde \mu_3)^{k'}{}_{k}Q_{i j k'}(\tilde \mu_3)^{k'}{}_{k}\,,
\end{equation} 
and more generally, for higher-rank tensors 
\begin{equation}
	\label{mQtens}
	\begin{array}{c}
		(\tilde{\mu}_1)^{[i_{1}}{}_{i'} Q^{[i'] i_{2} ... i_{r}] JK}   = (\tilde{\mu}_2)^{[j_{1}}{}_{j'} Q^{I[j'] j_{2} ... j_{r}]K}
		= (\tilde{\mu}_3)^{[k_{1}}{}_{k'} Q^{IJ[k'] k_{2} ... k_{r}]}\,,\\[.2cm]
		(\tilde{\mu}_1)^{i'}{}_{[i_{1}} Q_{[i'] i_{2} ... i_{r}]JK} = (\tilde{\mu}_2)^{j'}{}_{[j_{1}} Q_{I[j'] j_{2} ... j_{r}]K}
		= (\tilde{\mu}_3)^{k'}{}_{[k_{1}} Q_{IJ[k'] k_{2} ... k_{r}]}\,,
	\end{array}
\end{equation}
with identical formulas for contractions from the right.
As $\ve\to 0$, the shifted moment maps $\tilde \mu_a$ become ordinary moment maps $\mu_a$, and these relations reduce to the well known Eqs. (2.4)-(2.6) of \cite{TNRev}. As argued in Section \ref{Nkquivers} via diagonalization, relations \eqref{mQanti}-\eqref{mQtens} still hold if the moment maps are replaced by their $n$-th powers.

We also always have \eqref{TNk-Q=Q} relating fundamental and antifundamental tensors, namely
\be \frac{1}{(r!)^3} \epsilon_{I'I}\epsilon_{J'J}\epsilon_{K'K} Q^{IJK} = Q_{I'J'K'}\,,\qquad Q^{IJK} = \frac{1}{(N-r)!^3} Q_{I'J'K'} \epsilon^{I'I}\epsilon^{J'J}\epsilon^{K'K}\,, \label{TN-QQ}\ee
where $I,J,K$ are multi-indices of length $r$ and $I',J',K'$ are multi-indices of length $N-r$.

\subsection{Contractions and Casimirs}

One of the special relations we find for $\k=3$ is a quantized version of (2.7) in \cite{TNRev}. It takes the form
 \begin{equation}
	\label{QQCas}
	Q^{i j k} Q_{i' j' k} = (-1)^N \sum\limits_{\ell=0}^{N-1} c_{\ell,3} \sum \limits_{m = 0}^{N-l-1} (\tilde{\mu}_1^{N-l-m-1})^i_{i'} (\tilde{\mu}_2^{m})^j_{j'}
\end{equation} where $c_{\ell,a}$ are coefficients the characteristic equation of $\tilde{\mu}_a$ \begin{equation}
	P_a(t) = \det(t + \tilde{\mu}_a + (N-i)\delta_{ij} \varepsilon) = \sum\limits_{\ell=0}^{N} [t]^{\ell} c_{N-\ell,a}.
\end{equation} This Capelli-like determinant is used to ensure that the $c_{\ell,a}$ are central elements; they can be expressed in terms of the $\textrm{Tr}[(\tilde{\mu}_a)^n]$ just as with ordinary determinants. Several of the $c_\ell$ are listed in Appendix \ref{sec:Capelli}, as explicit elements of the abelianized algebra $\CA_\ve$.
  As noted in \cite{TNRev},  \eqref{TNtraces} implies that \begin{equation}
	c_{\ell,1} = c_{\ell,2} = c_{\ell,3} =: c_{\ell}
\end{equation} and so 
\begin{equation}
	\label{chareqns}
	P_{1}(t) = P_{2}(t) = P_{3}(t)\,.
\end{equation}

The verification of  \eqref{QQCas} will closely mirror \cite{TNRev}. Consider the operator $R^{ij}{}_{i'j'} = Q^{ijk}Q_{i'j'k}$. This operator clearly transforms as a bi-adjoint (with possible traces) of $SL(N,\mathbb{C})^2$ and trivially under the third $SL(N,\mathbb{C})$. From  \eqref{muQrel} and  \eqref{Qmurel} we have \begin{equation}
	(\tilde{\mu}_1)^i{}_{i''}R^{i'' j}{}_{i' j'} = (\tilde{\mu}_1)^{i''}{}_{i'} R^{i j}{}_{i'' j'} \qquad R^{i'' j}{}_{i' j'} (\tilde{\mu}_1)^i{}_{i''} = R^{i' j}{}_{i'' j'} (\tilde{\mu}_1)^{i''}{}_{i'}
\end{equation} and similarly for $\tilde{\mu}_2, \tilde{\mu}_3$. At the classical level this would imply that $R$ commutes with $\tilde \mu_1$ and $\tilde \mu_2$ as matrices but that is not necessarily the case for $\varepsilon \neq 0$. In order to show that this indeed is the case, note that since $Q^{i j k}$ is a fundamental and $Q_{i j k}$ is an antifundamental it follows that \begin{equation}
	\begin{array}{rl}
		[(\tilde{\mu}_1)^{i'}{}_{i''}, Q^{i j k}] & = \varepsilon \left[\delta^{i}{}_{i''} Q^{i'' j k} - \delta^{i'}{}_{i''} \frac{1}{N} Q^{i j k}\right]\,,\\
		
		[(\tilde{\mu}_1)^{i''}{}_{i'}, Q_{i j k}] & = -\varepsilon \left[\delta^{i''}{}_{i} Q_{i'' j k} - \delta^{i''}{}_{i'}\frac{1}{N} Q_{i j k}\right]\,.
	\end{array}
\end{equation} Putting these together we find that \begin{equation}
	\begin{array}{rl} 
		[(\tilde{\mu}_1)^{i}{}_{i''},R^{i'' j}{}_{i' j'}] & = [(\tilde{\mu}_1)^{i}{}_{i''}, Q^{i'' j k}Q_{i' j' k}]\\
		&  = \frac{1}{2}\left\{[(\tilde{\mu}_1)^{i}{}_{i''}, Q^{i'' j k}Q_{i' j' k}] + [(\tilde{\mu}_1)^{i''}{}_{i'}, Q^{i j k}Q_{i'' j' k}]\right\}\\
		& = \frac{\varepsilon(N^2-1)}{2N} \left\{ \left(Q^{i j k} Q_{i' j' k} - Q^{i' j k} Q_{i j' k}\right) + \left(Q^{i' j k} Q_{i j' k} - Q^{i j k} Q_{i' j' k} \right)\right\}\\
		& = 0\\
	\end{array}
\end{equation} as desired. An identical computation shows that $R$ commutes with $\tilde{\mu}_2$. From the fact that $R$ commutes with both $\tilde{\mu}_1$ and $\tilde{\mu}_2$ and its tensor structure, it must be that $R$ is a  polynomial in $\tilde{\mu}_1$, $\tilde{\mu}_2$ with some prefactors that commute with everything. Furthermore, the R-charge of $Q$ implies that this can be a polynomial of degree at most $N-1$. Equation \eqref{muQrel} can be written as $(\tilde{\mu}_1\otimes \mathds{1} - \mathds{1}\otimes \tilde{\mu}_2)R = 0$ and so, since $P(\tilde{\mu}_1) = P(\tilde{\mu}_2)$, one finds that such a polynomial is given by \begin{equation}
	R \propto \frac{P(\tilde{\mu}_1) - P(\tilde{\mu}_2)}{\tilde{\mu}_1 - \tilde{\mu}_2} \propto \sum\limits_{\ell = 0}^{N-1}c_{\ell} \sum\limits_{m = 0}^{N-l -1} (\tilde{\mu}_1^{N-l-1-m})\tilde{\mu}_2^m.
\end{equation} We claim the constant of proportionality is $(-1)^N$, which has been verified explicitly for low values of $N$, resulting in the proposed formula \eqref{QQCas}.
In the classical limit, the modified Capelli determinant becomes the usual determinant and so the coefficients $c_\ell$ take their classical values, in agreement with  (2.7) of \cite{TNRev}.

\subsection{Fundamentals and higher tensors}

At $\k=3$ there are also a beautiful set of relations between products of tri-fundamental operators and higher-rank tensors, which un-diagonalize \eqref{TNk-qq}.

One example of such a relation was proposed%
\footnote{The expressions we write here agree with (2.8) and (2.9) in \cite{TNRev} upon substituting $Q_{ijk} \mapsto - Q_{ijk}$.} %
in  (2.8) and  (2.9) in \cite{TNRev}; namely, at $\ve=0$,
\begin{equation}
	\label{fund2anti}
	\begin{array}{l}
		Q^{i_1 j_1 k_1} Q^{i_2 j_2 k_2} \dots Q^{i_{N-1} j_{N-1} k_{N-1}} \epsilon_{j_1 j_2 \dots j_{N-1} j} \epsilon_{k_1 k_2 \dots k_{N-1} k}\\[.1cm]
		\hspace{1in}= -(N-1)! Q_{i j k} (\mu_1^0)^{(i_1}_{i_1'}  (\mu_1)^{i_2}_{i_2'}\dots  (\mu_1^{N-2})^{i_{N-1})}_{i_{N-1}'} \epsilon^{i_1' i_2' \dots i_{N-1}' i}\\[.2cm]
		Q_{i_1 j_1 k_1} Q_{i_2 j_2 k_2} \dots Q_{i_{N-1} j_{N-1} k_{N-1}} \epsilon^{j_1 j_2 \dots j_{N-1} j} \epsilon^{k_1 k_2 \dots k_{N-1} k}\\[.1cm]
		\hspace{1in} =(-1)^{\frac{(N+1)(N-2)}{2}} (N-1)! Q^{i j k} (\mu_1^0)_{(i_1}^{i_1'}  (\mu_1)_{i_2}^{i_2'}\dots (\mu_1^{N-2})_{i_{N-1})}^{i_{N-1}'} \epsilon_{i_1' i_2' \dots i_{N-1}' i}\,,
	\end{array}
\end{equation}
where $\mu_a^0=\delta$ is just the identity matrix.
One can permute $\mu_1\to \mu_2\to \mu_3$ to obtain other, similar relations.

For the $N=4$ trinion theory, a generalization of \eqref{fund2anti} appears in Table 3 of \cite{LemosPeelaers} (discovered by the chiral-algebra analysis of \cite{chiral1, chiral2}.) With our normalization conventions, the relation takes the form
\be \label{LP-T4} \begin{array}{l} Q^{(i_1[j_1[k_1}Q^{i_2)j_2]k_2]} = -\tfrac{1}{2} (\mu_1)^{(i_1}{}_{i_1'} Q^{[i_2)i_1'][j_1j_2][k_1k_2]} \\[.2cm]
  Q_{(i_1[j_1[k_1}Q_{i_2)j_2]k_2]} =  \tfrac12 ({\mu_1})^{i_2'}{}_{(i_2}   Q_{[i_1) i_2'][j_1j_2][k_1k_2]} 
\end{array} \qquad (N=4)\,. \ee
Notice that if we were to read this as an N=3 relation, it would reduce to \eqref{fund2anti} after contractions with Levi-Civita tensors and an application of \eqref{TN-QQ}.

We propose a uniform generalization of \eqref{fund2anti} and \eqref{LP-T4}. Namely, from explicit computations at low values of $N$, we find that
\be \label{LP-gen}\begin{array}{l}  Q^{(i_1[j_1[k_1}Q^{i_2j_2k_2} ... Q^{i_r)j_r]k_r]}= \pm \tfrac{1}{r!} \delta^{(i_1}{}_{i_1'} (\tilde{\mu}_1)^{i_2}{}_{i_2'} ... (\tilde{\mu}_1^{r-1})^{i_r)}{}_{i_r'} Q^{[i_1' i_2' ... i_r'] [j_1 j_2 ... j_r] [k_1 k_2... k_r]} \\[.2cm]
 Q_{(i_1[j_1[k_1}Q_{i_2 j_2 k_2} ... Q_{i_r)j_3]k_r]} = \pm \tfrac{1}{r!} Q_{[i_1' i_2' ... i_r'] [j_1 j_2 ... j_r] [k_1 k_2 ... k_r]} \delta^{i_1'}{}_{(i_1}(\tilde{\mu}_1)^{i_2'}{}_{i_2} ... (\tilde{\mu}_1^{r-1})^{i_r'}{}_{i_r)}\,. \end{array}  \ee
Moreover, these are relations that hold in the quantized chiral ring $\C_\ve[\CM_C]$. Some explicit instances of \eqref{LP-gen} are listed in Appendix \ref{tensorcomps}. Appendix \ref{tensorcomps} holds several more examples of relations between products of fundamentals and higher antisymmetric tensors, particularly of the form
\be (\tilde\mu^{r-1} Q)^{[i_1[j_1[k_2}\ldots (\tilde \mu Q)^{i_{r-1}j_{r-1}k_{r-1}} Q^{i_r] j_r] k_r]} = (T_{(N,r)})^I{}_{I'} Q^{I'JK}\,.\ee
for appropriate tensors $T_{(N,r)}$ built out of the moment map $\tilde \mu_1$. We do not yet have a general formula for $T_{(N,r)}$, for all $N$ and $r$.

Some additional quantum relations among the $Q$'s with vanishing classical limits are summarized in Section \ref{sec:newq}, and detailed in Appendix \ref{app:comm}.


\section*{Acknowledgements}

We wish to thank David Ben-Zvi, Jacques Distler, Davide Gaiotto, Justin Hilburn, Yuji Tachikawa, and Ben Webster for enlightening discussions. 
The work of T.D. is partially supported by a Hellman Fellowship at UC Davis and by NSF CAREER Grant DMS 1753077.

\appendix

\section{$\C_\ve[\CM_C]$ and $\CA_\ve$ in the BFN construction}
\label{app:BFN}

The abelianized algebra $\CA_\ve$ is related to fixed-point localization in the mathematical definition of the Coulomb branch proposed by Braverman, Finkelberg, and Nakajima \cite{Nak2016, BFN2016}. 
To provide some context, we review some aspects of the physical interpretation of the BFN construction and the appearance of $\CA_\ve$ therein. For further discussion see \cite{VV} (where an analogous construction of $\C[\CM_C]$ via equivariant cohomology appears) and the upcoming~\cite{lineops}.

The analysis here is directly analogous to physical constructions of the category of 't Hooft (or Wilson-'t Hooft) operators in topologically twisted 4d $\CN=4$ (or $\CN=2$) super-Yang-Mills theory. Line operators in 4d are the dimensional lifts of local operators in 3d. The categories were studied in \cite{KapustinWitten} and \cite{KapustinSaulina} (generalized in \cite{GuW06}), and identified as categories of sheaves on particular versions of the space $\CM_{[\C\cup\C]}$ that appears below.

The BFN construction arises physically by considering a 3d $\CN=4$ gauge theory on spacetime $\C\times \R$, with a particular half-BPS boundary condition $\CB$ near spatial infinity on~$\C$
\be \CB\,:\qquad \left\{\begin{array}{l}\text{Neumann b.c. on the vectormultiplet,}\\[.1cm]
 \text{Dirichlet b.c. on the chiral half of the hypermultiplets in $R^*$} \end{array}\right.
\ee
We recall that the full hypermultiplets are in a quaternionic representation \eqref{RR} that is split (not necessarily uniquely) into two unitary representations  $R\oplus R^*$; this boundary condition thus sets half of the hypermultiplets to zero.

\begin{figure}[htb]
\centering
\includegraphics[width=5in]{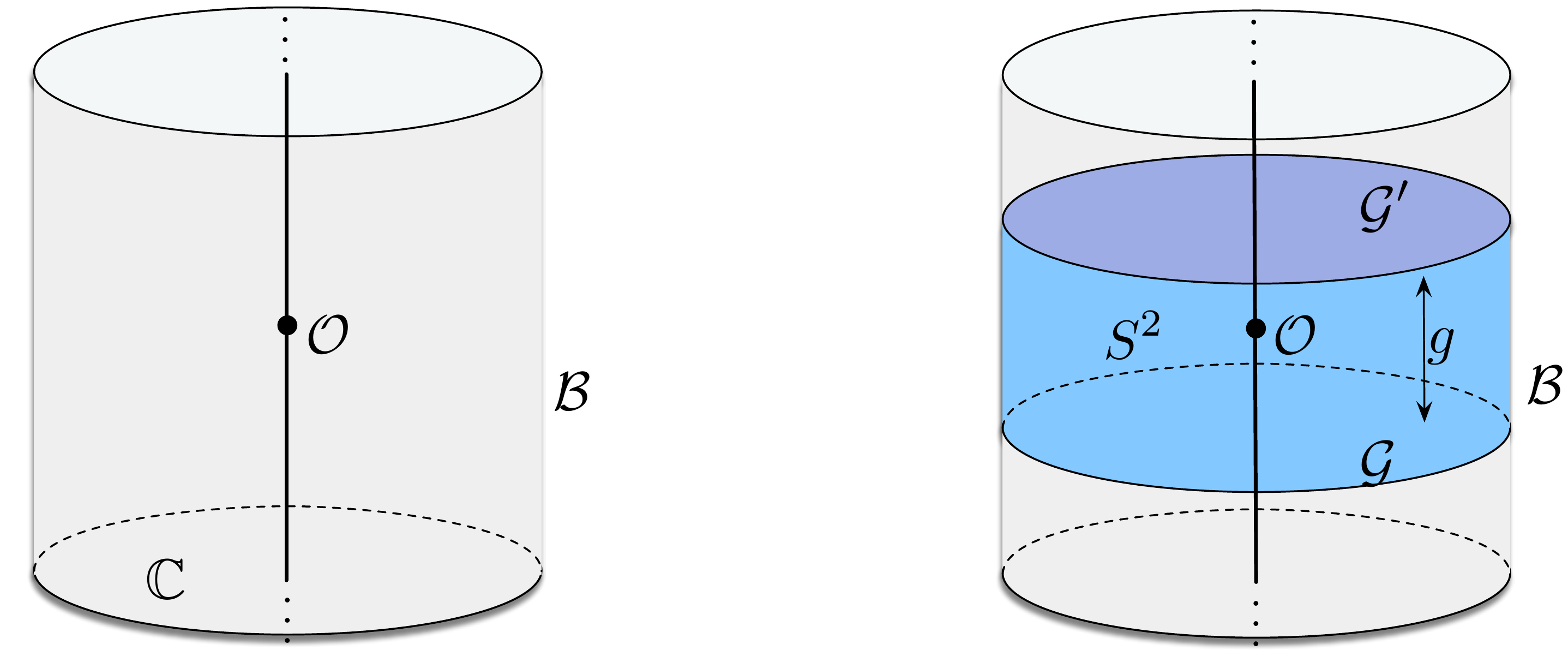}
\caption{Left: 3d $\CN=4$ theory on $\C\times \R$ with a boundary condition $\CB$ at infinity, reinterpreted as 1d $\CN=4$ quantum mechanics. Right: surrounding a local operator with a topological $S^2$ that has been deformed to two closely separated copies of $\C$ (the boundary of a cylindrical slab).}
\label{fig:cyl}
\end{figure}

This setup may roughly be imagined as a 3d theory on a solid cylinder (Figure \ref{fig:cyl}). 
We may also think of this 3d theory as a 1d $\CN=4$ quantum mechanics whose fields are various sections of bundles on $\C$, obeying the boundary condition. If we further work in the cohomology of the twisted Rozansky-Witten supercharge $Q_{RW}$ (whose local operators contain elements of $\C[\CM_C]$), we may restrict (localize) the fields in the quantum mechanics to sections of bundles that solve the $Q_{RW}$ BPS equations. After a bit of work, and a translation to algebraic language, this yields gauged $\CN=4$ QM on
\be \label{MD} \CM_{[\C]} = \left\{\begin{array}{l}
\text{pairs $(E,X)$, where $E$ is a holomorphic $G_\C$ bundle on $\C$} \\[.1cm]
\text{and $X$ is a holomorphic section of an associated $R$-bundle} \end{array}\right\}
\ee
with a gauge group%
\footnote{In the mathematics literature, $\C$ is usually replaced by a formal disc (so that holomorphic sections look like Taylor series rather than polynomials), and the gauge group is usually denoted $\CG=G_\C[\CO]$, where $\CO=\C[\![z]\!]$ is the ring of formal Taylor series. We will not be careful about such distinctions here, as we are merely trying to give an overview.}
\be  \CG = \text{holomorphic $G_\C$ gauge transformations on $\C$}\,. \ee
The Hilbert space of the quantum mechanics, in the $Q_{RW}$ twist, should be the $\CG$-equivariant de Rham cohomology of $\CM_{[\C]}$ \cite{Witten-Morse}.%
\footnote{Since $\CM_{[\C]}$ is infinite-dimensional, physics would dictate that $L^2$ de Rham cohomology be used. In the mathematics literature, Borel-Moore homology, \ie\ ``homology with closed support,'' is employed instead, as it is better behaved on spaces such as $\CM_{[\C\cup\C]}/\CG'$ encountered below, which are generalizations of affine Grassmannians.
It is fairly clear by now that mathematical computations in Borel-Moore homology, such as \eqref{BFN}, agree with physical expectations about the structure of local operators. Nevertheless, a fundamental understanding of why Borel-Moore homology is natural in QFT still seems to be missing.\label{foot:BM}}
Turning on the Omega background further corresponds to working equivariantly with respect to the $U(1)_\ve$ spatial rotation group of $\C$, which is an ordinary symmetry of the moduli space $\CM_{[\C]}$. We find a Hilbert space
\be \CH = H^*_{\CG\times U(1)_\ve}(\CM_{[\C]})\,. \ee
While this may look foreboding, a bit of thought shows that $\CM_{[\C]}$ is contractible to a point where $E$ is the trivial bundle and $X$ is the zero-section. Similarly, $\CG$ is contractible to $G$, so
\be \CH \simeq H^*_{G\times U(1)_\ve}(p) = \C[\varphi,\ve]^W \label{Hmod} \ee
just consists of Weyl-invariant polynomials in the equivariant weights $\varphi\in \ft_\C$ and $\ve$.

Now consider local operators of the 3d theory, inserted at a point $(0,0)\in \C\times \R$. A state-operator correspondence in TQFT would identify the vector space of local operators with the Hilbert space on a sphere $S^2$ surrounding the point $(0,0)\in \C\times \R$. From the perspective of $\CN=4$ QM, it is convenient to deform $S^2$ to the boundary of a cylindrical slab, as on the right side of Figure \ref{fig:cyl}. We then obtain a description of the Hilbert space as the cohomology of a moduli space of solutions to the $Q_{RW}$-BPS equations on this deformed $S^2$, \ie\
\be \CM_{[\C\cup \C]} = \left\{\begin{array}{l}
\text{pairs $(E,X)$ and $(E',X')$ as in \eqref{MD}, together with} \\[.1cm]
\text{a gauge transformation $g:(E,X)\big|_{\C^*}\overset{\sim}{\to}(E',X')\big|_{\C^*}$} \\[.1cm]
\text{that identifies these data away from the origin} \end{array}\right\}
\ee
Here $g$ is a holomorphic gauge transformation on $\C^*$, which is allowed to be singular at the origin, where a putative local operator is inserted. We then obtain the vector space of local operators in $Q_{RW}$-cohomology as
\begin{align} \C_\ve[\CM_C]  &= H^*_{\CG\times \CG'\times U(1)_\ve}(\CM_{[\C\cup \C]}) \notag \\[.1cm]
  &\simeq H^*_{G\times U(1)_\ve} (\CM_{[\C\cup\C]}/\CG')\,,  \label{BFN}
\end{align}
where $\CG,\CG'$ are the groups of regular (holomorphic) gauge transformations on the top and bottom copies of $\C$, and $U(1)_\ve$ as usual is the spatial rotation group.
In a nutshell, \eqref{BFN} is the BFN construction.%

There are several highly nontrivial aspects of the formal definition \eqref{BFN}. In contrast to \eqref{Hmod}, the space $\CM_{[\C\cup\C]}/\CG'$ is not contractible; it has highly nontrivial topology, as it must in order for \eqref{BFN} to contain monopole operators. Moreover, the OPE of local operators in the algebra $\C_\ve[\CM_C]$ does \emph{not} correspond to a cup product in cohomology. Rather, it naturally corresponds to an operation known as the \emph{convolution product}, which results from colliding and merging two of the ``spheres'' surrounding local operators in Figure \ref{fig:cyl}. Both the cohomology classes in \eqref{BFN} and their convolution product can be difficult to describe explicitly.

A useful tool in equivariant cohomology is fixed-point localization. Letting $T\subset G$ denote the maximal torus as usual, one finds that the $T\times U(1)_\ve$ fixed points of $\CM_{[\C\cup\C]}/\CG'$ are isolated and actually quite easy to describe: they are points $(E,X)$, $(E',X')$ where $X=X'=0$ are zero-sections,  $E$ is trivial, and $E'$ is obtained from $E$ by a gauge transformation
\be g(z) = z^A\,,\qquad A\in \text{cochar}(G)\,. \ee
(Here ``$z$'' is the local coordinate on $\C$, and we are using $A\in \text{Hom}(U(1),T)$ to define a meromorphic gauge transformation. For example, if $G=U(N)$, $z^A$ means $\text{diag}(z^{A_1},...,z^{A_N})$.)
Thus the fixed points are labelled by cocharacters --- just right to correspond to abelian monopole operators!

Let $\CF$ denote the fixed point set of the $T\times U(1)_\ve$ action on $\CM_{[\C\cup\C]}/\CG'$. 
We just explained that $\CF \simeq \text{cochar}(G)$ is isomorphic to the cocharacter lattice. 
The equivariant cohomology of the fixed point set just contains a copy of $H^*_{T\times U(1)_\ve}(\text{point}) = \C[\varphi,\ve]$ for every point in $\CF$, \ie\ $H^*_{T\times U(1)_\ve}(\CF)\simeq \C\big[\varphi,\ve,\{v_A\}_{A\in \text{cochar}(G)}\big]$. Its ``localized'' version inverts all weights $\langle \lambda,\varphi\rangle+n\ve$ (for any $\lambda$ in the weight lattice of $G$),
\be  H^*_{T\times U(1)_\ve}(\CF)^{\text{loc}} \simeq  \C\big[\varphi,\ve,\{v_A\}_{A\in \text{cochar}(G)}, \tfrac{1}{\langle \lambda,\varphi\rangle+n\ve}\big]\,,\ee
from which we see that our abelianized algebra $\CA_\ve$ from \eqref{Ae} sits inside
\be \CA_\ve \subset H^*_{T\times U(1)_\ve}(\CF)^{\text{loc}}\,. \ee
The only difference between $\CA_\ve$ and $H^*_{T\times U(1)_\ve}(\CF)^{\text{loc}}$ (as vector spaces) is that in $\CA_\ve$ we only inverted roots $M_\alpha+n\ve$ and weights $M_\lambda+n\ve$ where $\lambda\in \text{weights}(R)$; whereas the localized cohomology indiscriminately inverts all weights. The localization theorem provides the map
\be \C_\ve[\CM_C] = H^*_{G\times U(1)_\ve} (\CM_{[\C\cup\C]}/\CG') \;\hookrightarrow\; H^*_{T\times U(1)_\ve}(\CF)^{\text{loc}}\,. \label{loc} \ee
When this is carefully interpreted using Borel-Moore homology (see Footnote \ref{foot:BM}), one finds that the image actually lies inside $\CA_\ve$,
\be \C_\ve[\CM_C] \hookrightarrow \CA_\ve\,. \label{loc2} \ee
In other words, only the roots $M_\alpha+n\ve$ need to be inverted.

It is hardly obvious mathematically that the maps \eqref{loc}, \eqref{loc2} are embeddings of algebras (under the convolution product) rather than just vector spaces. The compatibility of fixed-point localization with the convolution product was proved by \cite{BFN2016}.

\section{$PSU(2)$ Coulomb branch via Demazure operators}
\label{app:PSU2}

Here we wish to give the simplest possible example of how the algebra $\CW_\ve$ from Section \ref{sec:image} fully reproduces the quantized Coulomb branch chiral ring. We consider 3d $\CN=4$ pure gauge theory with $G=PSU(2)$ (and $\CR=\oslash$). The mathematics of this example (in fact of any pure gauge theory) first appeared in \cite{BFM}, and were connected with the physics of 3d $\CN=4$ theories in \cite{Teleman}.

The Coulomb branch of $PSU(2)$ gauge theory is the centered moduli space of two $SU(2)$ monopoles \cite{SeibergWitten-3d}, known as the Atiyah-Hitchin manifold \cite{AtiyahHitchin}. As a holomorphic symplectic manifold, it is cut out of $\C^3$ by the equation
\be  U^2-\Phi V^2 = 4\,, \ee
and has a holomorphic symplectic form induced from the holomorphic 3-form on $\C^3$,
\be \Omega = \frac{d\Phi \wedge dV}{2U}\,.  \ee
In terms of the 3d $\CN=4$ theory, $\Phi = \frac12 \text{Tr}(\phi^2)$ is the generator of gauge-invariant polynomials in the complex scalar $\phi\in \fg_\C$; $V$ is a nonabelian monopole operator of fundamental charge (labelled by the fundamental cocharacter `1'), and $U$ is a dressed monopole operator.

The quantized ring of functions $\C_\ve[\CM_C]$ was described in \cite{BDG2015}. It is generated by operators $U,V,\Phi$ with commutation relations
\be [\Phi,V] = 2\ve U-\ve^2 V\,,\quad [\Phi,U]=2\ve \Phi V-\ve^2 V\,,\quad [U,V]=-\ve U^2\,, \label{AH-comm}\ee
and a central constraint 
\be U^2+\ve UV - \Phi V^2 = 4\,. \label{AH-rel}\ee
Let's see how these arise from abelianization.

The abelianized complex scalar $\phi$ is $\bsp \varphi & 0 \\ 0 & -\varphi\esp\in \ft_\C$, and the full abelianized algebra can be generated from $\varphi$, $\varphi^{-1}$, and the abelian monopole operators $v_+ = v_{1}$ and $v_-=v_{-1}$ of unit cocharacters,
\be \CA_\ve = \C\Big[\varphi, v_\pm, \frac{1}{\varphi}\Big]\,, \ee
with relations
\be [\varphi,v_\pm]=\pm \ve v_\pm\,,\qquad v_+v_- = \frac{-1}{\varphi(\varphi-\ve)}\,,\qquad v_-v_+ =  \frac{-1}{\varphi(\varphi+\ve)} \ee
following from \eqref{comm1}, \eqref{comm3}. Note that here the W-boson masses are $M_\alpha = \pm \varphi$, and it is sufficient to invert $\varphi$ because all other denominators $\frac{1}{\varphi+n\ve}$ can be obtained by commutation with $v_\pm$\,:
\be \frac{1}{\varphi+n\ve} = (-1)^n(\varphi v_-)^n \frac{1}{\varphi} (\varphi v_+)^n\,.\ee

The subalgebra $\CW_\ve$ is obtained by considering polynomials in $\varphi$ and the rescaled monopole operators \eqref{defu}
\be u_+ = -\varphi v_+\,,\qquad u_- = \varphi v_-\,,\ee
which obey the extremely simple algebra
\be [\varphi,u_\pm] = \pm \ve u_\pm\,,\qquad u_+u_- = u_-u_+ = 1\,.\ee
We then throw in the Demazure operator
\be \theta  =\frac{1}\varphi(s-1)\,, \ee
where $s$ is the simple reflection that generates the Weyl group, and acts as $su_\pm = u_{\mp}s$, $sf(\varphi) = f(-\varphi)s$. The definition of the $\CW_\ve$ algebra \eqref{We} says to take the Weyl-invariant part of the polynomials in $\varphi$, $u_\pm$, and $\theta$:
\be \CW_\ve = \C[\varphi,u_\pm,\theta]^W \simeq (s+1) \C[\varphi,u_\pm,\theta] (s+1) \,. \ee

Consider the sorts of elements this contains. The Weyl-invariant functions of $\varphi$ and $u_\pm$ include
\be \Phi := \varphi^2\,,\qquad  U:= u_+ + u_-\,.\ee
With the Demazure operator, we may also construct an operator
\be V := \big(\text{Weyl-invariant part of $\theta u_+$}\big) = \frac{1}{\varphi}(u_--u_+) = v_++v_-\,.  \ee
The Weyl-invariant part of the Demazure operator itself is zero; and we cannot apply the Demazure operator multiple times because it is nilpotent $\theta^2=0$. It turns out that the entire algebra $\CW_\ve$ is generated by the three Weyl-invariant operators above,
\be \CW_\ve = \C[\Phi,U,V] \subset \CA_\ve \ee
with relations among them induced from the relations in $\CA_\ve$. The reader may check that the these relations perfectly match \eqref{AH-comm} and \eqref{AH-rel} above, so that $\CW_\ve\simeq \C_\ve[\CM_C]$ as claimed.

\section{Some basic quantum identities}
\label{app:comm}

In this appendix, we collect some basic results on commutation relations between moment maps and the tensor operators $Q_{(r)}$, $Q^{(r)}$, as well as among the tensor operators themselves.  We work exclusively in the quantized chiral rings.

\subsection{Highest and lowest weight vectors}
\label{app:reps}

We first show, by means of  \eqref{comm1} and  \eqref{comm3}, that $Q^{NN...N}$ (resp. $Q_{NN...N}$) is a lowest (resp. highest) weight vector of a fundamental (resp. antifundamental) representation for each $SU(N)$ action, and then generalize to higher antisymmetric powers. Due to the permutation symmetry between the different legs, it suffices to consider the case of $\k = 1$.

As described in Section \ref{Nkquivers}, $Q^{N} = V^+_N$, \ie\ \begin{equation}
	Q^{N} = \sum \limits_{\alpha = 1}^{N} v^{+}_{N \alpha}\,,
\end{equation} where $v^{\pm}_{N \alpha} := v_{\pm e_{N\alpha}}$ is the abelian monopole with cocharacter $\pm e_{N\alpha}=((0...\overset{\alpha}{\pm1}...0);\vec 0;...;\vec 0)$.
Using  \eqref{comm1}, it follows that \begin{equation}
	[(\tilde{\mu})^{n}{}_n, v^{+}_{N \alpha}] = \varepsilon (\delta^N{}_n - \tfrac{1}{N})v^{+}_{N \alpha}
\end{equation} for all $\alpha$ and so \begin{equation}
	[(\tilde{\mu})^{n}{}_n, Q^{NN...N}] = \varepsilon (\delta^N{}_n - \tfrac{1}{N})Q^{NN...N}\,,
\end{equation} which precisely matches the weight of the lowest weight vector of a fundamental representation.

Now consider the lowering operators $V^{-}_n$, which can be expressed as \begin{equation}
	V^{-}_{n} = \sum \limits_{\alpha = 1}^{n} v^{-}_{n \alpha}
\end{equation} where $v^{\pm}_{n \alpha}:=v_{\pm e_{n\alpha}}$ is the abelian monopole with cocharacter $\pm e_{n\alpha}$, \emph{i.e.} a unit of fundamental magnetic charge (in the $\alpha$-th weight space) for the $U(n)$ node.
To see that $Q^{N}$ commutes with these operators, it suffices to notice that \begin{equation}
	[v^{-}_{n \beta}, v^{+}_{N \alpha}] = 0
\end{equation} for all $n, \beta, \alpha$. For $n < N$ this follows from  \eqref{comm1} by noting that the set of weights that pair nontrivially with both cocharacter is completely empty. For $n = N$, it is suffices to note that the only weight that pairs nontrivially with both cocharacters yield +1 for both cocharacters. Thus, $Q^{N}$ is a lowest weight vector of $SU(N)$. A completely analogous proof shows that $Q_{N}$ is a highest weight vector of a antifundamental representation.

Now consider the operator $Q^{[N N-1 ... N - r+1]}$ described in the main body of the text, we show that this operator is indeed a lowest weight vector of a rank $r$ antisymmetric tensor for each $SU(N)$ factor. We will express cocharacters of the gauge group as $A = (\vec{a}_N; \vec{a}_{N-1}, ..., \vec{a}_1)$ where $\vec{a}_n \in \mathbb{Z}^n$. Choose a subset $I^r_n \subseteq \{1, 2, ..., n\}$ of size $r \leq n$ and set $\vec{e}_{I^r_n} \in \mathbb{Z}^n$ to be the vector with 1 for each $i \in I^r_n$ and 0 otherwise. With this notation, we can write \begin{equation}
	Q^{[N N-1 ... N - r]} = \sum \limits_{\{I^r_N, I^{r-1}_{N-1}, ..., I^{1}_{N-r}\}} v_{(\vec{e}_{I^r_N}; \vec{e}_{I^{r-1}_{N-1}}, ..., \vec{e}_{I^1_{N-r}}, \vec{0}, ...., 0)}
\end{equation} It is worth noting that for $r = 1$ this reduces to $Q^N$ and for $r = N$ the only cocharacter is $A_{\textrm{diag}}$. From this presentation it should be clear that \begin{equation}
	[(\tilde{\mu})^{n}{}_n, Q^{[N N-1 ... N-r+1]}] = \varepsilon \Big(\sum \limits_{i = N - r + 1}^N \delta^i{}_n - \frac{r}{N}\Big) Q^{[N N-1 ... N-r+1]}
\end{equation} so this operator certainly has the proper weight to be a lowest weight vector of a rank $r$ antisymmetric tensor representation of $SU(N)$. To see that it is a lowest weight vector, we appeal to  \eqref{fund2tensor}; in particular we write \begin{equation}
	Q^{[N N-1 ... N-r+1]} = \pm r! \tilde{\mu}^{r-1} Q^{[N} \tilde{\mu}^{r-2}Q^{N-1} ... Q^{N-r + 1]}.
\end{equation} From this expression it is clear that \begin{equation}
	[V^-_n, Q^{[N N-1 ... N-r+1]} ] = 0
\end{equation} for all $n < N-r+1$ since the $Q^i$ furnish a fundamental representation. (By noticing all of the abelianized monopoles making up $Q^{[N N-1 ... N-r+1]}$ are either uncharged or have the same charge as the abelianized monopoles appearing in these $V^-_n$.) To see that this extends to the remaining $n$ we compute: \begin{equation}
	\begin{array}{rl}
		[V^-_n, Q^{[N N-1 ... N-r+1]}] & = (-1)^{N(r+1)} [V^-_n, \tilde{\mu}^r Q^{[N} \tilde{\mu}^{r-1}Q^{N-1} ... Q^{N-r + 1]}]\\
		& = (-1)^{N(r+1)} \varepsilon \tilde{\mu}^{r-1} Q^{[N} \tilde{\mu}^{r-2}Q^{N-1}... \tilde{\mu}^{r+n-N-1}Q^{n-1}  \tilde{\mu}^{r+n-N-2}Q^{n-1} ... Q^{N-r + 1]}\\
		& = 0\\
	\end{array}
\end{equation} where the last line follows because $n-1$ appears twice in the anti-symmetrization. In principle it should be possible to show that $Q^{[N N-1 ... N-r+1]}$ is a lowest weight vector directly from applying \eqref{comm3} but this would be a rather nontrivial process. It is worth noting that a completely analogous proof shows that $Q_{[N N-1 ... N-r+1]}$ is a highest weight vector of a rank $r$ antifundamental tensor representation, as desired.

\subsection{Index swaps}
\label{app:index}

We next investigate a series of identities of the form 
\begin{equation}
	\label{asymComm}
	[Q^{i_1 i_2 ... i_{\k}}, Q^{j_1 j_2 ... j_{\k}}] = - [Q^{j_1 i_2 ... i_{\k}}, Q^{i_1 j_2 ... j_{\k}}]\,,
\end{equation}
capturing the antisymmetry of the commutator of two fundamental tensors under a swap of any two indices. (Here it is shown for $i_1\leftrightarrow j_1$, but by symmetry of the legs of the star quiver, any swap $i_n\leftrightarrow j_n$ behaves the same way.)
For $\k= 3$ this will motivate a conjectured quantum relation between $Q_{(1)}$ and the antisymmetric tensor $Q_{(2)}$ in Appendix \ref{app:cyc}.
Whenever an identity \eqref{asymComm} holds, it should also hold upon replacing  $Q^{i_1 i_2 ... i_{\k}}$ with $Q_{i_1 i_2 ... i_{\k}}$; but we shall focus on the relations with $Q^{i_1 i_2 ... i_{\k}}$ for simplicity.

A quick computation shows \eqref{asymComm} relation cannot hold for all $\k$. For example in the $(N,\k) = (2,4)$ theory we have \begin{equation}
	[Q^{1111},Q^{2222}] + [Q^{2111},Q^{1222}] = \varepsilon(\Phi^1_1 - \Phi_2)\,.
\end{equation}
Nonetheless, we will prove that  \eqref{asymComm} does hold  for $\k \leq 3$.

We go about this in several steps. First we show that antisymmetry \eqref{asymComm} holds if \begin{equation}
	\label{asymstep1}
	[Q^{i_1 i_2 ... i_{\k}}, Q^{i_1 j_2 ... j_{\k}}] = 0
\end{equation} for all $i_1, i_2, ...., i_{\k}, j_2, ..., j_{\k}$. Then we show that \eqref{asymstep1} holds for all $i_1, i_2, ...., i_{\k}, j_2, ..., j_{\k}$ if \begin{equation}
	\label{asymstep2}
	[Q^{N i_2 ... i_{\k}}, Q^{N j_2 ... j_{\k}}] = 0
\end{equation} for all $i_2, ...., i_{\k}, j_2, ..., j_{\k}$. We finish the proof by showing that, indeed,  \eqref{asymstep2} holds for $\k \leq 3$. It is worth noting that   \eqref{asymComm},  \eqref{asymstep1}, and  \eqref{asymstep2} hold trivially for $\k = 1$. Furthermore, for $\k = 2$  \eqref{asymComm} implies that all of those commutators must vanish.

\cleardoublepage

\medskip
\noindent \textbf{Step 1:} 
\medskip

We show  \eqref{asymstep1} implies  \eqref{asymComm} via induction on $|i_1-j_1|$. Without loss of generality assume $i_1<j_1$, the base case is then $j_1 - i_1 = 0$ which follows from the assumption that $[Q^{i_1 i_2 ... i_{\k}}, Q^{i_1 j_2 ... j_{\k}}] = 0$. Now, for $j_1 - i_1 = n$ we compute: 
\begin{equation}
	\begin{array}{cl}
		[Q^{i_1 i_2 ... i_{\k}}, Q^{(i_1+n) j_2 ... j_{\k}}] &= [Q^{i_1 i_2 ... i_{\k}}, [V^{1+}_{i_1 +n},Q^{(i_1+n-1) j_2 ... j_{\k}}]]/\ve\\
		&= \left([[Q^{i_1 i_2 ... i_{\k}}, V^{1+}_{i_1+n}],Q^{(i_1+n-1) j_2 ... j_{\k}}] + [V^{1+}_{i_1+n}, [Q^{i_1 i_2 ... i_{\k}}, Q^{(i_1+n-1) j_2 ... j_{\k}}]]\right)/\ve\\
		&= -[V^{1+}_{i_1+n}, [Q^{(i_1+n-1) i_2 ... i_{\k}}, Q^{i_1 j_2 ... j_{\k}}]]/\ve,\\
	\end{array}
\end{equation} where the second line follows form the Jacobi identity and the third follows from the inductive hypothesis and the action of $V^{1+}_{i_1+n}$. Applying the Jacobi identity a second time yields\begin{equation}
	\begin{array}{cl}
		[Q^{i_1 i_2 ... i_{\k}}, Q^{(i_1+n) j_2 ... j_{\k}}] &= -\left([[V^{1+}_{i_1+n}, Q^{(i_1+n-1) i_2 ... i_{\k}}], Q^{i_1 j_2 ... j_{\k}}]] + [Q^{(i_1+n-1) i_2 ... i_{\k}}, [V^{1+}_{i_1+n}, Q^{i_1 j_2 ... j_{\k}}]]\right)/\epsilon\\
		& = -[Q^{(i_1+n) i_2 ... i_{\k}}, Q^{i_1 j_2 ... j_{\k}}],
	\end{array}
\end{equation} where the last line follows form the action of $V^{1+}_{i_1+n}$.

\medskip
\noindent \textbf{Step 2:}
\medskip

To show that  \eqref{asymstep2} implies  \eqref{asymstep1}, we use induction on $N - i_1$, the base case $i_1 = N$ holds assuming  \eqref{asymstep2} is true. We compute: \begin{equation}
	\begin{array}{rl}
		[Q^{i_1 i_2 ... i_{\k}}, Q^{i_1 j_2 ... j_{\k}}] &= [[V^{1+}_{i_1}, Q^{(i_1+1) i_2 ... i_{\k}}],Q^{i_1 j_2 ... j_{\k}}]/\ve\\
		&= \left([V^{1+}_{i_1}, [Q^{(i_1+1) i_2 ... i_{\k}},Q^{i_1 j_2 ... j_{\k}}]] - [Q^{(i_1+1) i_2 ... i_{\k}},[V^{1+}_{i_1}, Q^{i_1 j_2 ... j_{\k}}]]\right)/\ve\\
		&= [V^{1+}_{i_1}, [Q^{(i_1+1) i_2 ... i_{\k}},Q^{i_1 j_2 ... j_{\k}}]]/\ve = [V^{1+}_{i_1}, [Q^{(i_1+1) i_2 ... i_{\k}},[V^{1+}_{i_1}, Q^{(i_1+1) j_2 ... j_{\k}}]]]/\ve^2\\
		&= [V^{1+}_{i_1}, \left([[Q^{(i_1+1) i_2 ... i_{\k}},V^{1+}_{i_1}],Q^{(i_1+1) j_2 ... j_{\k}}] + [V^{1+}_{i_1},[Q^{(i_1+1) i_2 ... i_{\k}},Q^{(i_1+1) j_2 ... j_{\k}}]] \right)]/\ve^2\\
		&= [V^{1+}_{i_1}, [[Q^{(i_1+1) i_2 ... i_{\k}},V^{1+}_{i_1}],Q^{(i_1+1) j_2 ... j_{\k}}]]/\ve^2 = -[V^{1+}_{i_1}, [Q^{i_1 i_2 ... i_{\k}},Q^{(i_1+1) j_2 ... j_{\k}}]]/\ve\\
		&= -\left([[V^{1+}_{i_1}, Q^{i_1 i_2 ... i_{\k}}],Q^{(i_1+1) j_2 ... j_{\k}}] + [Q^{i_1 i_2 ... i_{\k}}, [V^{1+}_{i_1}, Q^{(i_1+1) j_2 ... j_{\k}}]] \right)/\ve\\
		&= -[Q^{i_1 i_2 ... i_{\k}}, Q^{i_1 j_2 ... j_{\k}}],\\
	\end{array}
\end{equation} where only the Jacobi identity (lines 1 $\rightarrow$ 2, 3 $\rightarrow$ 4, 5 $\rightarrow$ 6), the $SU(N)$ action (lines 1, 2 $\rightarrow$ 3, 3, 4 $\rightarrow$ 5, 6 $\rightarrow$ 7) and the inductive hypothesis (line 4 $\rightarrow$ 5) were used. Therefore $[Q^{i_1 i_2 ... i_{\k}}, Q^{i_1 j_2 ... j_{\k}}] = 0$.

\medskip
\noindent \textbf{Step 3:}
\medskip

We now move on to showing that  \eqref{asymstep2} holds for $\k = 2, 3$, starting with $\k = 2$. For $\k = 2$ we can assume $j_2 < N$ and since $Q^{NN}$ is only charged under the central node it suffices to check $j_2 = N-1$. We are then interested in the commutator $[Q^{NN}, Q^{N N-1}]$. This commutator takes the following (schematic) form: \begin{equation}
	[Q^{NN}, Q^{N N-1}] = \sum \limits_{\alpha_1, \alpha_2 = 1}^N \sum \limits_{\beta = 1}^{N-1} [v_{A_{\alpha_1}}, v_{A_{\alpha_2}+B^2_{\beta}}]
\end{equation} where $B^a_{\beta}$ is the analog of the $A_{\alpha}$ but for the $N-1$ node on the $a$-th leg of the quiver. When $\alpha_1 = \alpha_2$ this commutator must vanish, so we are interested in \begin{equation}
[v_{A_{\alpha_1}}, v_{A_{\alpha_2}+B^2_{\beta}}] + [v_{A_{\alpha_2}}, v_{A_{\alpha_1}+B^2_{\beta}}]
\end{equation} for $\alpha_1 \neq \alpha_2$. A quick application of  \eqref{comm3} yields \begin{equation}
v_{A_{\alpha_1}} v_{A_{\alpha_2}+B^2_{\beta}} - v_{A_{\alpha_1}+B^2_{\beta}} v_{A_{\alpha_2}} = \frac{v_{A_{\alpha_1} + A_{\alpha_2} + B^2_{\beta}}}{(\varphi_{N \alpha_2}-\varphi_{N \alpha_1})}
\end{equation} and therefore \begin{equation}
[v_{A_{\alpha_1}}, v_{A_{\alpha_2}+B^2_{\beta}}] + [v_{A_{\alpha_2}}, v_{A_{\alpha_1}+B^2_{\beta}}] = 0
\end{equation} for $\alpha_1 \neq \alpha_2$ and therefore $[Q^{NN}, Q^{N j_2}] = 0$ for all $j_2$. To see this extends to $[Q^{N j_1}, Q^{N j_2}]$, we induct on $N - j_1$. We can further assume $j_2 \leq j_1$, this yields \begin{equation}
	\begin{array}{rl}
		[Q^{N j_1}, Q^{N j_2}] & = [[V^{2+}_{j_1} Q^{N j_1 +1}], Q^{N j_2}]\\
		& = - [[Q^{N j_2},V^{2+}_{j_1}], Q^{N j_1 +1}] - [[Q^{N j_1 +1},Q^{N j_2}],V^{2+}_{j_1}]\\
		& = 0
	\end{array}
\end{equation} where the second line follows from the Jacobi identiy and the third by the $SU(N)$ action and the inductive hypothesis.

Now consider $\k = 3$, we hope to show that \begin{equation}
	[Q^{N j_1 k_1}, Q^{N j_2 k_2}] = 0
\end{equation} via induction in a similar fashion as the $\k = 2$ computation, it suffices to check the base cases as the induction will follow identically to the above. We can assume that $j_1 \leq j_2$ giving us two cases to check, depending on which of $k_1, k_2$ is larger. When $k_1 \leq k_2$, the base case of induction translates to \begin{equation}
	[Q^{NNN}, Q^{N j_2 k_2}] = 0\,,
\end{equation} which follows from the case where $j_2 = k_2 = N-1$. We compute: \begin{equation}
	[Q^{NNN}, Q^{N N-1 N-1}] = \frac{1}{2} \sum \limits_{\alpha_1 \neq \alpha_2 = 1}^N \sum \limits_{\beta_2, \beta_3 = 1}^{N-1} [v_{A_{\alpha_1}}, v_{A_{\alpha_2}+B^2_{\beta_2}+B^3_{\beta_3}}]+[v_{A_{\alpha_2}}, v_{A_{\alpha_1}+B^2_{\beta_2}+B^3_{\beta_3}}] \,.
\end{equation} Another quick application of  \eqref{comm3} yields \begin{equation}
	v_{A_{\alpha_1}} v_{A_{\alpha_2}+B^2_{\beta_2}+B^3_{\beta_3}}- v_{A_{\alpha_1}+B^2_{\beta_2}+B^3_{\beta_3}} v_{A_{\alpha_2}} = \frac{(\varphi_{N \alpha_1}+\varphi_{N \alpha_2}-\varphi^2_{N-1 \beta_2}-\varphi^3_{N-1 \beta_3})}{(\varphi_{N \alpha_2}-\varphi_{N \alpha_1})}v_{A_{\alpha_1} + A_{\alpha_2} + B^2_{\beta_2} + B^3_{\beta_3}}
\end{equation} therefore \begin{equation}
	[v_{A_{\alpha_1}}, v_{A_{\alpha_2}+B^2_{\beta_2}+B^3_{\beta_3}}]+[v_{A_{\alpha_2}}, v_{A_{\alpha_1}+B^2_{\beta_2}+B^3_{\beta_3}}] = 0
\end{equation} as desired.

When $k_2 \leq k_1$, the base case of induction similarly translates to \begin{equation}
	[Q^{NN k_1}, Q^{N j_2 N}] = 0
\end{equation} which follows from $k_1 = j_2 = 0$. We compute: \begin{equation}
	[Q^{NNN-1}, Q^{N N-1 N}] = \frac{1}{2} \sum \limits_{\alpha_1 \neq \alpha_2 = 1}^N \sum \limits_{\beta_2, \beta_3 = 1}^{N-1} [v_{A_{\alpha_1}+B^3_{\beta_3}}, v_{A_{\alpha_2}+B^2_{\beta_2}}]+[v_{A_{\alpha_2}+B^3_{\beta_3}}, v_{A_{\alpha_1}+B^2_{\beta_2}}].
\end{equation} and by applying  \eqref{comm3} we find that \begin{equation}
v_{A_{\alpha_1}+B^3_{\beta_3}} v_{A_{\alpha_2}+B^2_{\beta_2}}- v_{A_{\alpha_1}+B^2_{\beta_2}} v_{A_{\alpha_2}+B^3_{\beta_3}} = \frac{(\varphi^2_{N-1 \beta_2}-\varphi^3_{N-1 \beta_3})}{(\varphi_{N \alpha_2}-\varphi_{N \alpha_1})}v_{A_{\alpha_1} + A_{\alpha_2} + B^2_{\beta_2} + B^3_{\beta_3}}
\end{equation} which implies that \begin{equation}
	[v_{A_{\alpha_1}+B^3_{\beta_3}}, v_{A_{\alpha_2}+B^2_{\beta_2}}]+[v_{A_{\alpha_2}+B^3_{\beta_3}}, v_{A_{\alpha_1}+B^2_{\beta_2}}] = 0,
\end{equation} as desired.

\medskip

\noindent\textbf{The $\k\geq 4$ cases}
\medskip

The above analysis shows why antisymmetry \eqref{asymComm} cannot hold as written for $\k \geq 4$. Consider the commutator $[Q^{NN...N}, Q^{N N-1 ... N-1}]$, this has the schematic form \begin{equation}
	[Q^{NN...N}, Q^{N N-1 ... N-1}] = \frac{1}{2} \sum \limits_{\alpha_1 \neq \alpha_2 = 1}^N \sum \limits_{\beta_2, ...,  \beta_{\k} = 1}^{N-1} [v_{A_{\alpha_1}}, v_{A_{\alpha_2}+B^2_{\beta_2}+...+B^{\k}_{\beta_{\k}}}]+[v_{A_{\alpha_2}}, v_{A_{\alpha_1}+B^2_{\beta_2}+...+B^{\k}_{\beta_{\k}}}].
\end{equation} Applying  \eqref{comm3} yields \begin{equation}
	\begin{array}{c}
		v_{A_{\alpha_1}} v_{A_{\alpha_2}+B^2_{\beta_2}+...+B^{\k}_{\beta_{\k}}} - v_{A_{\alpha_1}+B^2_{\beta_2}+...+B^{\k}_{\beta_{\k}}} v_{A_{\alpha_2}}\\
	
		= \frac{\bigg(\prod \limits_{a = 2}^{\k} \varphi_{N \alpha_1} - \varphi_{N-1 \beta_a} - \varepsilon/2 \bigg) - \bigg(\prod \limits_{a = 2}^{\k} \varphi_{N \alpha_2} - \varphi_{N-1 \beta_a} + \varepsilon/2 \bigg)}{(\varphi_{N \alpha_2}-\varphi_{N \alpha_1})(\varphi_{N \alpha_1}-\varphi_{N \alpha_2}-\varepsilon)}v_{A_{\alpha_1} + A_{\alpha_2} + B^2_{\beta_2} + ... + B^{\k}_{\beta_{\k}}}\\
	\end{array}.
\end{equation} The numerator in the above expression vanishes when $\varphi_{N \alpha_1}-\varphi_{N \alpha_2} = \varepsilon$ and so the vanishing of $[Q^{NN...N}, Q^{N N-1 ... N-1}]$ hinges on the fact that \begin{equation}
	\frac{\bigg(\prod \limits_{a = 2}^{\k} \varphi_{N \alpha_1} - \varphi_{N-1 \beta_a} - \varepsilon/2 \bigg) - \bigg(\prod \limits_{a = 2}^{\k} \varphi_{N \alpha_2} - \varphi_{N-1 \beta_a} + \varepsilon/2 \bigg)}{(\varphi_{N \alpha_1}-\varphi_{N \alpha_2}-\varepsilon)}
\end{equation} is invariant under $\alpha_1 \leftrightarrow \alpha_2$. Unfortunately, for $\k \geq 4$ this need not be the case.

\medskip
\noindent\textbf{Triviality of $\k=1$}
\medskip

What happens for $\k = 1$? In this case, just as above, induction reduces the problem to computing one commutator: $[Q^N, Q^{N-1}]$. We compute: \begin{equation}
	v_{A_{\alpha_1}} v_{A_{\alpha_2}+B^1_{\beta}} - v_{A_{\alpha_1}+B^1_{\beta}} v_{A_{\alpha_2}} = \frac{v_{A_{\alpha_1} + A_{\alpha_2} + B^1_{\beta}}}{(\varphi_{N \alpha_2}-\varphi_{N \alpha_1})}.
\end{equation} After tracing through the above steps, it follows that $[Q^N, Q^{N-1}] = 0$ and therefore \begin{equation}
	[Q^{i_1}, Q^{i_2}] = 0
\end{equation} for all $i_1, i_2$! This fact can also be deduced by reducing $SU(N+1)$ to $SU(N)$, whereby the $Q^i$ are realized as raising operators which are required to commute with one another.

\subsection{Generating antisymmetric powers}
\label{app:cyc}

In the case of $\k=3$, the antisymmetry \eqref{asymComm} of the commutator $[Q^{i_1 j_1 k_1},Q^{i_2 j_2 k_2}]$ under an interchange of any pair of indices ($i_1\leftrightarrow i_2$, or $j_1\leftrightarrow j_2$, or $k_1\leftrightarrow k_2$) implies that the commutator itself transforms as a 3-fold 2-index antisymmetric tensor representation of $SU(N)^3$.
We conjecture that in fact
 \begin{equation}
	[Q^{i_1 j_1 k_1},Q^{i_2 j_2 k_2}] = \varepsilon Q^{[i_1 i_2] [j_1 j_2] [k_1 k_2]}
\end{equation} where $Q^{[i_1 i_2] [j_1 j_2] [k_1 k_2]}$ are the 2-index antisymmetric tensor operators described in the main text. This identity has been verified explicitly for $N \leq 4$.

From this data, it is rather suggestive to guess that all higher antisymmetric tensor operators could be constructed from iterated commutators but this simply cannot be the case. In particular, for $N = 3$ a short computation shows that \begin{equation}
	[Q^{2 1 1},[Q^{2 2 2}, Q^{3 3 3}]] = \varepsilon^2 V^{1-}_1
\end{equation} which clearly does not vanish despite having a repeated index.

Nonetheless, we can consider something slightly more complex. In particular, it is straightforward to check that \begin{equation}
	Q^{i_1 i_2 i_3} Q^{[j_1 k_1] [j_2 k_2] [j_3 k_3]} + Q^{j_1 i_2 i_3} Q^{[k_1 i_1] [j_2 k_2] [j_3 k_3]} + Q^{k_1 i_2 i_3} Q^{[i_1 j_1] [j_2 k_2] [j_3 k_3]}
\end{equation} is antisymmetric under the interchange $i_1 \leftrightarrow j_1$ and $i_1 \leftrightarrow k_1$. With this in mind, define \begin{equation}
	\textrm{Cyc}^3_a(Q^{i_1 i_2 i_3} Q^{[j_1 k_1] [j_2 k_2] [j_3 k_3]}) := \sum \limits_{n = 1}^3 \sigma^n_a(Q^{i_1 i_2 i_3} Q^{[j_1 k_1] [j_2 k_2] [j_3 k_3]})
\end{equation} where is the three cycle $\sigma$ is the 3-cycle (123) and $\sigma^n_a(Q^{i_1 j_1 k_1} Q^{[i_2 i_3] [j_2 j_3] [k_2 k_3]})$ means apply $\sigma^n$ to the set $\{i_a, j_a, k_a\}$. From the above, it follows that $\textrm{Cyc}^3_a(Q^{i_1 i_2 i_3} Q^{[j_1 k_1] [j_2 k_2] [j_3 k_3]})$ is totally antisymmetric in $\{i_a, j_a, k_a\}$ and, if we require this operation be linear, $\textrm{Cyc}^3_a$ and $\textrm{Cyc}^3_b$ commute with one another with $(\textrm{Cyc}^3_a)^2 = 3 \textrm{Cyc}^3_a$. This suggests the definition \begin{equation}
	\widetilde{Q}^{[i_1 j_1 k_1] [i_2 j_2 k_2] [i_3 j_3 k_3]} := \textrm{Cyc}^3_1 \circ \textrm{Cyc}^3_2 \circ \textrm{Cyc}^3_3(Q^{i_1 i_2 i_3} Q^{[j_1 k_1] [j_2 k_2] [j_3 k_3]})
\end{equation} which, by construction, furnishes a 3-fold 3-index antisymmetric tensor representation. This construction can be extended to any rank tensor by defining \begin{equation}
	\textrm{Cyc}^r_a(A^{i_1 i_2 i_3} B^{[j_1 ... k_1] [j_2 ... k_2] [j_3... k_3]}) := \sum \limits_{n = 1}^r(-1)^{n(r-1)} \sigma^n_a(A^{i_1 i_2 i_3} B^{[j_1 ... k_1] [j_2 ... k_2] [j_3... k_3]}),
\end{equation} where $\sigma$ is the $r$-cycle (12...r) and $\sigma^n_a$ means apply $\sigma^n$ to the set $\{i_a, j_a, ..., k_a\}$. Clearly, the resulting operator is totally antisymmetric in $\{i_a, j_a, ..., k_a\}$ and if we require this operation to be linear we come to the definition \begin{equation}
	\widetilde{Q}^{[i_1 j_1 ... k_1] [i_2 j_2 ... k_2] [i_3 j_3 ... k_3]} := \textrm{Cyc}^r_1 \circ \textrm{Cyc}^r_2 \circ \textrm{Cyc}^r_3(Q^{i_1 i_2 i_3} \widetilde{Q}^{[j_1 ... k_1] [j_2 ... k_2] [j_3 ... k_3]}).
\end{equation}

Although the above operators furnish the appropriate representation to agree with the 3-fold antisymmetric tensor operators found in the primary text, it is not clear whether these operators are identical, up a simple prefactor, or whether there is a nontrivial relationship between the two (e.g. some contraction of moment maps with the indices of $Q^{[i_1 j_1 ... k_1] [i_2 j_2 ... k_2] [i_3 j_3 ... k_3]}$). What can be said is that \begin{equation}
	\textrm{Cyc}^2_1 \circ \textrm{Cyc}^2_2 \circ \textrm{Cyc}^2_3(Q^{i_1 i_2 i_3} Q^{j_1 j_2 j_3}) = 4 [Q^{i_1 i_2 i_3}, Q^{j_1 j_2 j_3}] = 4 \varepsilon Q^{[i_1 j_1] [i_2 j_2] [i_3 j_3]},
\end{equation} which follows the antisymmetry of the commutators found above. The general form of this relationship could provide a systematic approach to constructing the higher tensor operators from the fundamentals. It would also be worthwhile to understand how this procedure compares to completely antisymmetrizing a product of fundamentals. This agreement is obvious for $r = 2$, since $\mathbb{Z}_2 \simeq S_2$ and so these processes are identical, but for higher $r$ the relationship is much less clear.

\section{Characteristic polynomials and higher tensors}
\label{tensorcomps}

In this appendix we recall the definition of the Capelli determinant that is used to define characteristic polynomials and Casimir operators in a non-commutative algebra. Then we collect some computations of relations among (quantum) $\k$-fold fundamental operators $Q_{(r)}$ and higher tensors that should correspond to un-diagonalizing relations of the form
\be q^{i_1} q^{i_2} ... q^{i_r} = (-1)^{r+1}\prod \limits_{1 \leq n < m \leq r} \left[\frac{(m_{i_m} - m_{i_n})^{\k-1}}{(m_{i_n} - m_{i_m} - \varepsilon)}\right] q^{[i_1 ... i_r]}\,, \ee
as in \eqref{qprod-q}, \eqref{qprod-q2}.

\subsection{Quantum characteristic polynomials}
\label{sec:Capelli}

Here we review the Capelli determinant that is used to compute the characteristic polynomial of quantum moment maps.

At $\ve=0$, the moment-map operators $\mu_a$ for the $SL(N,\C)^\k$ action on the Coulomb branch chiral ring have characteristic polynomials
\be \det(t+\mu_1) = \det(t+\mu_2) = \ldots =\det(t+\mu_\k) = \sum_{\ell=0}^N c_{N-\ell} t^\ell\,.\ee
Thinking of each moment map as an element of $\mathfrak{sl}(N,\C)^*$, the coefficients $c_i$ are polynomials in the Casimir operators of the enveloping algebra $U\mathfrak{sl}(N,\C)$. In particular, the $c_\ell$ are invariant under the $SL(N,\C)$ action; they Poisson-commute with all the individual components of the $\mu_a$.

In the quantum case, at $\ve\neq 0$, we would similarly like to identify a characteristic polynomial whose coefficients are $SL(N,\C)$ invariants, in that they commute with all individual components of the $\mu_a$. The naive determinant $\det(t+\mu_a)$ does not have this property. (It is also not well defined.) The solution, however, is well known: we must use a Capelli shifted determinant instead.
We recall how this works.

Given a matrix  $E = (E_{ij})$ whose elements are noncommutative operators that satisfy the $\mathfrak{gl}(N,\mathbb{C})$ commutation relations,
\begin{equation}
[E_{ij}, E_{kl}] = \varepsilon (\delta_{jk} E_{il} - \delta_{il}E_{jk}), \label{E-comm}
\end{equation} define the shifted matrix $\tilde{E} = E + \varepsilon(N-i)\delta_{ij}$. The Capelli determinant of $E$ is then given by \begin{equation}
\wt\det E = \tilde{E}_{i_1 1} \tilde{E}_{i_2 2} .... \tilde{E}_{i_N N} \epsilon^{i_1 i_2 ... i_N}, 
\end{equation} and the characteristic equation is then defined as usual: \begin{equation}
P_E(t) = \wt\det(t + E)\,. \label{def-Cap}
\end{equation} 
The coefficients of \eqref{def-Cap} are central in the algebra \eqref{E-comm}.

\begin{table}[htb]
	\centering
	\begin{tabular}{c | c c c}
		& $N = 2$ & $N = 3$ & $N = 4$\\ \hline
		$c_1$ &    $-\varepsilon$      &     $-3\varepsilon$     &    $-6 \varepsilon$\\
		$c_2$ &  $\prod \limits_{\alpha = 1}^2 \bigg(\varphi_{2 \alpha} - \Phi_{2}/2\bigg)$  &  $-\frac{1}{6}\bigg(\sum \limits_{\alpha = 1}^3 \sum \limits_{\beta > \alpha} \big(\varphi_{3\alpha} - \varphi_{3\beta}\big)^2 - 6\varepsilon^2\bigg)$   &     $-\frac{1}{8}\bigg(\sum \limits_{\alpha = 1}^4 \sum \limits_{\beta > \alpha} \big(\varphi_{4\alpha} - \varphi_{4\beta}\big)^2 -56\varepsilon^2\bigg)$\\
		$c_3$ & N/A & $\prod \limits_{\alpha = 1}^3 \bigg(\varphi_{3 \alpha} - \Phi_{3}/3\bigg)$  & \begin{tabular}{@{}c@{}}$\frac{1}{8}\bigg(\sum \limits_{\alpha = 1}^4 \prod \limits_{\beta \neq \alpha} \big(\varphi_{4\alpha} - \varphi_{4\beta}\big)$ \\ $+ \varepsilon \sum \limits_{\alpha = 1}^4 \prod \limits_{\beta > \alpha} \big(\varphi_{4\alpha} - \varphi_{4\beta}\big)^2 - 8\varepsilon^3 \bigg)$ \end{tabular} \\ 
		$c_4$ & N/A & N/A &  $\prod \limits_{\alpha = 1}^4 \bigg(\varphi_{4 \alpha} - \Phi_{4}/4\bigg)$   \\
	\end{tabular}
	\caption{Coefficients of (Capelli corrected) characteristic equation for the $SL(N,\C)$ moment maps, for $N = 2,3,4$, in terms the scalars $\varphi_{N \alpha}$ on the central node. Keeping with the notation of the main text, $\Phi_N$ is the sum of the $\varphi_{N \alpha}$.}
	\label{capelliCoeffs}
\end{table}

A particularly natural way to parameterize the coefficients $c_\ell$ of the quantum characteristic polynomial is as
\begin{equation} \label{def-cl}
\wt\det(t + E) = \sum \limits_{\ell = 0}^{N} [t]^{\ell} c_{N-\ell}
\end{equation} where, just as in the main body of the text, \begin{equation}
[a]^b := \begin{cases} 
\prod\limits_{k=0}^{b-1} (a+k\varepsilon) & b>0\\ 
\prod\limits_{k=1}^{|b|} (a-k\varepsilon) & b<0\\1 & b = 0\\ 
\end{cases}.
\end{equation}
Notice that   $c_0 = 1$ and $c_N = \wt{\det} E$.

We use \eqref{def-cl} to define the $c_\ell$'s for the quantum moment map operators $\tilde{\mu}_a$, namely $\wt\det(t+ \tilde{\mu}_a) = \sum \limits_{\ell = 0}^{N} c_{N-\ell} [t]^{\ell}$. Just as in the classical ($\ve=0$) case, the coefficients $c_\ell$ do not depend on which leg $a$ we choose. This is consistent with the fact that the eigenvalues of the quantum moment maps, described in the main text, are independent of $a$. 

We write down some explicit expressions for the $c_\ell$'s in $\TNk$ theories with $N\leq 4$ in Table~\ref{capelliCoeffs}. Note that they depend only on the scalars $\varphi_{N \alpha}$ associated to the central node of the star quiver.

\subsection{Fundamentals and higher tensors}
As described in the main text, it is expected that by taking appropriate antisymmetric combinations of the (anti)fundamental operators $Q^{i_1 i_2 ... i_{\k}}$ it should be possible to construct the higher rank tensor operators --- thereby, un-diagonalizing relations such as
\be q^{i_1} q^{i_2} ... q^{i_r} = (-1)^{r+1}\prod \limits_{1 \leq n < m \leq r} \left[\frac{(m_{i_m} - m_{i_n})^{\k-1}}{(m_{i_n} - m_{i_m} - \varepsilon)}\right] q^{[i_1 ... i_r]}\,, \label{qprod-app} \ee
\be q_{i_1} q_{i_2} ... q_{i_r} = (-1)^{(N-1)(r+1)} \prod \limits_{1 \leq n < m \leq r} \left[\frac{(m_{i_n} - m_{i_m} + \ve)^{\k-1}}{(m_{i_m} - m_{i_n})}\right] q_{[i_1 ... i_r]}\,, \label{qprod-2app}  \ee
from \eqref{qrels-q}.

Specifically, in the theory $\CT_{N,\k}$ we expect a relation of the form
\begin{equation}
\begin{array}{l}
(r!)^{\k} (\tilde{\mu}^{r-1}Q)^{[i_{(1,1)} [i_{(2,1)} ... [i_{(\k,1)}} \dots (\tilde{\mu}Q)^{i_{(1,r-1)} i_{(2,r-1)} ... i_{(\k,r-1)}} Q^{i_{(1,r)}] i_{(2,r)}] ... i_{(\k,r)}]}\\
= (T_{(N,\k,r)}^a)^{[i_{(a,1)} ... i_{(a,r-1)} i_{(a,r)}]}{}_{[i_{(a,1)}' ... i_{(a,r-1)}' i_{(a,r)}']}Q^{[i_{(1,1)} ... i_{(1,r-1)} i_{(1,r)}] ... [i_{(a,1)}' ... i_{(a,r-1)}' i_{(a,r)}'] ...  [i_{(\k,1)} ... i_{(\k,r-1)} i_{(\k,r)}]}\,,
\end{array}
\end{equation}
where the RHS involves a contraction on the $a$-th leg with some tensor  $T_{(N,\k,r)}^a$. (By symmetry of the legs, $T_{(N,\k,r)}^a$ should look essentially the same for any $a$.) Similarly, for antifundamental tensors, we expect
\begin{equation}
\begin{array}{l}
(r!)^{\k} Q_{[i_{(1,1)} [i_{(2,1)} ... [i_{(\k,r)}} \dots (Q\tilde{\mu}^{(r-2)})_{i_{(1,r-1)} i_{(2,r-1)} ... i_{(\k,r-1)}} (Q\tilde{\mu}^{r-1})^{i_{(1,r)}] i_{(2,r)}] ... i_{(\k,r)}]}\\
= Q_{[i_{(1,1)} ... i_{(1,r-1)} i_{(1,r)}] ... [i_{(a,1)}' ... i_{(a,r-1)}' i_{(a,r)}'] ...  [i_{(\k,1)} ... i_{(\k,r-1)} i_{(\k,r)}]} (\wt{T}_{(N,\k,r)}^a)^{[i_{(a,1)}' ... i_{(a,r-1)}' i_{(a,r)}']}{}_{[i_{(a,1)} ... i_{(a,r-1)} i_{(a,r)}]}
\end{array}
\end{equation} 
for some some tensors $\wt{T}_{(N,\k,r)}^a$.
This Appendix collects some direct computations of such relations. The computations suggest that, up to a numerical prefactor,
\be
(T_{(N,\k,r)}^a)^{[i_1 ... i_{r-1} i_r]}{}_{[i_1' ... i_{r-1}' i_r']} = (\wt{T}_{(N,\k,r)}^a)^{[i_1 ... i_{r-1} i_r]}{}_{[i_1' ... i_{r-1}' i_r']}\,.
\ee
Unfortunately, we have not yet identified an expression for $T_{(N,\k,r)}^a$ that is valid all $(N,\k,r)$.

We also include a set of similar relations obtained without inclusion of $\tilde{\mu}$, namely \begin{equation}
\begin{array}{l}
(r!)^{\k} Q^{[i_{(1,1)} [i_{(2,1)} ... [i_{(\k,1)}} \dots Q^{i_{(1,r-1)} i_{(2,r-1)} ... i_{(\k,r-1)}} Q^{i_{(1,r)}] i_{(2,r)}] ... i_{(\k,r)}]}\\
= (S_{(N,\k,r)}^a)^{[i_{(a,1)} ... i_{(a,r-1)} i_{(a,r)}]}{}_{[i_{(a,1)}' ... i_{(a,r-1)}' i_{(a,r)}']}Q^{[i_{(1,1)} ... i_{(1,r-1)} i_{(1,r)}] ... [i_{(a,1)}' ... i_{(a,r-1)}' i_{(a,r)}'] ...  [i_{(\k,1)} ... i_{(\k,r-1)} i_{(\k,r)}]}
\end{array}
\end{equation} and similarly \begin{equation}
\begin{array}{l}
(r!)^{\k} Q_{[i_{(1,1)} [i_{(2,1)} ... [i_{(\k,1)}} \dots Q_{i_{(1,r-1)} i_{(2,r-1)} ... i_{(\k,r-1)}} Q^{i_{(1,r)}] i_{(2,r)}] ... i_{(\k,r)}]}\\
= Q_{[i_{(1,1)} ... i_{(1,r-1)} i_{(1,r)}] ... [i_{(a,1)}' ... i_{(a,r-1)}' i_{(a,r)}'] ...  [i_{(\k,1)} ... i_{(\k,r-1)} i_{(\k,r)}]} (\wt{S}_{(N,\k,r)}^a)^{[i_{(a,1)}' ... i_{(a,r-1)}' i_{(a,r)}']}{}_{[i_{(a,1)} ... i_{(a,r-1)} i_{(a,r)}]}
\end{array}
\end{equation}
for tensors $S_{(N,\k,r)}^a$ and $\wt{S}_{(N,\k,r)}^a$.

The $\k = 3$ theories are known to be related to $T_N$ theories. There are many relations in the literature \cite{LemosPeelaers, TNRev} of the form \be
\label{LPrels}
\begin{array}{l}
	(r!)^{2} Q^{(i_{(1,1)} [i_{(2,1)} [i_{(3,1)}} \dots Q^{i_{(1,r-1)} i_{(2,r-1)} i_{(3,r-1)}} Q^{i_{(1,r)}) i_{(2,r)}] i_{(3,r)}]}\\
	= (P_{(N,r)}^1)^{[i_{(1,1)} ... i_{(1,r-1)} i_{(1,r)}]}{}_{[i_{(1,1)}' ... i_{(1,r-1)}' i_{(1,r)}']}  Q^{[i_{(1,1)}' ... i_{(1,r-1)}' i_{(1,r)}'] ...  [i_{(3,1)} ... i_{(3,r-1)} i_{(3,r)}]}\\
	\hspace{1cm}\\
	(r!)^{2} Q_{(i_{(1,1)} [i_{(2,1)} [i_{(3,1)}} \dots Q_{i_{(1,r-1)} i_{(2,r-1)} i_{(3,r-1)}} Q_{i_{(1,r)}) i_{(2,r)}] i_{(3,r)}]}\\
	= Q_{[i_{(1,1)}' ... i_{(1,r-1)}' i_{(1,r)}'] ...  [i_{(3,1)} ... i_{(3,r-1)} i_{(3,r)}]} (\wt{P}_{(N,r)}^1)^{[i_{(1,1)}' ... i_{(1,r-1)}' i_{(1,r)}']}{}_{[i_{(1,1)} ... i_{(1,r-1)} i_{(1,r)}]}
\end{array}
\ee for $r = 2, 3, ..., N$. There are identical relations when other indices are symmetrized (with the others antisymmetrized) and $P^1$, $\wt{P}^1$ are replaced by an appropriate tensor. Wherever possible, we will compare the form of $P$ with known or conjectured results.

\subsubsection{$N = 2$}
For $N = 2$ the only interesting case of the above relation is for $r = 2$ and so the result must be proportional to $\epsilon^{i_{(1,1)} i_{(1,2)}} ... \epsilon^{i_{(\k,1)} i_{(\k,2)}}$. The general form of the relation should be \be
\begin{array}{c}
	(2!)^{\k}(\tilde{\mu}Q)^{[i_{(1,1)} ... [i_{(\k,1)}} Q^{i_{(1,2)}] ... i_{(\k,2)}]} = (T^1_{(2,\k,2)})^{[i_{(1,1)} i_{(1,2)}]}{}_{[i_{(1,1)}' i_{(1,2)}']} \epsilon^{i_{(1,1)}' i_{(1,2)}'}  ...  \epsilon^{i_{(\k,1)} i_{(\k,2)}}\\
	
	(2!)^{\k}Q_{[i_{(1,1)} ... [i_{(\k,1)}} (Q\tilde{\mu})_{i_{(1,2)}] ... i_{(\k,2)}]} =  \epsilon_{i_{(1,1)}' i_{(1,2)}'}  ...  \epsilon_{i_{(\k,1)} i_{(\k,2)}} (\wt{T}^1_{(2,\k,2)})^{[i_{(1,1)}' i_{(1,2)}']}{}_{[i_{(1,1)} i_{(1,2)}]}\\
\end{array}
\ee and similarly \be
\begin{array}{c}
	(2!)^{\k} Q^{[i_{(1,1)} ... [i_{(\k,1)}} Q^{i_{(1,2)}] ... i_{(\k,2)}]} = (S^1_{(2,\k,2)})^{[i_{(1,1)} i_{(1,2)}]}{}_{[i_{(1,1)}' i_{(1,2)}']} \epsilon^{i_{(1,1)}' i_{(1,2)}'}  ...  \epsilon^{i_{(\k,1)} i_{(\k,2)}}\,\,\\
	
	(2!)^{\k}Q_{[i_{(1,1)} ... [i_{(\k,1)}} Q_{i_{(1,2)}] ... i_{(\k,2)}]} =  \epsilon_{i_{(1,1)}' i_{(1,2)}'}  ...  \epsilon_{i_{(\k,1)} i_{(\k,2)}} (\wt{S}^1_{(2,\k,2)})^{[i_{(1,1)}' i_{(1,2)}']}{}_{[i_{(1,1)} i_{(1,2)}]}\,.
\end{array}
\ee There are typically many expressions for the tensors of interest arising from relations amongst traces of powers of the moment map. We label the ambiguity by the numbers $w,z$. We do not do this for $N= 3$ or $N = 4$.

For $\k =1$ we find \be
\begin{array}{cc}
	2(\tilde{\mu}Q)^{[i_1} Q^{i_2]} = \epsilon^{i_1 i_2} & 2 Q_{[i_1} (Q\tilde{\mu})_{i_2]} = \epsilon_{i_1 i_2}\\
	2 Q^{[i_1} Q^{i_2]} = 0 & 2 Q_{[i_1} Q_{i_2]} = 0\\
\end{array}.
\ee which corresponds to the assignment \be
\begin{array}{cc}
	(T^1_{(2,1,2)})^{[i_1 i_2]}{}_{[i_1' i_2']} = \frac{1}{2}\epsilon^{i_1 i_2} \epsilon_{i_1' i_2'} &  (\wt{T}^2_{(2,1,2)})^{[i_1' i_2']}{}_{[i_1 i_2]} = \frac{1}{2} \epsilon^{i_1' i_2'} \epsilon_{i_1 i_2}\\
	(S^1_{(2,1,2)})^{[i_1 i_2]}{}_{[i_1' i_2']} = 0 &  (\wt{S}^2_{(2,1,2)})^{[i_1' i_2']}{}_{[i_1 i_2]} = 0\,,
\end{array}
\ee whose one independent component is 1 for both $T^1_{(2,1)}$ and $\wt{T}^1_{(2,1)}$. For $\k = 2$ these relations start to become nontrivial: \be
\begin{array}{rl}
	4(\tilde{\mu}Q)^{[i_{(1,1)} [i_{(2,1)}} Q^{i_{(1,2)}] i_{(2,2)}]} & = \ve \epsilon^{i_{(1,1)} i_{(1,2)}} \epsilon^{i_{(2,1)} i_{(2,2)}}\\
	4Q_{[i_{(1,1)} [i_{(2,1)}} (Q\tilde{\mu})_{i_{(1,2)}] i_{(2,2)}]} & = \ve \epsilon_{i_{(1,1)} i_{(1,2)}} \epsilon_{i_{(2,1)} i_{(2,2)}}\\
\end{array}
\ee as well as \be
\begin{array}{c}
	4 Q^{[i_{(1,1)} [i_{(2,1)}} Q^{i_{(1,2)}] i_{(2,2)}]} = -2 \epsilon^{i_{(1,1)} i_{(1,2)}} \epsilon^{i_{(2,1)} i_{(2,2)}}\,\,\\
	4 Q_{[i_{(1,1)} [i_{(2,1)}} Q_{i_{(1,2)}] i_{(2,2)}]} = -2 \epsilon_{i_{(1,1)} i_{(1,2)}} \epsilon_{i_{(2,1)} i_{(2,2)}}\,.
\end{array}
\ee Thus we have relations with \be
\begin{array}{c}
	(T^{1}_{(2,2,2)})^{[i_1 i_2]}{}_{[i_1' i_2']} = (2z-2)\delta^{[i_1}{}_{[i_1'} (\tilde{\mu}_1)^{i_2]}{}_{i_2']} + z \delta^{[i_1}{}_{[i_1'} \delta^{i_2]}{}_{i_2']}= \frac{1}{2} \ve \epsilon^{i_1 i_2} \epsilon_{i_1' i_2'}\\
	(\wt{T}^{1}_{(2,2,2)})^{[i_1' i_2']}{}_{[i_1 i_2]} = (2z-2)\delta^{[i_1'}{}_{[i_1} (\tilde{\mu}_1)^{i_2']}{}_{i_2]} +z \delta^{[i_1'}{}_{[i_1} \delta^{i_2']}{}_{i_2]} = \frac{1}{2} \ve \epsilon^{i_1' i_2'} \epsilon_{i_1 i_2}\,,
\end{array} \ee for any number $z$ as well as \be
\begin{array}{c}
	(S^{1}_{(2,2,2)})^{[i_1 i_2]}{}_{[i_1' i_2']} = - \epsilon^{i_1 i_2} \epsilon_{i_1' i_2'}\\
	(\wt{S}^{1}_{(2,2,2)})^{[i_1' i_2']}{}_{[i_1 i_2]} = - \epsilon^{i_1' i_2'} \epsilon_{i_1 i_2}\,.
\end{array}\ee

For $\k = 3$ there are similar relations with \be
\begin{array}{rl}
	(T^{1}_{(2,3,2)})^{[i_1 i_2]}{}_{[i_1' i_2']} & = \big(4\delta^{[i_1}{}_{[i_1'} (\tilde \mu_1^2)^{i_2]}{}_{i_2']} + (2z-2) \ve \delta^{[i_1}{}_{[i_1'} (\tilde \mu_1)^{i_2]}{}_{i_2']}+ z \ve^2 \delta^{[i_1}{}_{[i_1'} \delta^{i_2]}{}_{i_2']}\big)\\
	& = \frac{1}{2}(-4 c_2 + \ve^2) \epsilon^{i_1 i_2} \epsilon_{i_1' i_2'}\\
	(\wt{T}^{1}_{(2,3,2)})^{[i_1' i_2']}{}_{[i_1 i_2]} & = \big(4\delta^{[i_1'}{}_{[i_1} (\tilde \mu_1^2)^{i_2']}{}_{i_2]}  + (2z-2) \ve \delta^{[i_1'}{}_{[i_1} (\tilde \mu_1)^{i_2']}{}_{i_2]} + z \ve^2 \delta^{[i_1'}{}_{[i_1} \delta^{i_2']}{}_{i_2]}\big)\\
	& = \frac{1}{2}(-4 c_2 + \ve^2) \epsilon^{i_1' i_2'} \epsilon_{i_1 i_2}\\
\end{array}
\ee for any number $z$ and \be
\begin{array}{c}
	(S^{1}_{(2,3,2)})^{[i_1 i_2]}{}_{[i_1' i_2']} = -4\big((2z-2)\delta^{[i_1}{}_{[i_1'} (\tilde{\mu}_1)^{i_2]}{}_{i_2']} +z \delta^{[i_1}{}_{[i_1'} \delta^{i_2]}{}_{i_2']}\big) = -2 \ve \epsilon^{i_1 i_2} \epsilon_{i_1' i_2'}\,\,\\
	(\wt{S}^{1}_{(2,3,2)})^{[i_1' i_2']}{}_{[i_1 i_2]} = -4\big((2z-2)\delta^{[i_1}{}_{[i_1'} (\tilde{\mu}_1)^{i_2]}{}_{i_2']} + z \delta^{[i_1}{}_{[i_1'} \delta^{i_2]}{}_{i_2']}\big) = - 2 \ve \epsilon^{i_1' i_2'} \epsilon_{i_1 i_2}\,.
\end{array}
\ee

For $\k = 4$ the form of the tensors becomes \be
\begin{array}{rl}
	(T^{1}_{(2,4,2)})^{[i_1 i_2]}{}_{[i_1' i_2']} & = \big( (2w-24)\delta^{[i_1}{}_{[i_1'} (\tilde \mu_1^3)^{i_2]}{}_{i_2']} +w \ve \delta^{[i_1}{}_{[i_1'} (\tilde \mu_1^2)^{i_2]}{}_{i_2']}-2 \ve^2 \delta^{[i_1}{}_{[i_1'} (\tilde \mu_1)^{i_2]}{}_{i_2']}\big)\\
	& = -\frac{1}{2} \ve (12 c_2 - \ve^2) \epsilon^{i_1 i_2} \epsilon_{i_1' i_2'} \\
	(\wt{T}^{1}_{(2,4,2)})^{[i_1' i_2']}{}_{[i_1 i_2]} & = \big( (2w-24)\delta^{[i_1'}{}_{[i_1} (\tilde \mu_1^3)^{i_2']}{}_{i_2]}+w \ve \delta^{[i_1'}{}_{[i_1} (\tilde \mu_1^2)^{i_2']}{}_{i_2]}-2 \ve^2 \delta^{[i_1'}{}_{[i_1} (\tilde \mu_1)^{i_2']}{}_{i_2]}\big)\\
	& = -\frac{1}{2} \ve (12 c_2 - \ve^2) \epsilon^{i_1' i_2'} \epsilon_{i_1 i_2} \\
\end{array}
\ee for any numbers $w,z$ and \be 
\begin{array}{rl}
	(S^{1}_{(2,4,2)})^{[i_1 i_2]}{}_{[i_1' i_2']} & = \big(-4\delta^{[i_1}{}_{[i_1'} (\tilde \mu_1^2)^{i_2]}{}_{i_2']} + (2z+6) \ve \delta^{[i_1}{}_{[i_1'} (\tilde \mu_1)^{i_2]}{}_{i_2']}+ z \ve^2 \delta^{[i_1}{}_{[i_1'} \delta^{i_2]}{}_{i_2']}\big)\\
	& = \frac{1}{2}(4 c_2 - 3 \ve^2) \epsilon^{i_1 i_2} \epsilon_{i_1' i_2'} \,\, \\
	(\wt{S}^{1}_{(2,4,2)})^{[i_1' i_2']}{}_{[i_1 i_2]} & = \big(-4\delta^{[i_1'}{}_{[i_1} (\tilde \mu_1^2)^{i_2']}{}_{i_2]}  + (2z+6) \ve \delta^{[i_1'}{}_{[i_1} (\tilde \mu_1)^{i_2']}{}_{i_2]} + z \ve^2 \delta^{[i_1'}{}_{[i_1} \delta^{i_2']}{}_{i_2]}\big)\\
	& = \frac{1}{2}(4 c_2 - 3 \ve^2) \epsilon^{i_1' i_2'} \epsilon_{i_1 i_2} \,.
\end{array}
\ee

For $\k = 5$ the form of the tensors becomes \be
\begin{array}{rl}
	(T^{1}_{(2,5,2)})^{[i_1 i_2]}{}_{[i_1' i_2']} & = 16\delta^{[i_1}{}_{[i_1'}(\tilde{\mu}_1^4)^{i_2]}{}_{i_2']} + (2 w - 48)\ve \delta^{[i_1}{}_{[i_1'}(\tilde{\mu}_1^3)^{i_2]}{}_{i_2']} + w \ve ^2 \delta^{[i_1}{}_{[i_1'}(\tilde{\mu}_1^2)^{i_2]}{}_{i_2']}\\ 
	& + (2 z - 2)\ve^3 \delta^{[i_1}{}_{[i_1'}(\tilde{\mu}_1)^{i_2]}{}_{i_2']} + z \ve ^4 \delta^{[i_1}{}_{[i_1'}\delta^{i_2]}{}_{i_2']}\\ 
	& = \frac{1}{2}(16 c_2^2 -24 \ve^2 c_2 + \ve^4) \epsilon^{i_1 i_2} \epsilon_{i_1' i_2'} \\
	\hspace{1cm}\\
	(\wt{T}^{1}_{(2,5,2)})^{[i_1' i_2']}{}_{[i_1 i_2]}&  16\delta^{[i_1'}{}_{[i_1}(\tilde{\mu}_1^4)^{i_2']}{}_{i_2]} + (2 w - 48)\ve \delta^{[i_1'}{}_{[i_1}(\tilde{\mu}_1^3)^{i_2']}{}_{i_2]} + w \ve ^2 \delta^{[i_1'}{}_{[i_1}(\tilde{\mu}_1^2)^{i_2']}{}_{i_2]}\\ 
	& + (2 z - 2)\ve^3 \delta^{[i_1'}{}_{[i_1}(\tilde{\mu}_1)^{i_2']}{}_{i_2]} + z \ve ^4 \delta^{[i_1'}{}_{[i_1}\delta^{i_2']}{}_{i_2]}\\ 
	& = \frac{1}{2}(16 c_2^2 -24 \ve^2 c_2 + \ve^4) \epsilon^{i_1' i_2'} \epsilon_{i_1 i_2} \\
\end{array}
\ee for any numbers $w,z$ and \be 
\begin{array}{rl}
	(S^{1}_{(2,5,2)})^{[i_1 i_2]}{}_{i_1' i_2'} & =\big( (2w+32)\delta^{[i_1}{}_{[i_1'} (\tilde \mu_1^3)^{i_2]}{}_{i_2']} +w \ve \delta^{[i_1}{}_{[i_1'} (\tilde \mu_1^2)^{i_2]}{}_{i_2']} + (2z+8) \ve^2 \delta^{[i_1}{}_{[i_1'} (\tilde \mu_1)^{i_2]}{}_{i_2']}+z \ve^3 \delta^{[i_1}{}_{[i_1'} \delta^{i_2]}{}_{i_2']}\big)\\
	& = \ve(16 c_2 - 4 \ve^2) \epsilon^{i_1 i_2} \epsilon_{i_1' i_2'} \,\, \\
	(\wt{S}^{1}_{(2,5,2)})^{i_1' i_2'}{}_{[i_1 i_2]} & =\big( (2w+32)\delta^{[i_1'}{}_{[i_1} (\tilde \mu_1^3)^{i_2']}{}_{i_2]} +w \ve \delta^{[i_1'}{}_{[i_1} (\tilde \mu_1^2)^{i_2']}{}_{i_2]} + (2z+8) \ve^2 \delta^{[i_1'}{}_{[i_1} (\tilde \mu_1)^{i_2']}{}_{i_2]}+z \ve^3 \delta^{[i_1'}{}_{[i_1} \delta^{i_2']}{}_{i_2]}\big)\\
	& = \ve(16 c_2 - 4 \ve^2) \epsilon^{i_1' i_2'} \epsilon_{i_1 i_2} \,.
\end{array}
\ee

Finally, the form of \eqref{LPrels} for $N = 2$ is given by \be
\label{LPrels23}
\begin{array}{l}
	(2)^{2} Q^{(i_{(1,1)} [i_{(2,1)} [i_{(3,1)}}  Q^{i_{(1,2)}) i_{(2,2)}] i_{(3,2)}]}	= -2 \delta^{(i_{(1,1)}}{}_{[i_{(1,1)'}} (\tilde{\mu}_1)^{i_{(1,2)})}{}_{i_{(1,2)}']} \epsilon^{i_{(1,1)}' i_{(1,2)}'} \epsilon^{i_{(2,1)} i_{(2,2)}} \epsilon^{i_{(3,1)} i_{(3,2)}}\\
	\hspace{1cm}\\
	(2)^{2} Q_{(i_{(1,1)} [i_{(2,1)} [i_{(3,1)}} Q_{i_{(1,2)}) i_{(2,2)}] i_{(3,2)}]} = 2 \delta^{[i_{(1,1)}'}{}_{(i_{(1,1)}} (\tilde{\mu}_1)^{i_{(1,2)}']}{}_{i_{(1,2)})}\epsilon_{i_{(1,1)}' i_{(1,2)}'} \epsilon_{i_{(2,1)} i_{(2,2)}} \epsilon_{i_{(3,1)} i_{(3,2)}}
\end{array}.
\ee

We also find a similar relation for $\k = 5$: \be
\label{PLrels25}
\begin{array}{l}
	(2)^{4} Q^{(i_{(1,1)} [i_{(2,1)} [i_{(3,1)} [i_{(4,1)} [i_{(5,1)}}  Q^{i_{(1,2)}) i_{(2,2)}] i_{(3,2)}]i_{(4,2)}] i_{(5,2)}]}\\
	= -\delta^{(i_{(1,1)}}{}_{[i_{(1,1)'}} (8\tilde{\mu}_1^3 + 6 \ve^2 \tilde{\mu}_1)^{i_{(1,2)})}{}_{i_{(1,2)}']} \epsilon^{i_{(1,1)}' i_{(1,2)}'} \epsilon^{i_{(2,1)} i_{(2,2)}} \epsilon^{i_{(3,1)} i_{(3,2)}} \epsilon^{i_{(4,1)} i_{(4,2)}} \epsilon^{i_{(5,1)} i_{(5,2)}}\\
	\hspace{1cm}\\
	(2)^{4} Q_{(i_{(1,1)} [i_{(2,1)} [i_{(3,1)} [i_{(4,1)} [i_{(5,1)}}  Q_{i_{(1,2)}) i_{(2,2)}] i_{(3,2)}]i_{(4,2)}] i_{(5,2)}]}\\
	= \delta^{[i_{(1,1)}'}{}_{(i_{(1,1)}} (8\tilde{\mu}_1^3 + 6 \ve^2 \tilde{\mu}_1)^{i_{(1,2)}']}{}_{i_{(1,2)})} \epsilon_{i_{(1,1)}' i_{(1,2)}'} \epsilon_{i_{(2,1)} i_{(2,2)}} \epsilon_{i_{(3,1)} i_{(3,2)}} \epsilon_{i_{(4,1)} i_{(4,2)}} \epsilon_{i_{(5,1)} i_{(5,2)}}
\end{array}.
\ee

\subsubsection{$N = 3$}
For $N = 3$ there are two nontrivial $r$'s of interest.

For $r = 2$ the general relation is given by \be
\begin{array}{c}
	(2!)^{\k}(\tilde{\mu}Q)^{[i_{(1,1)} ... [i_{(\k,1)}} Q^{i_{(1,2)}] ... i_{(\k,2)}]} = (T^1_{(3,\k,2)})^{[i_{(1,1)} i_{(1,2)}]}{}_{[i_{(1,1)}' i_{(1,2)}']} Q^{[i_{(1,1)}' i_{(1,2)}'] ... [i_{(\k,1)} i_{(\k,2)}]}\\
	
	(2!)^{\k}Q_{[i_{(1,1)} ... [i_{(\k,1)}} (Q\tilde{\mu})_{i_{(1,2)}] ... i_{(\k,2)}]} =  Q_{[i_{(1,1)}' i_{(1,2)}'] ... [i_{(\k,1)} i_{(\k,2)}]} (\wt{T}^1_{(3,\k,2)})^{[i_{(1,1)}' i_{(1,2)}']}{}_{[i_{(1,1)} i_{(1,2)}]}\\
\end{array}
\ee

\noindent We find the following relations at $\k = 1$: \be
\begin{array}{cc}
	2 (\tilde{\mu}Q)^{[i_1} Q^{i_2]} = - Q^{[i_1 i_2]} & 2Q_{[i_1} (Q\tilde{\mu})_{i_2]} = Q_{[i_1 i_2]}\\
	2 Q^{[i_1} Q^{i_2]} = 0 & 2Q_{[i_1} Q_{i_2]} = 0\,,
\end{array}
\ee which corresponds to the tensors \be
(T^1_{(3,1,2)})^{[i_1 i_2]}{}_{[i_1' i_2']} = -\delta^{[i_1}{}_{[i_1'} \delta^{i_2]}{}_{i_2']} \hspace{1cm} (\wt{T}^1_{(3,1,2)})^{[i_1' i_2']}{}_{[i_1 i_2]} = \delta^{[i_1'}{}_{[i_1} \delta^{i_2']}{}_{i_2]}
\ee and \be
(S^1_{(3,1,2)})^{[i_1 i_2]}{}_{[i_1' i_2']} = 0 \hspace{1cm} (\wt{S}^1_{(3,1,2)})^{[i_1' i_2']}{}_{[i_1 i_2]} = 0\,.
\ee

\noindent The $\k = 2$ versions of these tensors are \be
\begin{array}{c}
	(T^1_{(3,2,2)})^{[i_1 i_2]}{}_{[i_1' i_2']} = 2\delta^{[i_1}{}_{[i_1'} (\tilde{\mu}_1)^{i_2]}{}_{i_2']}\\
	(\wt{T}^1_{(3,2,2)})^{[i_1' i_2']}{}_{[i_1 i_2]} = 2 \delta^{[i_1'}{}_{[i_1} (\tilde{\mu}_1)^{i_2']}{}_{i_2]} \\
\end{array}
\ee and \be
\begin{array}{c}
	(S^1_{(3,2,2)})^{[i_1 i_2]}{}_{[i_1' i_2']} = -2\delta^{[i_1}{}_{[i_1'} \delta^{i_2]}{}_{i_2']}\,\,\\
	(\wt{S}^1_{(3,2,2)})^{[i_1' i_2']}{}_{[i_1 i_2]} = -2 \delta^{[i_1'}{}_{[i_1} \delta^{i_2']}{}_{i_2]} \,.
\end{array}
\ee

\noindent For $\k = 3$ they become
\be
\begin{array}{c}
	(T^1_{(3,3,2)})^{[i_1 i_2]}{}_{[i_1' i_2']} = \big( 2\delta^{[i_1}{}_{[i_1'} (\tilde{\mu}_1^2)^{i_2]}{}_{i_2']} - 2 (\tilde{\mu}_1)^{[i_1}{}_{[i_1'} (\tilde{\mu}_1)^{i_2]}{}_{i_2']} -4\varepsilon\delta^{[i_1}{}_{[i_1'} (\tilde{\mu}_1)^{i_2]}{}_{i_2']} \big)\\
	(\wt{T}^1_{(3,3,2)})^{[i_1' i_2']}{}_{[i_1 i_2]} = \big( 2\delta^{[i_1'}{}_{[i_1} (\tilde{\mu}_1^2)^{i_2']}{}_{i_2]} - 2 (\tilde{\mu}_1)^{[i_1'}{}_{[i_1} (\tilde{\mu}_1)^{i_2']}{}_{i_2]} -4\varepsilon\delta^{[i_1'}{}_{[i_1} (\tilde{\mu}_1)^{i_2']}{}_{i_2]} \big)\\
\end{array}
\ee and \be \label{Stensor3}
\begin{array}{c}
	(S^1_{(3,3,2)})^{[i_1 i_2]}{}_{[i_1' i_2']} = 4 \ve \delta^{[i_1}{}_{[i_1'} \delta^{i_2]}{}_{i_2']}\,\,\\
	(\wt{S}^1_{(3,3,2)})^{[i_1' i_2']}{}_{[i_1 i_2]} = 4 \ve \delta^{[i_1'}{}_{[i_1} \delta^{i_2']}{}_{i_2]} \,.
\end{array}
\ee
The relations encoded by \eqref{Stensor3} reduce to the $T_3$ relations \eqref{muQI} in the classical $\ve\to 0$ limit, after contracting with an additional Levi-Civita tensor.

The general $r = 3$ relations can be written as \be
\begin{array}{l}
	(3!)^{\k} (\tilde{\mu}^2Q)^{[i_{(1,1)} ... [i_{(\k,1)}} (\tilde{\mu}Q)^{i_{(1,2)} ... i_{(\k,2)}} Q^{i_{(1,3)}] ... i_{(\k,3)}]} \\
	\hspace{1cm} = (T^1_{(3,\k,3)})^{[i_{(1,1)} i_{(1,2)} i_{(1,3)}]}{}_{[i_{(1,1)}' i_{(1,2)}' i_{(1,3)}']} \epsilon^{[i_{(1,1)}' i_{(1,2)}' i_{(1,3)}']} ... \epsilon^{[i_{(\k,1)} i_{(\k,2)} i_{(\k,3)}]}\\
	(3!)^{\k} Q_{[i_{(1,1)} ... [i_{(\k,1)}} (Q\tilde{\mu})_{i_{(1,2)} ... i_{(\k,2)}} (Q \tilde{\mu}^2)^{i_{(1,3)}] ... i_{(\k,3)}]} \\
	\hspace{1cm} = \epsilon_{[i_{(1,1)}' i_{(1,2)}' i_{(1,3)}']} ... \epsilon_{[i_{(\k,1)} i_{(\k,2)} i_{(\k,3)}]} (\wt{T}^1_{(3,\k,3)})^{[i_{(1,1)}' i_{(1,2)}' i_{(1,3)}']}{}_{[i_{(1,1)} i_{(1,2)} i_{(1,3)}]} 
\end{array}
\ee and \be
\begin{array}{l}
	(3!)^{\k} Q^{[i_{(1,1)} ... [i_{(\k,1)}} Q^{i_{(1,2)} ... i_{(\k,2)}} Q^{i_{(1,3)}] ... i_{(\k,3)}]} \\
	\hspace{1cm} = (S^1_{(3,\k,3)})^{[i_{(1,1)} i_{(1,2)} i_{(1,3)}]}{}_{[i_{(1,1)}' i_{(1,2)}' i_{(1,3)}']} \epsilon^{[i_{(1,1)}' i_{(1,2)}' i_{(1,3)}']} ... \epsilon^{[i_{(\k,1)} i_{(\k,2)} i_{(\k,3)}]}\\
	(3!)^{\k} Q_{[i_{(1,1)} ... [i_{(\k,1)}} Q_{i_{(1,2)} ... i_{(\k,2)}} Q_{i_{(1,3)}] ... i_{(\k,3)}]} \\
	\hspace{1cm} = \epsilon_{[i_{(1,1)}' i_{(1,2)}' i_{(1,3)}']} ... \epsilon_{[i_{(\k,1)} i_{(\k,2)} i_{(\k,3)}]} (\wt{S}^1_{(3,\k,3)})^{[i_{(1,1)}' i_{(1,2)}' i_{(1,3)}']}{}_{[i_{(1,1)} i_{(1,2)} i_{(1,3)}]} 
\end{array}.
\ee

\noindent At $\k = 1$, the tensors are given by \be
\begin{array}{c}
	(T^1_{(3,1,3)})^{[i_1 i_2 i_3]}{}_{[i_1' i_2' i_3']} = \frac{1}{3!}\epsilon^{i_1 i_2 i_3} \epsilon_{i_1' i_2' i_3'}\\
	(\wt{T}^1_{(3,1,3)})^{[i_1' i_2' i_3']}{}_{[i_1 i_2 i_3]} = \frac{1}{3!}\epsilon^{i_1' i_2' i_3'} \epsilon_{i_1 i_2 i_3}\\
\end{array}
\ee and \be 
(S^1_{(3,1,3)})^{[i_1 i_2 i_3]}{}_{[i_1' i_2' i_3']} = 0 \hspace{1cm} (\wt{S}^1_{(3,1,3)})^{[i_1' i_2' i_3']}{}_{[i_1 i_2 i_3]} = 0\,.
\ee

\noindent At $\k = 2$, the tensors are given by \be
\begin{array}{c}
	(T^1_{(3,2,3)})^{[i_1 i_2 i_3]}{}_{[i_1' i_2' i_3']} = \frac{1}{3!}(3 c_3 + \frac{7}{3} \ve c_2 -2 \ve^3)\epsilon^{i_1 i_2 i_3} \epsilon_{i_1' i_2' i_3'}\\
	(\wt{T}^1_{(3,2,3)})^{[i_1' i_2' i_3']}{}_{[i_1 i_2 i_3]} = -\frac{1}{3!}(3 c_3 + \frac{7}{3} \ve c_2 -2 \ve^3)\epsilon^{i_1' i_2' i_3'} \epsilon_{i_1 i_2 i_3}\\
\end{array}
\ee and \be
\begin{array}{c}
	(S^1_{(3,2,3)})^{[i_1 i_2 i_3]}{}_{[i_1' i_2' i_3']} = -\epsilon^{i_1 i_2 i_3} \epsilon_{i_1' i_2' i_3'}\\
	(\wt{S}^1_{(3,2,3)})^{[i_1' i_2' i_3']}{}_{[i_1 i_2 i_3]} = -\epsilon^{i_1' i_2' i_3'} \epsilon_{i_1 i_2 i_3}\,.
\end{array}
\ee

\noindent At $\k = 3$, the tensors are given by \be
\begin{array}{c}
	(T^1_{(3,3,3)})^{[i_1 i_2 i_3]}{}_{[i_1' i_2' i_3']} = -\frac{1}{3!}(27 c_3^2 + 4 c_2^3)\epsilon^{i_1 i_2 i_3} \epsilon_{i_1' i_2' i_3'}\\
	(\wt{T}^1_{(3,3,3)})^{[i_1' i_2' i_3']}{}_{[i_1 i_2 i_3]} = \frac{1}{3!}(27 c_3^2 + 4 c_2^3)\epsilon^{i_1' i_2' i_3'} \epsilon_{i_1 i_2 i_3}\\
\end{array}
\ee and \be
\begin{array}{c}
	(S^1_{(3,3,3)})^{[i_1 i_2 i_3]}{}_{[i_1' i_2' i_3']} = \frac{1}{3!}(12 \ve c_2 +72 \ve^3)\epsilon^{i_1 i_2 i_3} \epsilon_{i_1' i_2' i_3'}\\
	(\wt{S}^1_{(3,3,3)})^{[i_1' i_2' i_3']}{}_{[i_1 i_2 i_3]} = \frac{1}{3!}(12 \ve c_2 +72 \ve^3)\epsilon^{i_1' i_2' i_3'} \epsilon_{i_1 i_2 i_3}\,.
\end{array}
\ee

The relations akin to \eqref{LPrels} occur at $r = 2, 3$. We find that for $r = 2$ \be
\label{LPrels33a}
\begin{array}{l}
	(2)^{2} Q^{(i_{(1,1)} [i_{(2,1)} [i_{(3,1)}}  Q^{i_{(1,2)}) i_{(2,2)}] i_{(3,2)}]}	= -2 \delta^{(i_{(1,1)}}{}_{[i_{(1,1)'}} (\tilde{\mu}_1)^{i_{(1,2)})}{}_{i_{(1,2)}']} Q^{[i_{(1,1)}' i_{(1,2)}'] [i_{(2,1)} i_{(2,2)}] [i_{(3,1)} i_{(3,2)}]}\\
	\hspace{1cm}\\
	(2)^{2} Q_{(i_{(1,1)} [i_{(2,1)} [i_{(3,1)}} Q_{i_{(1,2)}) i_{(2,2)}] i_{(3,2)}]} = 2 Q_{[i_{(1,1)}' i_{(1,2)}'] [i_{(2,1)} i_{(2,2)}] [i_{(3,1)} i_{(3,2)}]} \delta^{[i_{(1,1)}'}{}_{(i_{(1,1)}} (\tilde{\mu}_1)^{i_{(1,2)}']}{}_{i_{(1,2)})}
\end{array}.
\ee and for $r = 3$ we find \be
\label{LPrels33b}
\begin{array}{l}
	(6)^{2} Q^{(i_{(1,1)} [i_{(2,1)} [i_{(3,1)}}  Q^{i_{(1,2)} i_{(2,2)} i_{(3,2)}} Q^{i_{(1,3)}) i_{(2,3)}] i_{(3,3)}]}\\
	= -6 \delta^{(i_{(1,1)}}{}_{[i_{(1,1)'}} (\tilde{\mu}_1)^{i_{(1,2)}}{}_{i_{(1,2)}'} (\tilde{\mu}_1^2)^{i_{(1,3)})}{}_{i_{(1,3)}']} \epsilon^{i_{(1,1)}' i_{(1,2)}' i_{(1,3)}'} \epsilon^{i_{(2,1)} i_{(2,2)} i_{(2,3)}} \epsilon^{i_{(3,1)} i_{(3,2)} i_{(3,3)}}\\
	\hspace{1cm}\\
	(6)^{2} Q_{(i_{(1,1)} [i_{(2,1)} [i_{(3,1)}} Q_{i_{(1,2)} i_{(2,2)} i_{(3,2)}} Q^{i_{(1,3)}) i_{(2,3)}] i_{(3,3)}]}\\
	= -6 \delta^{[i_{(1,1)}'}{}_{(i_{(1,1)}} (\tilde{\mu}_1)^{i_{(1,2)}'}{}_{i_{(1,2)}} (\tilde{\mu}_1^2)^{i_{(1,3)}']}{}_{i_{(1,3)})} \epsilon_{i_{(1,1)}' i_{(1,2)}' i_{(1,3)}'} \epsilon_{i_{(2,1)} i_{(2,2)} i_{(2,3)}} \epsilon_{i_{(3,1)} i_{(3,2)} i_{(3,3)}}\\
\end{array}.
\ee

\subsubsection{$N = 4$}
\label{app:oneleg}

There are three ranks of tensor relations of interest for $N = 4$ theories, $r = 2, 3, 4$. We only have complete data for $\k = 1$. In this case, at $r = 2$ we find \be
2! (\tilde{\mu}Q)^{[i_1} Q^{i_2]} = -Q^{[i_1 i_2]} \hspace{1cm} 2! Q_{[i_1} (Q\tilde{\mu})_{i_2]} = -Q_{[i_1 i_2]}
\ee and \be
2! Q^{[i_1} Q^{i_2]} =  0 \hspace{1cm} 2! Q_{[i_1} Q_{i_2]} = 0\,.
\ee The $r = 3$ relations are given by \be
3! (\tilde{\mu}^2 Q)^{[i_1} (\tilde{\mu}Q)^{i_2} Q^{i_3]} = Q^{[i_1 i_2 i_3]} \hspace{1cm} 3! Q_{[i_1} (Q\tilde{\mu})_{i_2} (Q\tilde{\mu}^2)_{i_3]} = Q_{[i_1 i_2 i_3]}
\ee and \be
3! Q^{[i_1} Q^{i_2} Q^{i_3]} = 0 \hspace{1cm} 3! Q_{[i_1} Q_{i_2} Q_{i_3]} = 0\,.
\ee Finally, the $r = 4$ relations are given by \be
4! (\tilde{\mu}^3 Q)^{[i_1} (\tilde{\mu}^2 Q)^{i_2} (\tilde{\mu}Q)^{i_3} Q^{i_4]} =  \epsilon^{i_1 i_2 i_3 i_4} \hspace{1cm} 4! Q_{[i_1} (Q\tilde{\mu})_{i_2} (Q\tilde{\mu}^2)_{i_3} (Q\tilde{\mu}^3)_{i_4]} =  \epsilon_{i_1 i_2 i_3 i_4}
\ee and \be
4! Q^{[i_1} Q^{i_2} Q^{i_3} Q^{i_4]} =  0 \hspace{1cm} 4! Q_{[i_1} Q_{i_2} Q_{i_3} Q_{i_4]} =  0\,.
\ee

For $\k = 2$ we have found that \be
(2!)^2 Q^{[i_1 [i_2} Q^{j_1] j_2]} = -2 Q^{[i_1 j_1] [i_2 j_2]} \hspace{1cm} (2!)^2 Q_{[i_1 [i_2} Q_{j_1] j_2]} = -2 Q_{[i_1 j_1] [i_2 j_2]}
\ee
as well as \be
(3!)^2 Q^{[i_1 [i_2} Q^{j_1 j_2} Q^{k_1] k_2]} = -6 Q^{[i_1 j_1 k_1] [i_2 j_2 k_2]} \hspace{1cm} (3!)^2 Q_{[i_1 [i_2} Q_{j_1 j_2} Q_{k_1] k_2]} = -6 Q_{[i_1 j_1 k_1] [i_2 j_2 k_2]}
\ee and \be
(4!)^2 Q^{[i_1 [i_2} Q^{j_1 j_2} Q^{k_1 k_2} Q^{l_1] l_2]} = 24 \epsilon^{i_1 j_1 k_1 l_1} \epsilon^{i_2 j_2 k_2 l_2} \hspace{1cm} (4!)^2 Q_{[i_1 [i_2} Q_{j_1 j_2} Q_{k_1 k_2} Q_{l_1] l_2]}= 24 \epsilon_{i_1 j_1 k_1 l_1} \epsilon_{i_2 j_2 k_2 l_2}.
\ee

Additionally, we find the relations corresponding to the tensors:\be
\begin{array}{c}
	(T^1_{(4,2,2)})^{[i_1 i_2]}{}_{[i_1' i_2']} = 2\delta^{[i_1}{}_{[i_1'} (\tilde{\mu}_1)^{i_2]}{}_{i_2']}\\
	(\wt{T}^1_{(4,2,2)})^{[i_1' i_2']}{}_{[i_1 i_2]} = 2 \delta^{[i_1'}{}_{[i_1} (\tilde{\mu}_1)^{i_2']}{}_{i_2]} \\
\end{array}
\ee
Finally, for $\k = 3$ we find relations with\be
\begin{array}{c}
	(S^1_{(4,3,2)})^{[i_1 i_2]}{}_{[i_1' i_2']} =4 \delta^{[i_1}{}_{[i_1'} \delta^{i_2]}{}_{i_2']}\\
	(\wt{S}^1_{(4,3,2)})^{[i_1' i_2']}{}_{[i_1 i_2]} = 4 \delta^{[i_1'}{}_{[i_1} \delta^{i_2']}{}_{i_2]} \\
\end{array}
\ee

We also find the relations \be
\label{LPrels43a}
\begin{array}{l}
	(2)^{2} Q^{(i_{(1,1)} [i_{(2,1)} [i_{(3,1)}}  Q^{i_{(1,2)}) i_{(2,2)}] i_{(3,2)}]}	= -2 \delta^{(i_{(1,1)}}{}_{[i_{(1,1)'}} (\tilde{\mu}_1)^{i_{(1,2)})}{}_{i_{(1,2)}']} Q^{[i_{(1,1)}' i_{(1,2)}'] [i_{(2,1)} i_{(2,2)}] [i_{(3,1)} i_{(3,2)}]}\\
	\hspace{1cm}\\
	(2)^{2} Q_{(i_{(1,1)} [i_{(2,1)} [i_{(3,1)}} Q_{i_{(1,2)}) i_{(2,2)}] i_{(3,2)}]} = 2 Q_{[i_{(1,1)}' i_{(1,2)}'] [i_{(2,1)} i_{(2,2)}] [i_{(3,1)} i_{(3,2)}]} \delta^{[i_{(1,1)}'}{}_{(i_{(1,1)}} (\tilde{\mu}_1)^{i_{(1,2)}']}{}_{i_{(1,2)})}
\end{array}
\ee and \be
\label{LPrels43b}
\begin{array}{l}
	(3!)^{2} Q^{(i_{(1,1)} [i_{(2,1)} [i_{(3,1)}}  Q^{i_{(1,2)} i_{(2,2)} i_{(3,2)}} Q^{i_{(1,3)}) i_{(2,3)}] i_{(3,3)}]}\\
	= -6 \delta^{(i_{(1,1)}}{}_{[i_{(1,1)'}} (\tilde{\mu}_1)^{i_{(1,2)}}{}_{i_{(1,2)}'} (\tilde{\mu}_1^2)^{i_{(1,3)})}{}_{i_{(1,3)}']} Q^{[i_{(1,1)}' i_{(1,2)}' i_{(1,3)}'] [i_{(2,1)} i_{(2,2)} i_{(2,3)}] [i_{(3,1)} i_{(3,2)} i_{(3,3)}]}\\
	\hspace{1cm}\\
	(3!)^{2} Q_{(i_{(1,1)} [i_{(2,1)} [i_{(3,1)}} Q_{i_{(1,2)} i_{(2,2)} i_{(3,2)}} Q_{i_{(1,3)}) i_{(2,3)}] i_{(3,3)}]}\\
	= - 6 Q_{[i_{(1,1)}' i_{(1,2)}' i_{(1,3)}'] [i_{(2,1)} i_{(2,2)} i_{(2,3)}] [i_{(3,1)} i_{(3,2)} i_{(3,3)}]} \delta^{[i_{(1,1)}'}{}_{(i_{(1,1)}} (\tilde{\mu}_1)^{i_{(1,2)}'}{}_{i_{(1,2)}} (\tilde{\mu}_1^2)^{i_{(1,3)}']}{}_{i_{(1,3)})}
\end{array}.
\ee

\cleardoublepage

\bibliographystyle{JHEP_TD}
\bibliography{Coulomb}

\end{document}